\documentclass[aps,floatfix,twocolumn,prl]{revtex4}
\usepackage[latin9]{inputenc}
\usepackage{amsmath}
\usepackage{amssymb}
\usepackage{graphicx}
\usepackage{esint}
\usepackage{epstopdf}
\usepackage{braket}
\usepackage{color}
\usepackage{hyperref}
\usepackage{graphicx}
\usepackage{stackengine}
\usepackage{multibib}
\newcites{Supp}{Supp}
\makeatletter
\@ifundefined{textcolor}{}
{
 \definecolor{BLACK}{gray}{0}
 \definecolor{WHITE}{gray}{1}
 \definecolor{RED}{rgb}{1,0,0}
 \definecolor{GREEN}{rgb}{0,1,0}
 \definecolor{BLUE}{rgb}{0,0,1}
 \definecolor{CYAN}{cmyk}{1,0,0,0}
 \definecolor{MAGENTA}{cmyk}{0,1,0,0}
 \definecolor{YELLOW}{cmyk}{0,0,1,0}
}

\makeatother

\begin{document}

\title{Continuous time crystal from a spontaneous many-body Floquet state}

\author{J. R. M. de Nova}
\affiliation{Departamento de F\'isica de Materiales, Universidad Complutense de
Madrid, E-28040 Madrid, Spain}

\author{F. Sols}
\affiliation{Departamento de F\'isica de Materiales, Universidad Complutense de Madrid, E-28040 Madrid, Spain}

\begin{abstract}
Floquet driven systems represent an extremely interesting arena to study out-of-equilibrium phenomena. For instance, they provide realizations of discrete time crystals, where the discrete time translation symmetry of the periodic Hamiltonian is spontaneously broken by a subharmonic response of the system. However, the continuous presence of an external periodic driving is required within the current Floquet paradigm. We propose here the concept of spontaneous many-body Floquet state. This is a state that, in the absence of external periodic driving, self-oscillates like in the presence of a periodic Hamiltonian, this behavior being spontaneously induced by many-body interactions. In addition, its quantum fluctuations are described by regular Floquet theory. Furthermore, it is also a time crystal, presenting long-range time-periodic order. However, this crystalline behavior is very different to that of conventional Floquet discrete time crystals: here, there is no external periodic driving, energy is conserved, and the nature of the spontaneous symmetry breaking is continuous instead of discrete. We demonstrate that spontaneous many-body Floquet states can emerge in a variety of canonical many-body problems, ranging from interacting fermions to Bose-Hubbard models. We specifically show that a spontaneous many-body Floquet state is a universal intrinsic state of a one-dimensional flowing atom condensate, both subsonic and supersonic, resulting from a dynamical phase transition and robust against external perturbations and quantum fluctuations, proposing also realistic experimental scenarios for its observation. A spontaneous many-body Floquet state not only represents a realization of a continuous time crystal, but also a novel paradigm in Floquet physics.
\end{abstract}

\date{\today}

\maketitle

\textit{Introduction.}---Floquet driven systems \cite{Shirley1965,Sambe1973,Grifoni1998} provide rich scenarios to study out-of-equilibrium phenomena such as prethermalization \cite{Peng2021}, topological insulation \cite{Lindner2011}, dynamical phase transitions \cite{Prosen2011}, high-harmonic generation \cite{Murakami2018} or protected cat states \cite{Pieplow2019}. An important application of Floquet physics arises in the context of time crystals, which were originally proposed as systems displaying a non-trivial periodic motion at equilibrium \cite{Wilczek2012}. However, it was later shown that the presented example of time crystal, consisting of a moving soliton in a superconducting ring under a non-trivial magnetic flux \cite{Wilczek2012}, was not the actual ground state \cite{Bruno2013}. Eventually, a no-go theorem ruling out time crystals as first conceived was proven by Watanabe and Oshikawa \cite{Watanabe2015}, using a more precise definition of time crystal: an equilibrium state that exhibits long-range time-periodic order. 

As a result, time crystals require out-of-equilibrium scenarios, like Floquet driven systems, where the spontaneous continuous symmetry breaking is reduced to a discrete one, manifested as a subharmonic response to the periodic driving \cite{Sacha2015,Else2016}. Floquet discrete time crystals have been observed in nitrogen-vacancy centers \cite{Choi2017,Randall2021}, ion chains \cite{Zhang2017,Kyprianidis2021}, dipolar crystals \cite{Rovny2018}, atom condensates \cite{Smits2018}, and very recently in superconducting quantum computers \cite{Google2022,Frey2022}. A related phenomenon is the time quasicrystal observed in magnon condensates, signaled by an incommensurate periodic response to the external driving \cite{Autti2018,Kreil2019,Trager2021}.




Proposals for continuous time crystals evade the no-go theorem by considering long-range interactions \cite{Kozin2019}, or excited eigenstates \cite{Syrwid2017}. Dissipative \cite{Buca2019,Booker2020} or boundary time crystals are also alternatives \cite{Iemini2018}. A recent proposal of time crystal in a condensate within an interacting gauge model  \cite{Ohberg2019} has raised some ongoing debate \cite{Syrwid2020Comment,Syrwid2020,Ohberg2020comment}. An alternative definition of time crystal is based on the observation of periodic oscillations of many-body observables when single-body observables have already relaxed to stationarity \cite{Medenjak2020}. Moreover, classical time crystals are possible \cite{Shapere2012,Hurtado2020}. 




Here we propose the concept of spontaneous many-body Floquet (SMBF) state, a variational state whose original Hamiltonian is time-independent, but in which many-body interactions spontaneously set the effective Hamiltonian to be periodic, and self-consistently the quantum state oscillates as a Floquet state. This Floquet nature survives at the level of quantum fluctuations, in turn described by usual quasi-energy bands. Furthermore, we prove that an SMBF state presents time crystalline order, although in a manner completely different from conventional discrete Floquet time crystals, since there energy is not conserved, the period is imposed by external driving, and the symmetry breaking is discrete, not continuous. 

We show that the concept of SMBF state emerges in a variety of canonical many-body problems. In particular, SMBF states can arise within the MultiConfiguration Time-Dependent Hartree (MCTDH) method for bosons and fermions \cite{Caillat2005,Alon2008}, a genuine many-body ansatz beyond mean-field. We explicitly discuss paradigmatic mean-field limits of the MCTDH method, the celebrated Gross-Pitaevskii (GP) and Hartree-Fock (HF) equations for bosons and fermions, respectively, proving that their spectrum of quantum fluctuations, in turn described by the Bogoliubov-de Gennes (BdG) equations and the Time-Dependent Hartree-Fock approximation (TDHFA), are given in terms of quasi-energy bands within conventional Floquet theory. We also show that SMBF states can arise in discrete models like Bose-Hubbard \cite{Jaksch1998}, via the discrete GP equation or the Gutzwiller ansatz, with their quantum fluctuations described by another linear Floquet problem.

For illustrative purposes, as well as for definiteness, we focus on the particular case of Bose-Einstein condensates close to zero temperature. Specifically, we show that an SMBF state can arise in a one-dimensional (1D) flowing condensate, the so-called CES state \cite{deNova2016,deNova2021a}. Furthermore, we prove explicitly its time crystalline character by examining the robustness of the CES state. We show that the CES time crystal results from a dynamical phase transition, and survives for thermodynamically long times, scaling with the size of the system. The reached CES time crystal is independent of the initial state and of the transient details, and universal in the sense that it emerges regardless of the particular form of the background Hamiltonian. We explicitly prove that the proposed SMBF state is a genuinely many-body effect that, even though it can emerge in highly supersonic flows, necessarily requires the presence of interactions. In addition, it is robust against external perturbations and quantum fluctuations. Finally, we describe realistic experimental scenarios in cold atoms for the observation of the CES time crystal.

\textit{Spontaneous many-body Floquet state.}---In many-body systems, it is extremely difficult to compute exactly the dynamics due to the exponentially large size of the Hilbert space, and resorting to approximate methods becomes unavoidable. One of the most common approaches is the use of variational principles that look for extremizing a certain functional using a trial wave function characterized by a reduced number of parameters. For instance, in the Dirac-Frenkel variational principle, the \textit{ansatz} state $\ket{\Psi(t)}$ is required to extremize 
\begin{equation}\label{eq:DiracFrenkel}
    L(t)\equiv\bra{\Psi(t)}i\hbar\frac{d}{dt}-\hat{H}\ket{\Psi(t)}
\end{equation}
With no restrictions on the specific form of $\ket{\Psi(t)}$, the Dirac-Frenkel variational principle simply yields the \textit{exact} Schr\"odinger equation of the system. Other variational methods are based on the Lagrangian formalism, where the trial wave function extremizes some action. In all cases, the resulting variational equations of motion are typically described by effective self-consistent Hamiltonians. In this context, we define an SMBF state as a variational state of a time-independent many-body Hamiltonian that, due to interactions, spontaneously oscillates like a Floquet state, self-consistently setting the effective Hamiltonian to be periodic.

For illustrating the definition of SMBF state, we consider the following many-body Hamiltonian for bosons:
\begin{eqnarray}\label{eq:HamiltonianManyBody}
    \hat{H}&=&\int\mathrm{d}\mathbf{x}~\hat{\Psi}^{\dagger}(\mathbf{x})\left[-\frac{\hbar^2}{2m}\nabla^2+V(\mathbf{x})\right]\hat{\Psi}(\mathbf{x})\\
\nonumber&+&\frac{g}{2}\hat{\Psi}^{\dagger}(\mathbf{x})\hat{\Psi}^{\dagger}(\mathbf{x})\hat{\Psi}(\mathbf{x})\hat{\Psi}(\mathbf{x})
\end{eqnarray}
$\hat{\Psi}(\mathbf{x})$ being the field operator, $m$ the mass of the particles, $V(\mathbf{x})$ some \textit{time-independent} potential and $g$ the coupling constant. The Heisenberg equation of motion for the field operator reads
\begin{align}\label{eq:HeisenbergEquationOfMotion}
    &i\hbar\partial_t\hat{\Psi}(\mathbf{x},t)=[\hat{\Psi}(\mathbf{x},t),\hat{H}]\\
    \nonumber &=\left[-\frac{\hbar^2}{2m}\nabla^2+V(\mathbf{x})+g\hat{\Psi}^{\dagger}(\mathbf{x},t)\hat{\Psi}(\mathbf{x},t)\right]\hat{\Psi}(\mathbf{x},t)
\end{align}
Close to $T=0$, the dynamics is accurately described by a mean-field approximation where the field operator is replaced by the macroscopic wave function $\Psi(\mathbf{x},t)$ describing the condensate, $\hat{\Psi}(\mathbf{x},t) \rightarrow \Psi(\mathbf{x},t)$. This gives rise to the well-known Gross-Pitaevskii equation
\begin{eqnarray}\label{eq:HamiltonianEffective}
    i\hbar\partial_t\Psi(\mathbf{x},t)&=&H_{\rm{GP}}(\mathbf{x},t)\Psi(\mathbf{x},t)\\
    \nonumber H_{\rm{GP}}(\mathbf{x},t)&=&-\frac{\hbar^2}{2m}\nabla^2+V(\mathbf{x})+g|\Psi(\mathbf{x},t)|^2
\end{eqnarray}
The same equation could have been derived from the Dirac-Frenkel variational principle by using an ansatz where all bosons occupy the same quantum state. 

The operator $H_{\rm{GP}}(\mathbf{x},t)$ above is the effective self-consistent Hamiltonian governing the dynamics, similar to the usual Schr\"odinger operator but with a nonlinear term arising from interactions, responsible for the possible time-dependence of $H_{\rm{GP}}(\mathbf{x},t)$. Hence, if the density $|\Psi(\mathbf{x},t)|^2$ oscillates periodically with period $T$, $H_{\rm{GP}}(\mathbf{x},t)$ becomes periodic. Self-consistently, $\Psi(\mathbf{x},t)$ behaves like a Floquet state:
\begin{equation}\label{eq:FloquetWaveMu}
\Psi(\mathbf{x},t)=u(\mathbf{x},t)e^{-i\tilde{\mu}t/\hbar},~u(\mathbf{x},t)=\sum^{\infty}_{n=-\infty}u_n(\mathbf{x})e^{-in\omega_0t}
\end{equation}
with $u(\mathbf{x},t+T)=u(\mathbf{x},t)$, $\omega_0=2\pi/T$ and $\tilde{\mu}$ the \textit{quasi-chemical potential}, defined as usual modulo $\hbar\omega_0$. By inserting this expansion into the GP equation, we get a system of self-consistent equations for the Floquet components $u_n(x)$:
\begin{eqnarray}\label{eq:FloquetEquation}
    n\hbar\omega_0u_n(\mathbf{x})&=& \left[-\frac{\hbar^2}{2m}\nabla^2+V(\mathbf{x})-\tilde{\mu}\right]u_n(\mathbf{x})\\
    \nonumber&+&
    g\sum^{\infty}_{m=-\infty}\sum^{\infty}_{k=-\infty} u^*_{k}(\mathbf{x})u_{k+n-m}(\mathbf{x})u_{m}(\mathbf{x})
\end{eqnarray}




This is an SMBF state since a) the effective Hamiltonian $H_{\rm{GP}}(\mathbf{x},t)$ is periodic, and the wave function $\Psi(\mathbf{x},t)$ is self-consistently a Floquet state; b) this periodicity is spontaneously induced by many-body interactions, and not tuned by some external field or symmetry transformation. This last feature excludes from our definition more trivial examples of periodic motion such as those involving harmonic oscillators, solitons moving in a ring (as in the original proposal of time crystal \cite{Wilczek2012}), or traveling spatially periodic cnoidal waves describing trains of solitons. Periodic motion in a harmonic oscillator is not a consequence of interactions but instead of the harmonic character, which in addition does not involve any Floquet physics. On the other hand, a soliton moving in a ring or a traveling cnoidal wave can be reduced to a stationary solution by switching to the appropriate rotating or Galilean frame, respectively. 

The Floquet character of an SMBF state is inherited by its quantum fluctuations. In a condensate, quantum fluctuations are described by the Bogoliubov-de Gennes equations, which arise by considering the fluctuations of the field operator $\hat{\phi}(\mathbf{x},t)$, taken here as $\hat{\Psi}(\mathbf{x},t)\equiv [u(\mathbf{x},t)+\hat{\phi}(\mathbf{x},t)]e^{-i\tilde{\mu}t/\hbar}$. By expanding Eq. (\ref{eq:HeisenbergEquationOfMotion}) to linear order, we get the BdG equations
\begin{equation}\label{eq:BdGfieldequation}
i\hbar\partial_t\hat{\Phi}=M(t)\hat{\Phi},~\hat{\Phi}=\left[\begin{array}{c}\hat{\phi}\\ \hat{\phi}^{\dagger}\end{array}\right],~M(t)=\left[\begin{array}{cc} N(t) & A(t)\\
-A^*(t) &-N^*(t)\end{array}\right]
\end{equation}
where
\begin{equation}
N(t)=-\frac{\hbar^2}{2m}\nabla^2+V(\mathbf{x})+2g|u(\mathbf{x},t)|^2-\tilde{\mu},~ A(t)=gu^{2}(\mathbf{x},t)
\end{equation}
are periodic operators. Therefore, the whole BdG matrix $M(t)$ is a periodic linear operator, $M(t+T)=M(t)$, and consequently the spectrum of quantum fluctuations follows the conventional Floquet theory, described in terms of quasi-energy bands. As a result, the field operator admits an expansion in Floquet solutions:
\begin{eqnarray}\label{eq:BdGmodesexpansion}
\hat{\Phi}(\mathbf{x},t)&=&\sum_{\varepsilon,\nu}  z_{\varepsilon,\nu}(\mathbf{x},t)\hat{\alpha}_{\varepsilon,\nu}+\bar{z}_{\varepsilon,\nu}(\mathbf{x},t)\hat{\alpha}^{\dagger}_{\varepsilon,\nu}\\
\nonumber z_{\varepsilon,\nu}(\mathbf{x},t)&=&e^{-i
\varepsilon t/\hbar}\left[\begin{array}{c}u_{\varepsilon,\nu}(\mathbf{x},t)\\ v_{\varepsilon,\nu}(\mathbf{x},t)\end{array}\right]\\
\nonumber \bar{z}_{\varepsilon,\nu}(\mathbf{x},t)&=&e^{i
\varepsilon t/\hbar}\left[\begin{array}{c}v^*_{\varepsilon,\nu}(\mathbf{x},t)\\ u^*_{\varepsilon,\nu}(\mathbf{x},t)\end{array}\right]
\end{eqnarray}
where $i\hbar\partial_t z_{\varepsilon,\nu}(\mathbf{x},t)=M(t) z_{\varepsilon,\nu}(\mathbf{x},t)$, with $u_{\varepsilon,\nu}(\mathbf{x},t)$, $v_{\varepsilon,\nu}(\mathbf{x},t)$ periodic functions, $\varepsilon$ the quasi-energy, and $\nu$ the discrete index labeling the quasi-energy band, while the operators $\hat{\alpha}_{\varepsilon,\nu}, \hat{\alpha}^{\dagger}_{\varepsilon,\nu}$ are bosonic annihilation, creation operators, respectively. We stress that the period $T$ of the BdG problem is still spontaneously set by the SMBF state, not externally driven.

In contrast to Floquet driven systems, as the original Hamiltonian is time-independent, an SMBF state has a conserved energy, computed here by replacing $\hat{\Psi}(\mathbf{x})$ by $\Psi(\mathbf{x},t)$ in Eq. (\ref{eq:HamiltonianManyBody}).

Since it breaks continuous time translation symmetry, reduced to a discrete one, an SMBF state is also a time crystal. In the condensate case, the one-body correlation function already displays off-diagonal long-range order, so 
\begin{equation}\label{eq:OneBodyCorrelation}
    G(\mathbf{x},\mathbf{x}',t,t')\equiv\braket{\hat{\Psi}^{\dagger}(\mathbf{x},t)\hat{\Psi}(\mathbf{x}',t')}\simeq\Psi^{*}(\mathbf{x},t)\Psi(\mathbf{x}',t')
\end{equation}
exhibits a time-periodic behavior for $|\mathbf{x}-\mathbf{x}'|\rightarrow\infty$ [except for a trivial phase $e^{i\tilde{\mu}(t-t')/\hbar}$], precisely the definition of long-range time-periodic order. 

\begin{figure}[!tb]
\begin{tabular}{@{}cc@{}}
    \stackinset{r}{10pt}{t}{10pt}{\textbf{(a)}}{\includegraphics[width=0.5\columnwidth]{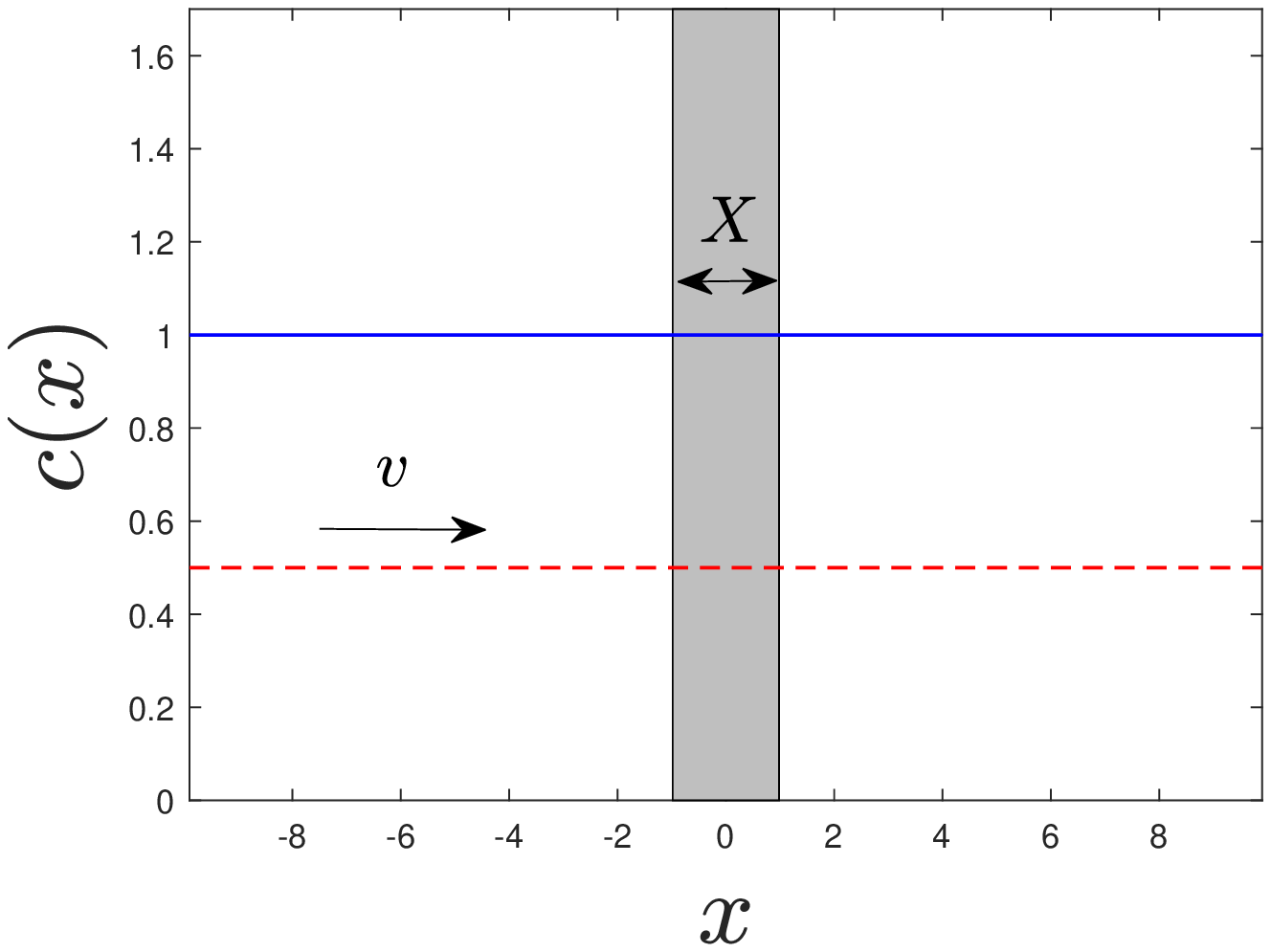}}
     &
    \stackinset{r}{10pt}{t}{10pt}{\textbf{(b)}}{\includegraphics[width=0.48\columnwidth]{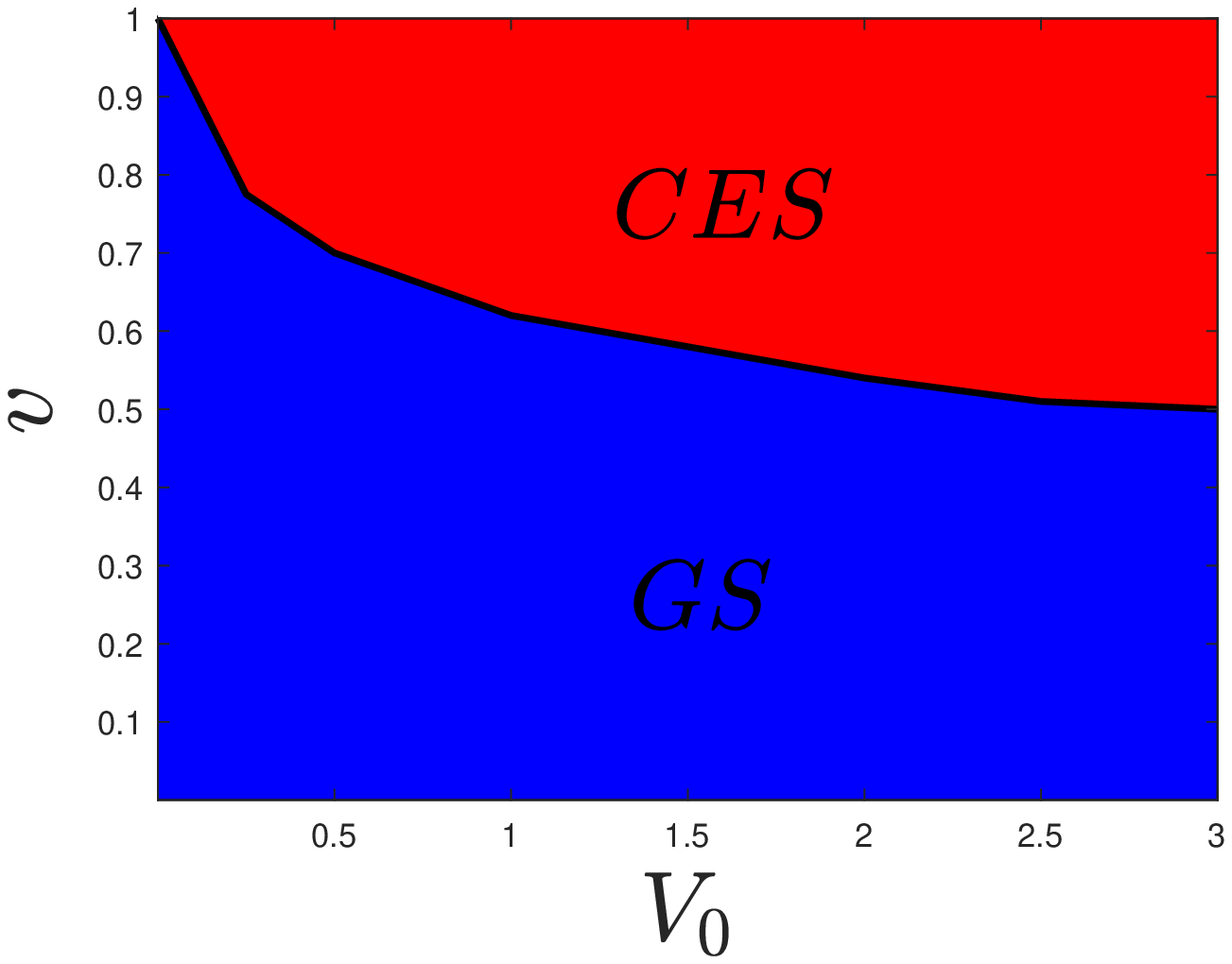}}
\end{tabular}
\caption{(a) Spatial profile of sound (solid blue) and flow (dashed red) velocities of the IHFC. The shaded area represents the region where the attractive constant potential $V(x)=-V_0$ is present. (b) Phase diagram for the final state of (a) in the $(V_0,v)$ plane for fixed $X=2$.}
\label{fig:PhaseDiagram}
\end{figure}

It is known that a time crystal should contain some many-body features that make it robust in order not to be a trivial example of periodic motion. In the case of an SMBF state, we can expect similar robustness to arise because precisely the periodicity is spontaneously set by interactions. Indeed, this robustness is already reflected by the Floquet character of the spectrum of quantum fluctuations discussed above. However, the time crystal resulting from an SMBF state is in stark contrast with conventional discrete Floquet time crystals since it is continuous, its period is spontaneous and not fixed by the external driving, and it has a well-defined energy.

While we have focused on condensates for simplicity, all these ideas can be extended to other canonical many-body problems \cite{SuppMatTimeCrystal}, like the MCTDH method for bosons and fermions \cite{Caillat2005,Alon2008}, the HF equations for fermions \cite{Giuliani2005}, or the Gutzwiller ansatz in Bose-Hubbard models \cite{Jaksch1998}.




\textit{CES state.}---The question is now: does a nontrivial SMBF state exist? We identify here one realization starting from a model previously studied in the literature \cite{deNova2021a}, consisting of a 1D initially homogeneous flowing condensate (IHFC) with density $n_0$ and velocity $v$. At $t=0$, an attractive square well of amplitude $-V_0$ and size $X$ is suddenly introduced. A schematic representation of the IHFC at $t=0$ is given in Fig. \ref{fig:PhaseDiagram}a. Hereafter we set $\hbar=m=c_0=1$ and rescale the GP wave function as $\Psi(x,t)\rightarrow \sqrt{n_0}\Psi(x,t)$, with $c_0=\sqrt{gn_0/m}$ the initial speed of sound.

\begin{figure}[!tb]
\begin{tabular}{@{}cc@{}}
    \stackinset{l}{22pt}{t}{5pt}{\textbf{(a)}}{\includegraphics[width=0.5\columnwidth]{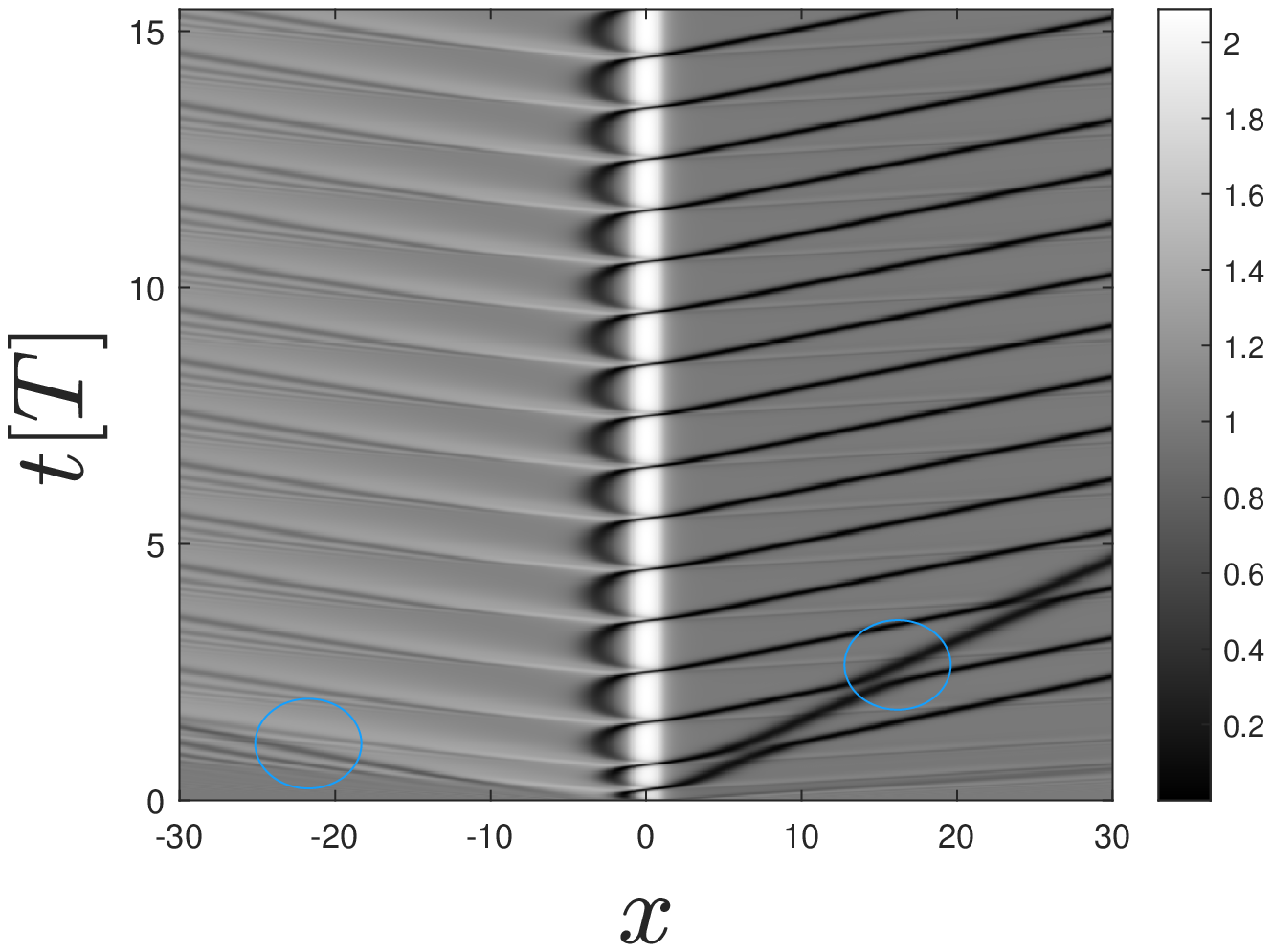}} & 
    \stackinset{l}{22pt}{t}{5pt}{\textbf{(b)}}{\includegraphics[width=0.5\columnwidth]{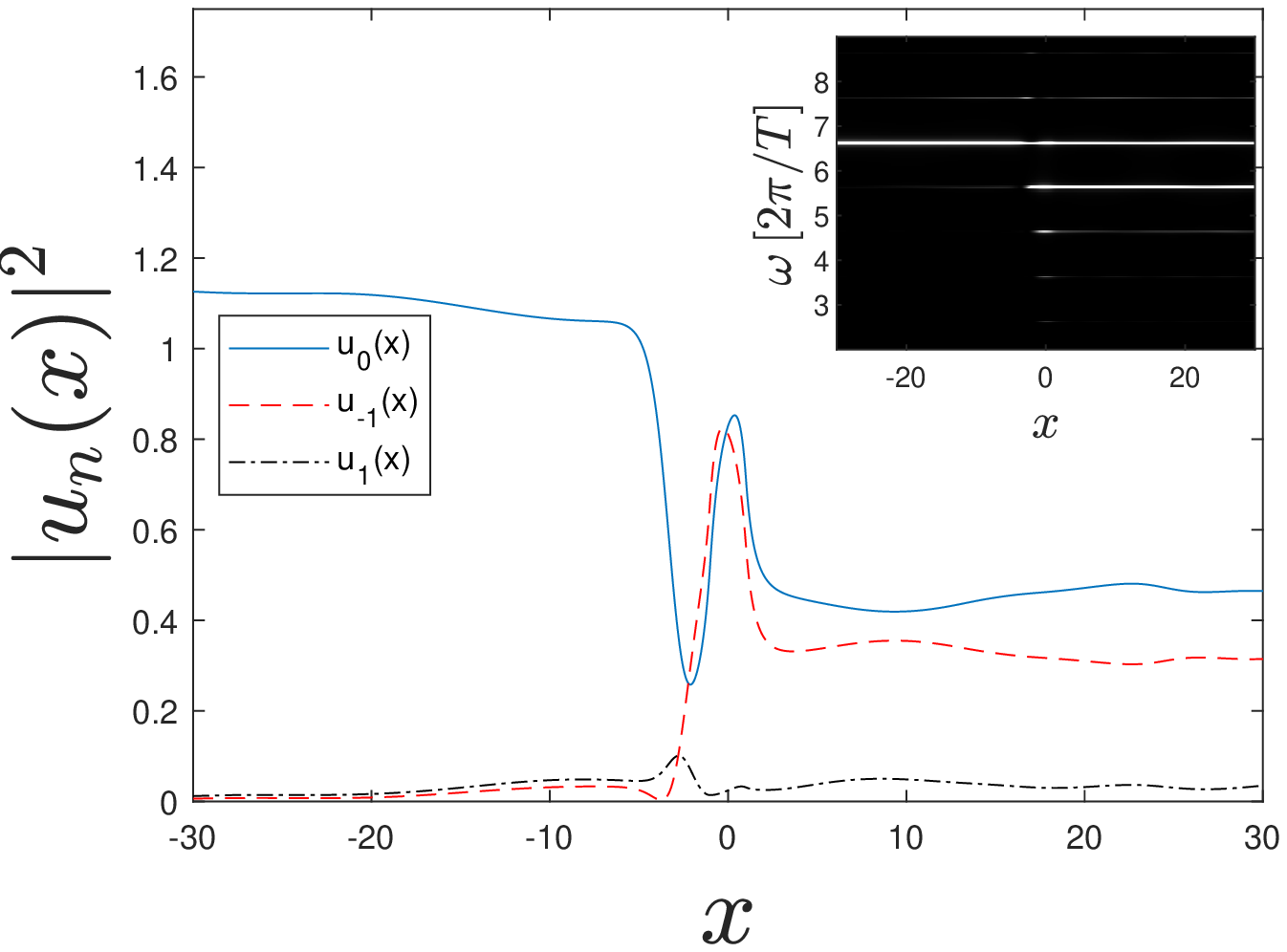}} \\ \\ \\
    \stackinset{l}{22pt}{t}{5pt}{\textbf{(c)}}{\includegraphics[width=0.5\columnwidth]{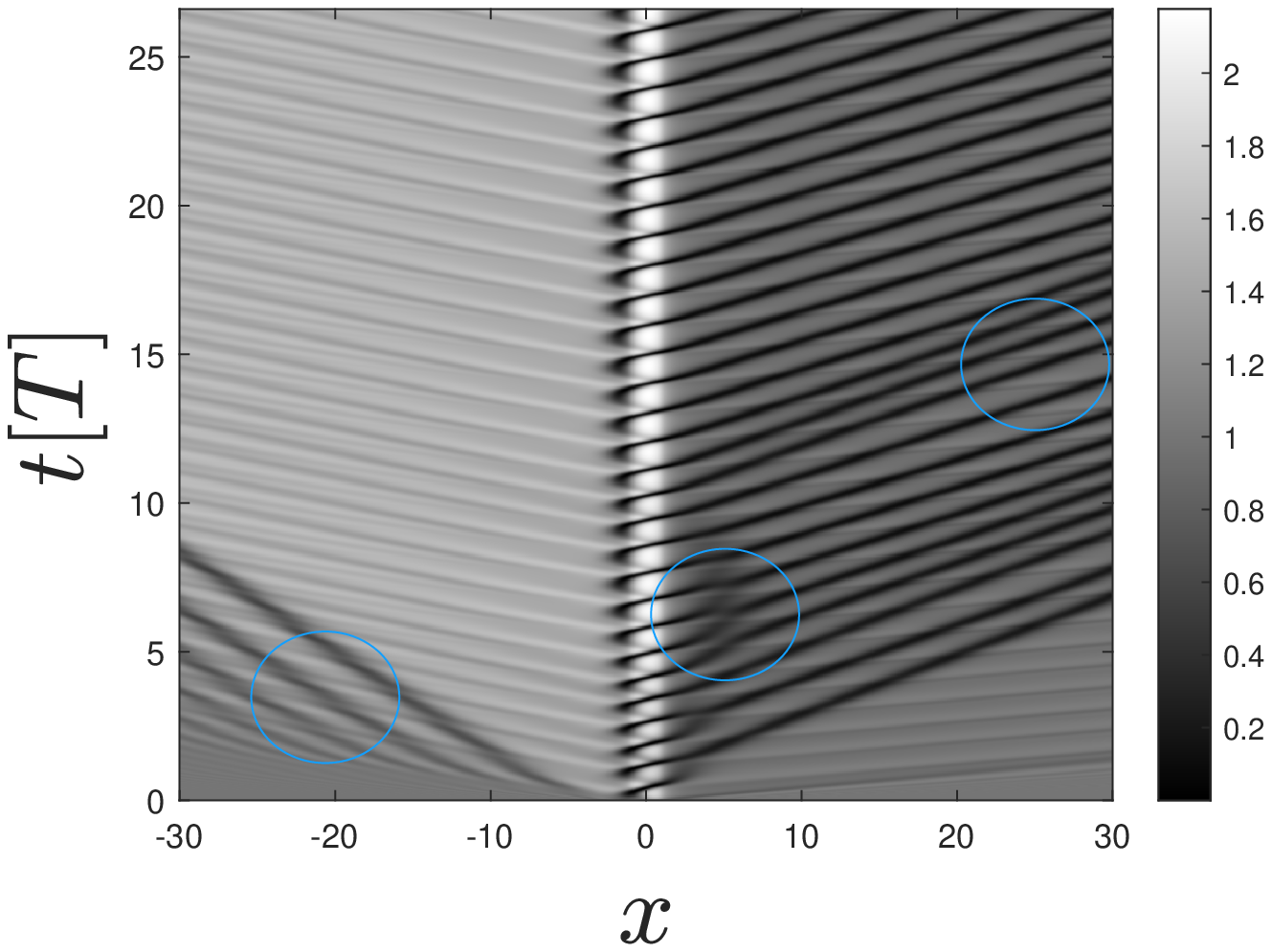}} & 
    \stackinset{l}{22pt}{t}{5pt}{\textbf{(d)}}{\includegraphics[width=0.5\columnwidth]{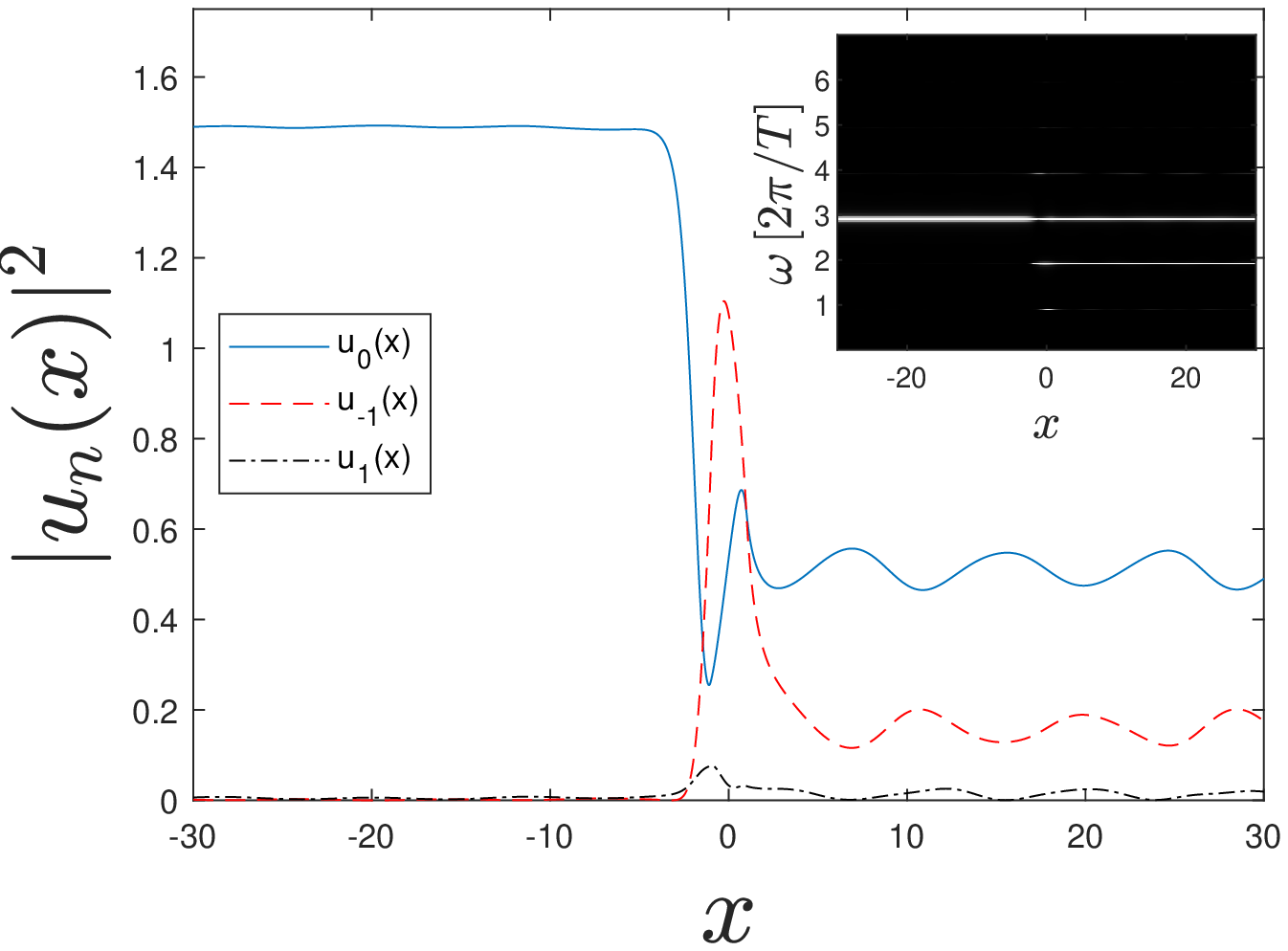}}
\end{tabular}
\caption{Analysis of the CES state. Upper row: simulation with $v=0.65$, $V_0=1$, and $X=2$. (a) 2D plot of $|\Psi(x,t)|^2$. Blue circles highlight transient features. Time is in units of the CES period $T$.  (b) $|u_n(x)|^2$ for $n=0$ (solid blue), $n=-1$ (dashed red) and $n=1$ (dashed-dotted black). Inset: Spectrum $|\Psi(x,\omega)|^2$ of the CES state, with $\omega$ in units of $2\pi/T$. (c)-(d) Same as first row but for $v=0.95$.}
\label{fig:Floquet}
\end{figure}

The quench in the external potential induces a deterministic dynamics in the condensate, numerically computed by integrating the time-dependent GP equation. It is seen that there are only two possible choices for the final state of the system at long times as a function of $(v,X,V_0)$: the ground state (GS), or a regime of periodic continuous emission of solitons (CES) \cite{deNova2021a}. A typical phase diagram is represented in Fig. \ref{fig:PhaseDiagram}b. 

We prove here that the CES state is actually an SMBF state. Figures \ref{fig:Floquet}a,c show the time evolution of the density. We observe that, after some transient features (marked by blue circles), particles are accumulated in the well in order to reach GS (vertical white stripe within the well, centered at $x=0$). In order to conserve particle number, a soliton is emitted upstream ($x<0)$; however, in the CES state such a soliton is dragged back to the well (half dark rings at the left of the white strip), passing to the downstream ($x>0$) region and traveling with the flow (diagonal black lines downstream). The process is accompanied by the emission of waves (diagonal white lines upstream) to ensure conservation of total particle number and energy.

The passage of the dragged soliton through the well leaves the system in the same configuration, restarting the process described above. The resulting density pattern is periodic for \textit{every} $x$, and not just in the downstream soliton trains, whose periodicity was already noticed in the literature \cite{Hakim1997,deNova2016,deNova2021a}. Consequently, $H_{\rm{GP}}(x,t)$ becomes periodic and, self-consistently, $\Psi(x,t)$ behaves like a Floquet state, as seen from the inset of Figs. \ref{fig:Floquet}b,d, where the Fourier transform $\Psi(x,\omega)$ exhibits a discrete spectrum $\Psi(x,\omega_n)=u_n(x),~\omega_n=\tilde{\mu}+n\omega_0$. While the discrete lines are separated $2\pi/T$, there is some offset that reveals a non-trivial quasi-chemical potential. By taking the inverse Fourier transform around the spectrum peaks, we recover each Floquet component $u_n(x)$, some of them depicted in main Figs. \ref{fig:Floquet}b,d. Second row of Fig. \ref{fig:Floquet} analyzes the case of larger flow velocity, resulting in smaller period $T$ and fewer dominant Floquet components. 

\begin{figure}[!tb]
\begin{tabular}{@{}cc@{}}
    \includegraphics[width=0.5\columnwidth]{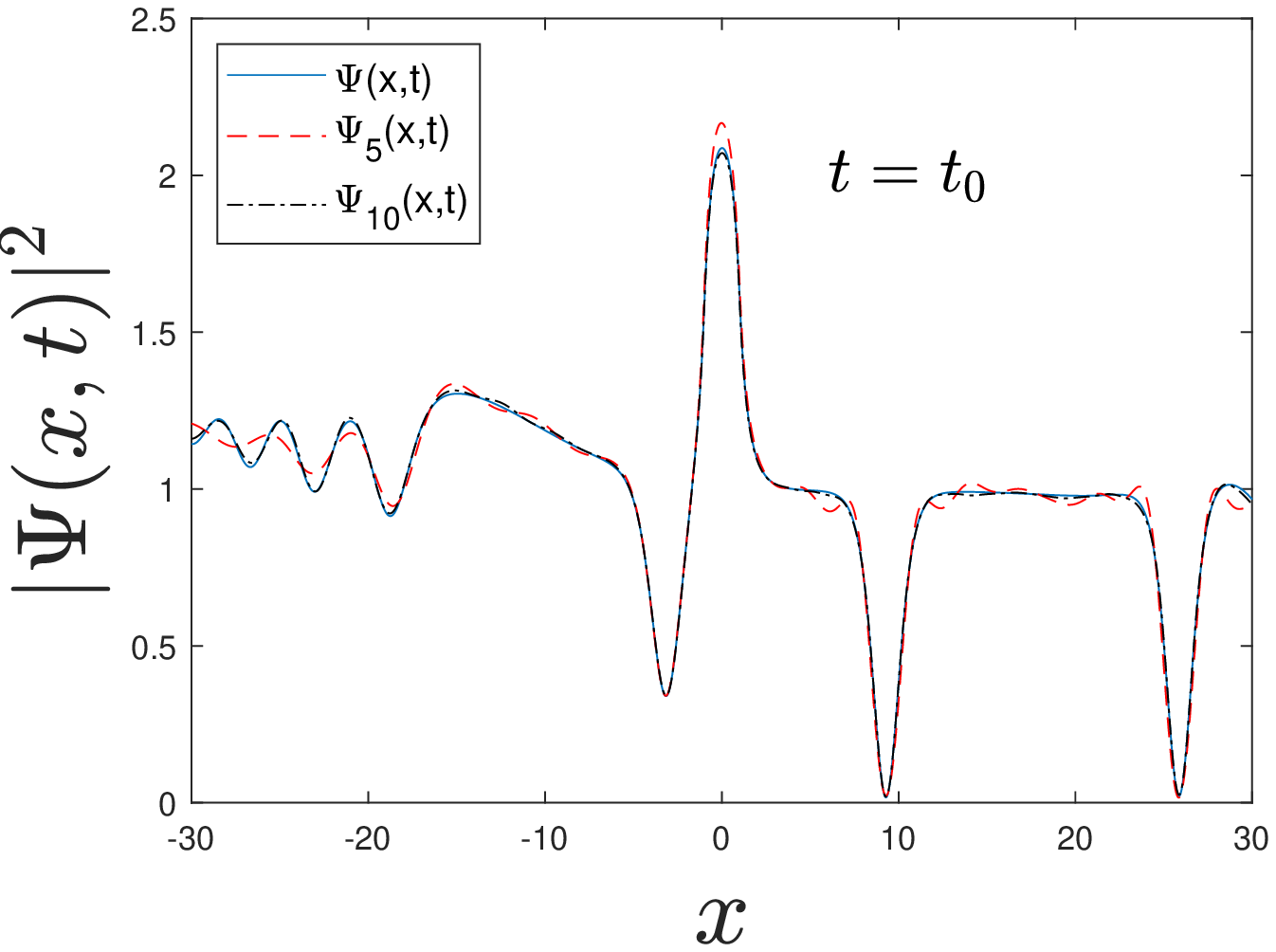} & \includegraphics[width=0.5\columnwidth]{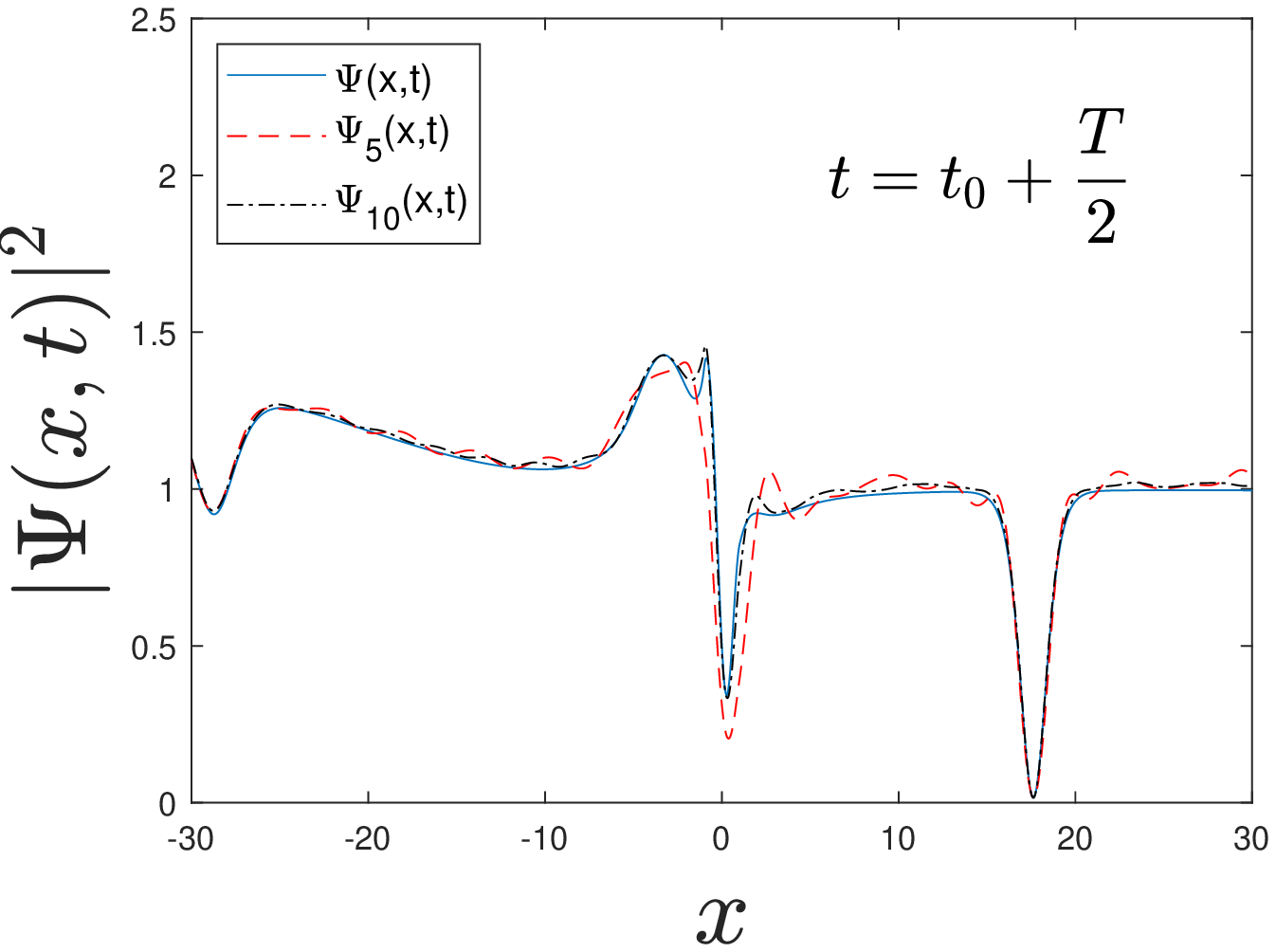} \\ \\ \\
    \includegraphics[width=0.5\columnwidth]{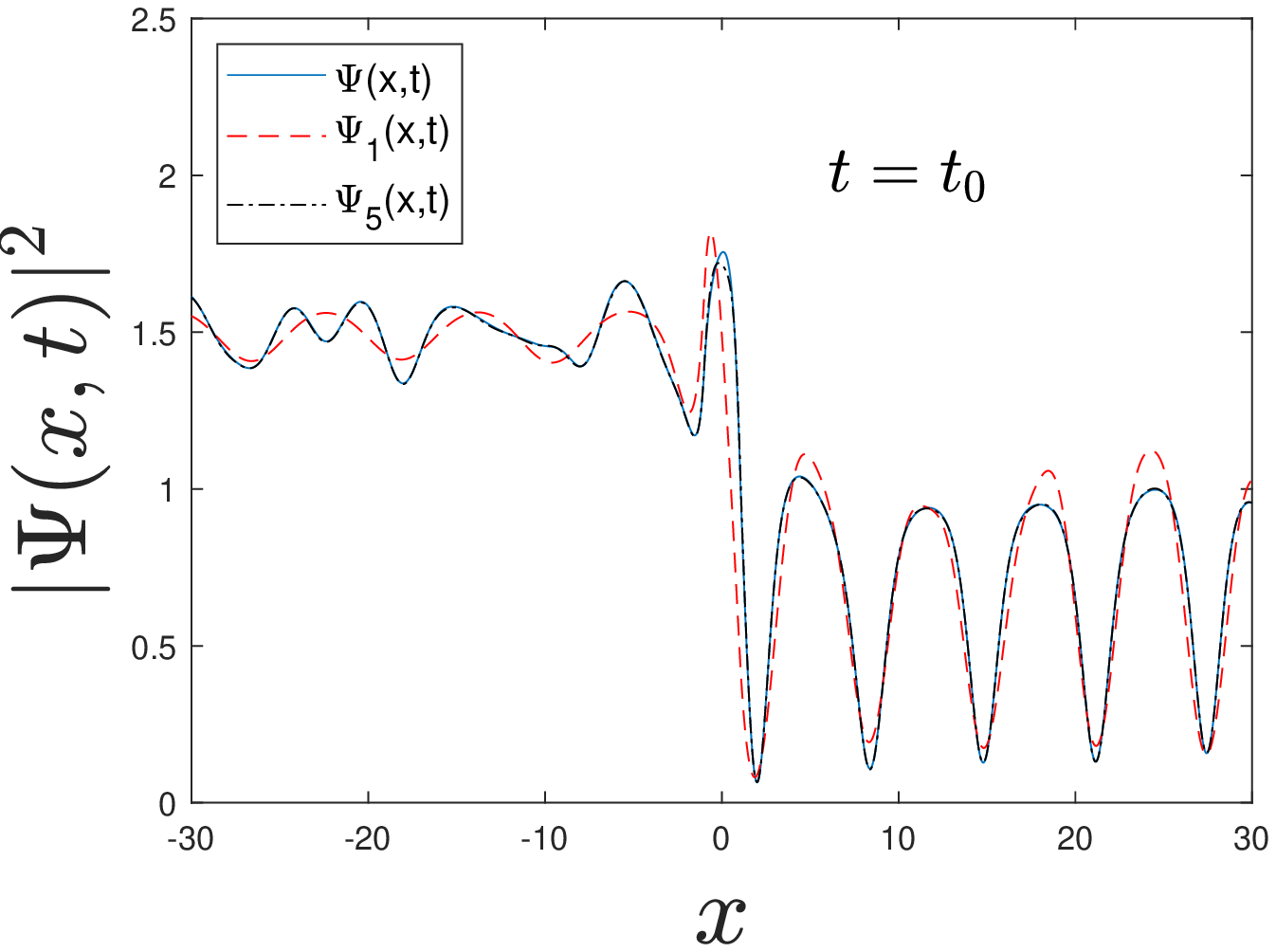} &
    \includegraphics[width=0.5\columnwidth]{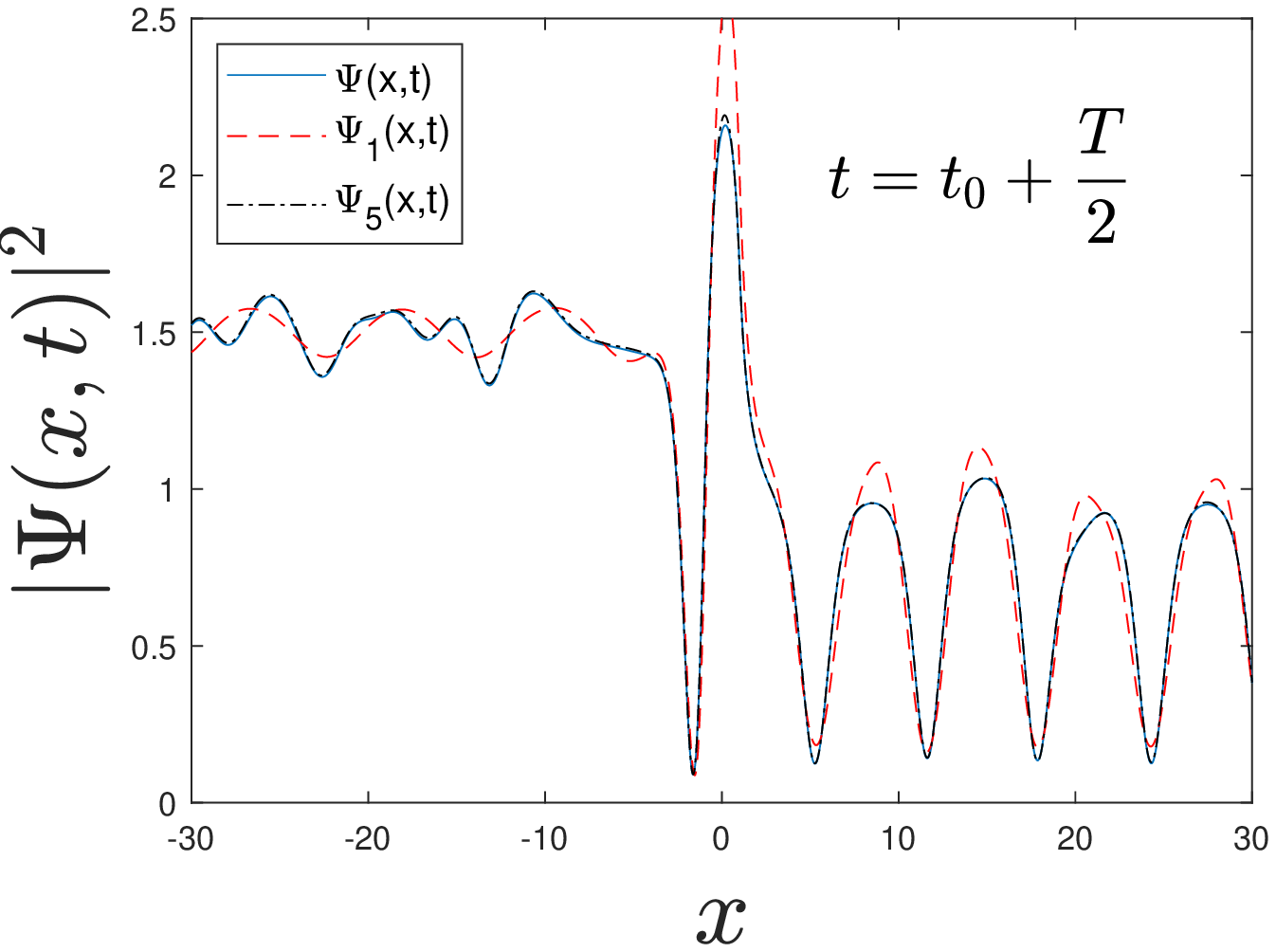}
\end{tabular}
\caption{Snapshots of the density $|\Psi(x,t)|^2$ during a period of the CES state (solid blue) and the truncated Floquet expansions $\Psi_{N}(x,t)$ (dashed red and dashed-dotted black) from Eq. (\ref{eq:FloquetWaveMuFinite}) for the simulations of upper and lower row of Fig. \ref{fig:Floquet}, respectively. The origin of times $t_0$ is chosen arbitrarily.}
\label{fig:Floquetnxt}
\end{figure}

We can perform a Floquet tomography of the CES state by using the Floquet components to reconstruct the wave function, truncating the expansion of Eq. (\ref{eq:FloquetWaveMu}) as
\begin{equation}\label{eq:FloquetWaveMuFinite}
\Psi_N(x,t)=e^{-i\tilde{\mu}t}\sum^{N}_{n=-N}u_n(x)e^{-in\omega_0t}
\end{equation}
where we fix the definition of $\tilde{\mu}$ so that the dominant component is $u_{n=0}(x)$. Figure \ref{fig:Floquetnxt} compares the oscillation within a period of the actual CES wave function $\Psi(x,t)$ with the reconstructed Floquet wave function $\Psi_{N}(x,t)$, which rapidly converges for small $N$. For lower flow velocities, more Floquet components are needed, as already expected from Fig. \ref{fig:Floquet}. This reconstruction explicitly demonstrates that the CES state is indeed an SMBF state.

\textit{CES time crystal.}---Since the CES state is an SMBF state, it is also a continuous time crystal. This allows us to study in detail the expected robustness of an SMBF time crystal using a specific realization. In particular, we prove the time crystalline character of the CES state by analyzing its thermodynamic properties, its independence from the initial condition and the transient as well as from the particular details of the background Hamiltonian, and its robustness against external perturbations and quantum fluctuations. Furthermore, we explicitly demonstrate the critical role played by interactions in the formation of the time crystal.

In discrete time crystals, the rigidity of the subharmonic response is a signature of crystalline behavior because it demonstrates that the state is not due to some particular fine-tuning \cite{Choi2017,Zhang2017}. Here, the analogue phenomenon is the robustness against variations of the parameters $(v,X,V_0)$. The phase diagram in Fig. \ref{fig:PhaseDiagram}b shows that the SMBF state survives in a wide region of parameter space. The period varies continuously with $(v,X,V_0)$, but this is quite a natural feature arising from the continuous character of the symmetry breaking. 

The GS/CES phase transition is an example of dynamical phase transition, typical of setups in which a strong quench in an external parameter is introduced \cite{Moeckel2008,Sciolla2010,Lang2018}. The GS is the symmetry unbroken phase, with continuous time translation symmetry, and the CES is the time crystal phase, with discrete time translation symmetry. The energy of the CES state satisfies $E_{\rm{CES}}=E_{\rm{GS}}+O(1)$, so in the thermodynamic limit $\lim_{N\rightarrow\infty}E_{\rm{CES}}/N=\lim_{N\rightarrow\infty}E_{\rm{GS}}/N$. However, separately in both upstream and downstream regions, the difference in number of particles and energy is $O(N)$. 

\begin{figure}[t]
\begin{tabular}{@{}cc@{}}
    \stackinset{l}{22pt}{t}{10pt}{\textbf{(a)}}{\includegraphics[width=0.5\columnwidth]{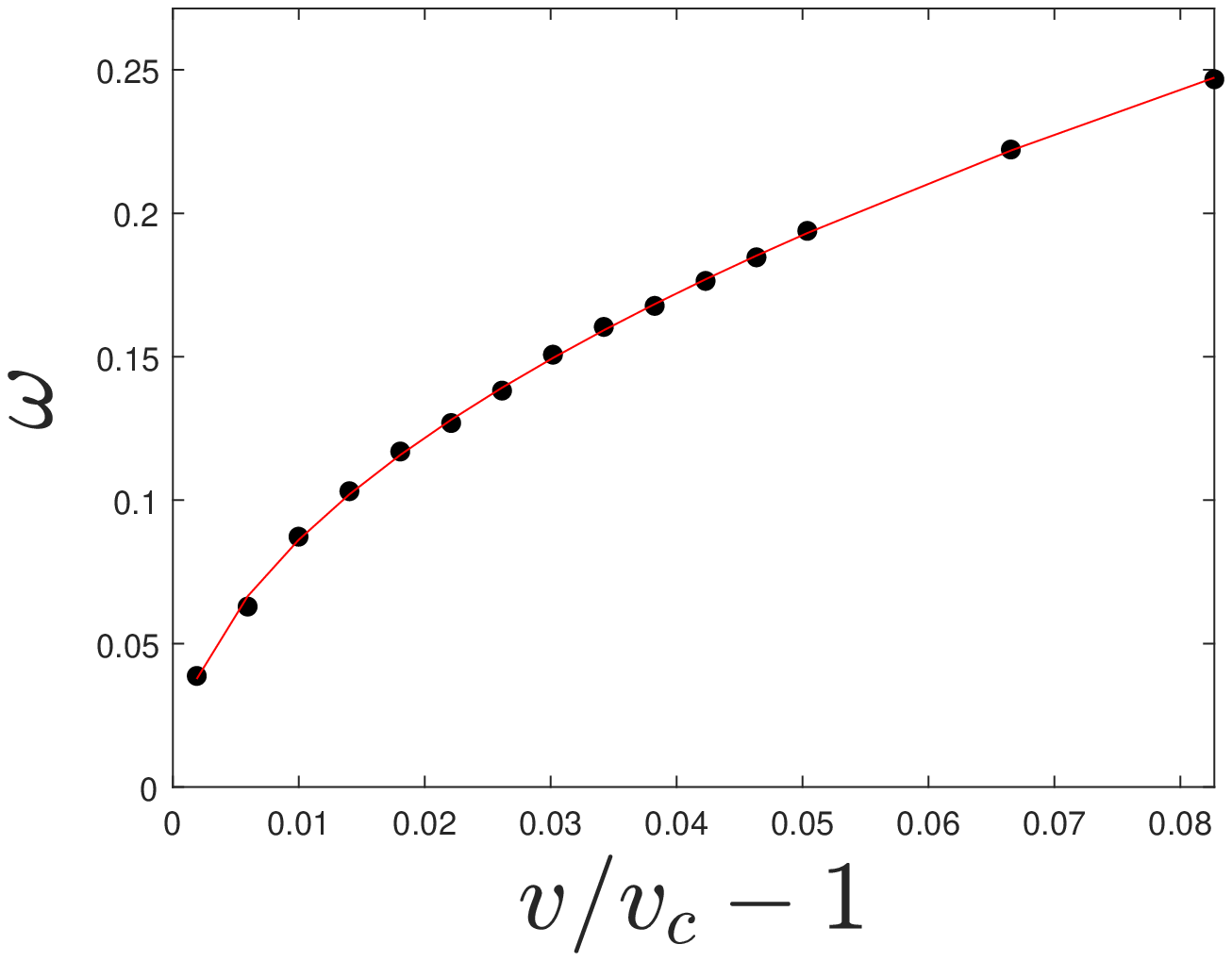}} &  
    \stackinset{l}{22pt}{t}{10pt}{\textbf{(b)}}
    {\includegraphics[width=0.5\columnwidth]{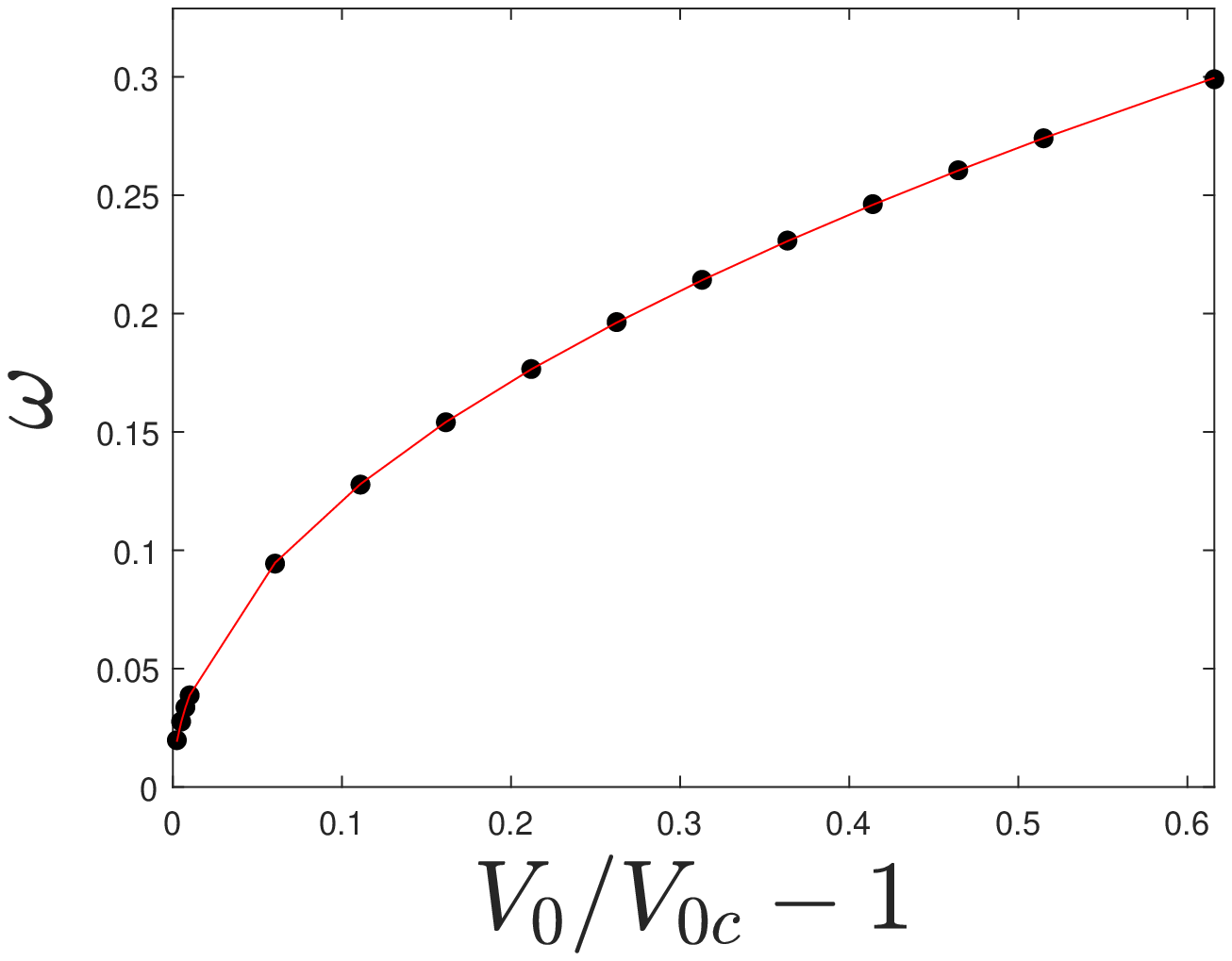}}
\end{tabular}
\caption{Critical behavior of the CES frequency $\omega$ close to the phase transition. The red line represents a fit to a power law. (a) Dependence on $v$ for $V_0=1,~X=2$. (b) Dependence on $V_0$ for $v=0.62,~X=2$.} 
\label{fig:Criticality}
\end{figure}

In Fig. \ref{fig:Criticality}, the CES frequency is shown to follow a power law close to the phase transition, produced at the critical values $v_c,V_{0c}$. We numerically find that the critical exponents associated to $v,V_0$ are $\alpha\simeq\beta\simeq0.50$, respectively, potentially suggesting a possible analytical derivation. Intuitively, the CES frequency vanishes at the phase transition because the intrinsic velocity of the upstream emitted soliton equals that of the background flow, so it takes an infinite time to return. 

In numerical simulations, the collapse of the time crystal is only observed due to finite-size effects, when reflections at the boundaries return and distort the dynamics. As a result, the system eventually relaxes locally to GS. This indicates that the CES state is an extended state that persists indefinitely in the thermodynamic limit, with a lifetime that scales linearly with the system size, similarly to Ref. \cite{Syrwid2017}. However, the nature of this lifetime is completely different to that of the stationary solutions of the GP spectrum, which are dynamically unstable (except GS) with an intrinsic lifetime \cite{Michel2015,deNova2021a}. The survival of the CES time crystal for thermodynamically long times is similar to that of prethermal discrete time crystals in Floquet driven systems, which survive exponentially long times in the frequency \cite{Pizzi2021,Ye2021,Kyprianidis2021}.

A quench is not the only way to reach the CES state. For fixed parameters $(v,V_0,X)$, we check that, when reached, the CES state is independent of the initial state and the transient details. Specifically, we consider three types of initial condition: 1) the IHFC model, 2) an unstable stationary black-hole laser (BHL) solution \cite{Michel2015,deNova2016,deNova2017a,deNova2021a}, and 3) the stable GS, but adding an upstream soliton headed downstream that will destabilize the GS. In all cases, the \textit{same} CES state is eventually reached, even though the transient dynamics can be vastly different \cite{SuppMatTimeCrystal}.

This independence with respect to the initial condition and the transient details further demonstrates that the CES state is an intrinsic SMBF state of the system, satisfying Eq. (\ref{eq:FloquetEquation}). Consequently, the CES period is a unique function of the system parameters, $T=T(v,V_0,X)$. 

The proposed SMBF state is not restricted to an \textit{idealized} square well. We have considered a wide range of quenched obstacles, including potentials with realistic shapes or delta barriers, both attractive and repulsive, and even spatial modulations of the coupling constant. In all cases, an SMBF state is eventually achieved provided the flow velocity is high enough \cite{SuppMatTimeCrystal}. 



The CES time crystal is also robust against the presence of time-dependent perturbations. We consider a stochastic modulation $f(t)=1+\epsilon h(t)$ of the external potential and of the coupling constant, with $h(t)$ normally distributed white noise and $\epsilon$ the modulation strength \cite{SuppMatTimeCrystal}. In both cases, even for appreciable perturbations $\epsilon \sim 0.1$, the time crystal survives. However, for large enough perturbations $\epsilon\sim 1$, the crystal is destroyed.

Since the induced dynamics in the IHFC is deterministic, the time crystal is also robust against quantum and thermal fluctuations. Technically speaking, off-diagonal long-range order is destroyed in 1D by phase fluctuations, and one must speak of a 1D quasi-condensate. Specifically, at $T=0$ there is an algebraic decay with distance, becoming exponential for $T>0$ \cite{Pitaevskii2003,Pethick2008}. Long-range order in time behaves as that in space in any dimensions \cite{SuppMatTimeCrystal}. However, the space-time decay can be neglected in typical setups if density and temperature are high and low enough, respectively, and the GP mean-field approximation becomes accurate. Actually, it is exact in the limit $g,T\rightarrow0$, $n_0\rightarrow\infty$, with $gn_0$ constant, and is considered in other works on time crystals \cite{Wilczek2012,Syrwid2017,Ohberg2019,Syrwid2020}. Here we go beyond mean-field and explicitly check the role of quantum fluctuations by using the Truncated Wigner method \cite{Sinatra2002,Carusotto2008}, finding that the periodicity of $G(x,x',t,0)$ survives for long times, allowing for a potential observation \cite{SuppMatTimeCrystal}.

Interactions are critical in the formation of the time crystal from the very beginning. Indeed, the non-interacting Schr\"odinger equation does not display any SMBF state. We study the role of interactions by gradually introducing them in the Schr\"odinger equation as $\lambda|\Psi|^2$, with $\lambda$ a dimensionless control parameter. We find that an SMBF state is restored even for weak interactions $\lambda\sim 0.1$, still deep in the supersonic regime, demonstrating the critical role of interactions to stabilize the time crystal \cite{SuppMatTimeCrystal}. Therefore, an SMBF state reflects a genuinely non-perturbative effect of interactions. This contrasts with other phenomena, as for instance nonlinear Bloch states for the GP equation in a lattice potential \cite{Wu2003,deNova2014a}, which have a well-defined non-interacting Schr\"odinger limit.


\textit{Experimental implementation.}---We propose a simple and feasible experimental scheme for the characterization of the CES time crystal using cold atoms. We consider an elongated quasi-1D Bose-Einstein condensate that is essentially homogeneous far from the edges of the trap, in which a localized obstacle is swept with velocity $v$. By Galilean invariance, this is equivalent to launching the condensate against the obstacle. If the velocity is high enough, the system will reach the CES state. Indeed, soliton trains have already been reported in similar setups \cite{Engels2007}, something that hints at an underlying SMBF state. 

We suggest to extend this type of experiment by further localizing the obstacle to shorten the transient, and following closely the periodic time evolution, in both upstream and downstream regions. This can be done by employing high-resolution imaging to measure the density profile in the obstacle frame, as successfully done in recent experiments in analogue gravity \cite{Steinhauer2016,deNova2019,Kolobov2021}. Another possibility is to confine a condensate in a long ring \cite{Eckel2018} and rotate a localized potential.

\textit{Conclusions and outlook.}---We propose a novel type of quantum phase: an SMBF state, which oscillates as a Floquet state without the need of external driving, with this periodic behavior spontaneously self-induced through interactions. This Floquet character is also translated into the spectrum of quantum fluctuations, where the full linear Floquet physics is recovered. Since it spontaneously breaks time translation symmetry, an SMBF state is a continuous time crystal. Indeed, as a phenomenon induced by interactions, an SMBF state is expected to present robustness typical of time crystals. Nevertheless, the time crystal arising from an SMBF state is very different from those arising in conventional Floquet driven systems, since it is continuous instead of discrete, its periodicity is spontaneously self-induced and not fixed by the external driving, and it has a well-defined energy.

We illustrate all these ideas by successfully identifying the CES state, which appears in 1D flowing condensates, as a particular example of SMBF state. We note that, while soliton trains in condensates are well-known phenomena \cite{Hakim1997,Pavloff2002,Engels2007,deNova2016,Nguyen2017,Sels2020,deNova2021a}, this is the first time that the concept of SMBF state is formulated. 

We prove the time crystal character of the CES state by examining in detail its properties. We argue that its lifetime is thermodynamically long, scaling linearly with the system size. We show that, when reached, it is independent of the initial state and transient details. We find that a CES state emerges for a variety of different Hamiltonians, regardless of the specific details of the model, suggesting some form of universality that goes beyond the BHL paradigm of Ref. \cite{deNova2021a}. The CES time crystal is robust against the introduction of time-dependent stochastic perturbations or the presence of quantum fluctuations. Importantly, we explicitly prove that the CES state is a genuine nonperturbative effect of interactions, finding that it can also arise in supersonic flows. Finally, we design realistic experimental scenarios for the observation of the CES time crystal.

Gathering our results, the CES state 1) is an intrinsic state, independent of the transient details and initial state;  2) also arises in supersonic regimes, stabilized by many-body interactions; 3) appears in a variety of configurations. Therefore, we conclude that an SMBF state is indeed a universal feature of a flowing 1D condensate. 

The concept of SMBF state represents a new paradigm in Floquet systems and, in general, in out-of-equilibrium physics. It extends the field of nonlinear Floquet waves \cite{Kreil2019,Trager2021} to scenarios without external driving. Apart from its intrinsic conceptual interest and potential applications, an SMBF state provides a specific realization of continuous time crystal. 

Based on our results, the concept of SMBF state is expected to be ubiquitous. In particular, nonlinear optical fibers \cite{Drummond2014}, quantum fluids of light \cite{Carusotto2013}, or superconductors \cite{Langer1967} are natural candidates to exhibit SMBF states, as they obey equations of motion similar to the GP equation. Of special interest would be the identification of an SMBF state in a Bose-Hubbard model, since that could provide a direct comparison with exact diagonalization results. Fermionic systems which can be effectively described by variational time-dependent ans\"atze, like the HF equations or the more general MCTDH method, are also candidates. Indeed, self-consistent BCS-type theories are known to display periodic oscillations in momentum space \cite{Barankov2004,Foster2014,Perfetto2020}.

We are very grateful to I. Carusotto for stimulating discussions. We also thank useful comments from C. Creffield, I. Zapata, S. Finazzi, M. P\l{}odzie\'n, D. Wild, F. Michel and, especially, R. Parentani, to whose memory we devote this work. This project has received funding from Grant FIS2017-84368-P from Spain's MINECO, and from European Union's Horizon 2020 research and innovation programme under the Marie Sk\l{}odowska-Curie grant agreement No 847635.  

\bibliographystyle{apsrev4-1}
\bibliography{Hawking}

\pagebreak
\widetext
\begin{center}
\textbf{\large Supplemental Material for ``Continuous time crystal from a spontaneous many-body Floquet state''}
\end{center}
\setcounter{equation}{0}
\setcounter{figure}{0}
\setcounter{table}{0}
\makeatletter

\renewcommand{\theequation}{S\arabic{equation}}
\renewcommand{\thefigure}{S\arabic{figure}}
\renewcommand{\bibnumfmt}[1]{[S#1]}
\renewcommand{\citenumfont}[1]{S#1}

We provide here the technical details about the following topics discussed in the main text:

\begin{itemize}

    \item[-] Emergence of SMBF states and Floquet character of the spectrum of quantum fluctuations in canonical examples of variational ans\"atze and many-body Hamiltonians.
    
    \item[-] Numerical analysis of the robustness of the CES state.
    
    \item[-] Analytical calculation of the long-range behavior in time of the one-body correlation function.
    
    \item[-] Truncated Wigner computation of the quantum fluctuations of the CES state.
    
\end{itemize}

\section{SMBF states and quantum fluctuations}

We show here the emergence of SMBF states as well as the Floquet character of its spectrum of quantum fluctuations for several canonical examples of variational ans\"atze and many-body Hamiltonians. 

\subsection{MultiConfiguration Time-Dependent Hartree method}

We consider the following \textit{time-independent} general many-body Hamiltonian in second quantization, valid for both bosons and fermions:
\begin{equation}\label{eq:HamiltonianManyBodyGeneral}
    \hat{H}=\hat{h}+\hat{V}=\sum_{lk} h_{lk}\hat{c}_l^{\dagger}\hat{c}_k+
    \frac{1}{2}\sum_{lkjm} V_{lk,jm}\hat{c}_l^{\dagger}\hat{c}_j^{\dagger}\hat{c}_m\hat{c}_k
\end{equation}
Here, $\hat{h}$ is the single-particle part of the Hamiltonian (typically kinetic energy plus some time-independent external field) and $\hat{V}$ the many-body interacting part, $h_{lk}$ and $V_{lk,jm}$ being their respective matrix elements:
\begin{eqnarray}
    h_{lk}&=&\int\mathrm{d}\mathbf{x}~\left[\phi^*_l(\mathbf{x})\right]_\alpha h_{\alpha\beta}(\mathbf{x})\left[\phi_k(\mathbf{x})\right]_\beta \\
    \nonumber V_{lk,jm}&=&\sum_{\alpha\beta\lambda\mu}\int\mathrm{d}\mathbf{x}\int\mathrm{d}\mathbf{x}'~\left[\phi^*_l(\mathbf{x})\right]_\alpha \left[\phi_k(\mathbf{x})\right]_\beta V_{\alpha\beta,\lambda\mu}(\mathbf{x}-\mathbf{x}'')\left[\phi^*_j(\mathbf{x}')\right]_\lambda \left[\phi_m(\mathbf{x}')\right]_\mu
\end{eqnarray}
where Latin indices label single-particle states and Greek indices label possible discrete (spin, pseudospin, etc.) degrees of freedom of the wave functions. 

The MCTDH method [37,38] restricts the dynamics of $N$ particles (bosons or fermions) to a fixed subspace of $M$ time-dependent single-particle states $\{\phi_i(\mathbf{x},t)\}^M_{i=1}$. This results in the following many-body trial wave function:
\begin{equation}\label{eq:MBAnsatz}
    \ket{\Psi(t)}=\sum_{\mathbf{n}}C_\mathbf{n}(t)\ket{\mathbf{n}~t},~\ket{\mathbf{n}~t}=\left(\prod^{M}_{i=1} \frac{\left[\hat{c}^{\dagger}_i(t)\right]^{n_i}}{\sqrt{n_i!}}\right)\ket{\mathbf{0}},~\sum^M_{i=1}n_i=N
\end{equation}
with $\ket{\mathbf{n}~t}$ a time-dependent Fock state, and $\mathbf{n}$ a $M$-dimensional vector with components $n_i$ that represent the occupation numbers of each orbital $\phi_i(\mathbf{x},t)$, whose associated creation operator is $\hat{c}^{\dagger}_i(t)$. For fermions, $n_i=0,1$, which implies $M\geq N$. We note that, in the limit $M\rightarrow\infty$, the MCTDH approximation becomes exact as it spans the complete single-particle Hilbert space. 



By inserting this ansatz in the Dirac-Frenkel variational principle (\ref{eq:DiracFrenkel}), one arrives at the MCTDH equations, which take the same form for both bosons and fermions [37,38]
\begin{eqnarray}\label{eq:MCTDH}
i\hbar\frac{dC_\mathbf{n}}{dt}&=&\sum_{\mathbf{n}'}\mathcal{H}_{\mathbf{n}\mathbf{n}'}(t)C_{\mathbf{n}'},~\mathcal{H}_{\mathbf{n}\mathbf{n}'}(t)=\bra{\mathbf{n}~t}\hat{H}-\hat{f}\ket{\mathbf{n}'~t}\\
\nonumber i\hbar\partial_t \phi_{k}&=&f\phi_{k}+Q(t)\left[(h-f)\phi_{k}+\sum^{M}_{i,j,l,m=1}[\rho^{-1}]_{kj}(t)\rho_{ji,lm}(t)W_{lm}(t)\phi_i\right],~Q=1-P, ~P=\sum^M_{k=1}\ket{\phi_k(t)}\bra{\phi_k(t)}
\end{eqnarray}
In this equation, $P$ is the projector onto the restricted single-particle subspace, $\rho_{lk}(t)=\bra{\Psi(t)}\hat{c}_l^{\dagger}(t)\hat{c}_k(t)\ket{\Psi(t)}$ and $\rho_{lk,jm}(t)=\bra{\Psi(t)}\hat{c}_l^{\dagger}(t)\hat{c}_j^{\dagger}(t)\hat{c}_m(t)\hat{c}_k(t)\ket{\Psi(t)}$ are the projected one-body and two-body density matrices, respectively, and $W_{lm}(t)$ is the non-local time-dependent potential
\begin{equation}
    \left[W_{lm}(\mathbf{x},t)\right]_{\alpha\beta}=\sum_{\lambda\mu}\int\mathrm{d}\mathbf{x}'~ V_{\alpha\beta,\lambda\mu}(\mathbf{x}-\mathbf{x}')\left[\phi^*_l(\mathbf{x}',t)\right]_\lambda \left[\phi_m(\mathbf{x}',t)\right]_\mu
\end{equation}
The operators $h,f$ are the single-particle versions of the second-quantized operators $\hat{h},\hat{f}$ acting on the wave functions $\phi_i$. The operator $f$ is characterized by the matrix elements $f_{lk}(t)=\braket{\phi_{l}(t)|i\hbar\partial_t \phi_{k}(t)}$ and can be actually chosen to be any Hermitian operator in the restricted single-particle subspace since the ansatz of Eq. (\ref{eq:MBAnsatz}) is invariant under unitary transformations within the single-particle basis. In particular, we can choose $f$ as the projection of the single-particle Hamiltonian, $f=PhP$, and hence Eqs. (\ref{eq:MCTDH}) are reduced to
\begin{eqnarray}\label{eq:MCTDHSP}
i\hbar\frac{dC_\mathbf{n}}{dt}&=&\sum_{\mathbf{n}'}\mathcal{H}_{\mathbf{n}\mathbf{n}'}(t)C_{\mathbf{n}'},~\mathcal{H}_{\mathbf{n}\mathbf{n}'}(t)=\bra{\mathbf{n}~t}\hat{V}\ket{\mathbf{n}'~t}\\
\nonumber i\hbar\partial_t \phi_{k}&=&h\phi_{k}+Q(t)\sum^{M}_{i,j,l,m=1}[\rho^{-1}]_{kj}(t)\rho_{ji,lm}(t)W_{lm}(t)\phi_i
\end{eqnarray}
We remark that the matrix $\mathcal{H}(t)$ only depends on the time-dependent single-particle orbitals, while the one-body and two-body density matrices only depend on the coefficients $C_\mathbf{n}(t)$.

We now assume that all single-particle wave functions behave as a Floquet wave,
$\phi_{k}(\mathbf{x},t)=e^{-i\varepsilon_kt/\hbar}u_{k}(\mathbf{x},t),~u_{k}(\mathbf{x},t+T)=u_{k}(\mathbf{x},t)$. This implies $\ket{\mathbf{n}~t}=e^{-i\varepsilon_\mathbf{n}t/\hbar}\ket{\bar{\mathbf{n}}~t}$, with $\varepsilon_\mathbf{n}=\sum^{M}_{i=1} n_i\varepsilon_i$ and $\ket{\bar{\mathbf{n}}~t}$ a periodic Fock state. In turn, the time-dependent phase can be absorbed as $C_\mathbf{n}=e^{i\varepsilon_\mathbf{n}t/\hbar}\tilde{C}_\mathbf{n}$, which yields
\begin{equation}
    i\hbar\frac{d\tilde{C}_\mathbf{n}}{dt}=\sum_{\mathbf{n}'}\bar{\mathcal{H}}_{\mathbf{n}\mathbf{n}'}(t)\tilde{C}_{\mathbf{n}'},~\bar{\mathcal{H}}_{\mathbf{n}\mathbf{n}'}(t)=\bra{\bar{\mathbf{n}}~t}\hat{V}\ket{\bar{\mathbf{n}}'~t}+\varepsilon_\mathbf{n}\delta_{\mathbf{n}\mathbf{n}'}
\end{equation}
Since $\bar{\mathcal{H}}(t)$ is a linear periodic operator with respect to $\tilde{C}_\mathbf{n}$, there are solutions of the form $\tilde{C}_\mathbf{n}=e^{-i\varepsilon t/\hbar}\bar{C}_\mathbf{n}(t)$, with $\bar{C}_\mathbf{n}(t)$ periodic. In this case, the total many-body wave function is indeed an SMBF state,
\begin{equation}
        \ket{\Psi(t)}=e^{-i\varepsilon t/\hbar}\ket{u(t)},~\ket{u(t)}=\sum_{\mathbf{n}}\bar{C}_\mathbf{n}(t)\ket{\bar{\mathbf{n}}~t}
\end{equation}
$\ket{u(t)}=\ket{u(t+T)}$ being periodic. This also implies that the one-body and two-body density matrices behave as
\begin{equation}
    \rho_{lk}(t)=e^{-i(\varepsilon_l-\varepsilon_k)t/\hbar}\bar{\rho}_{lk}(t),~
    \rho_{lk,jm}(t)=e^{-i(\varepsilon_l+\varepsilon_j-\varepsilon_m-\varepsilon_k)t/\hbar}\bar{\rho}_{lk,jm}(t)
\end{equation}
In addition, $Q(t)$ is periodic, and $W_{lm}(t)=e^{-i(\varepsilon_m-\varepsilon_l)t/\hbar}\bar{W}_{lm}(t)$, where the functions with a bar are periodic. Therefore, in the equation for the $u_k$,
\begin{equation}
    i\hbar\partial_t u_{k}=(h-\varepsilon_k)u_{k}+Q(t)\sum^{M}_{i,j,l,m=1}[\bar{\rho}^{-1}]_{kj}(t)\bar{\rho}_{ji,lm}(t)\bar{W}_{lm}(t)u_i
\end{equation}
the rightmost term is self-consistently periodic. After expanding in Floquet components each $u_{k}(\mathbf{x},t)$ as in the main text, a complicated system of nonlinear equations can be derived. For a better understanding of the emergence of SMBF states and the behavior of their quantum fluctuations, we consider some well-known limits of the MCTDH equations for both bosons and fermions.

\subsection{Gross-Pitaevskii and Bogoliubov-de Gennes equations}

The GP equation (\ref{eq:HamiltonianEffective}) is obtained by considering the MCTDH equations for $N$ spin-0 bosons condensed in a single orbital $\phi_0(\mathbf{x},t)$. First, we note that the Hamiltonian (\ref{eq:HamiltonianManyBody}) is in fact equivalent to that in Eq. (\ref{eq:HamiltonianManyBodyGeneral}) as the field operator of spin-0 bosons reads
\begin{equation}
    \hat{\Psi}(\mathbf{x})=\sum_k \phi_k(\mathbf{x})\hat{c}_k
\end{equation}
where $k$ here runs over the complete single-particle basis of the Hilbert space.

For $M=1$, all density matrices are just scalars, $\rho_{00}=N$ and $\rho_{00,00}=N(N-1)$, and there is only one coefficient $C_\mathbf{n}$ in the expansion of Eq. (\ref{eq:MBAnsatz}), whose time evolution is just a trivial time-dependent phase. We define a single-particle mean-field operator
\begin{equation}
    h^{\rm{MF}}\equiv  h+(N-1)W_{00},~W_{00}(\mathbf{x},t)=\int\mathrm{d}\mathbf{x}'~ V(\mathbf{x}-\mathbf{x}')|\phi_0(\mathbf{x}',t )|^2
\end{equation}
By taking $f=Ph^{\rm{MF}}P$ in Eq. (\ref{eq:MCTDH}), we get $ i\partial_t\phi_0=h^{\rm{MF}}\phi_0$. The usual form of the GP equation $h^{\rm{MF}}=H_{\rm{GP}}$ is retrieved after defining $\Psi(\mathbf{x},t)\equiv \sqrt{N}\phi_0(\mathbf{x},t)$, where $N\simeq N-1$ for large $N$, and by replacing the actual interaction by a contact pseudopotential $g\delta(\mathbf{x}-\mathbf{x}')$.

A simple derivation of the BdG equations for the quantum fluctuations $\hat{\phi}(\mathbf{x},t)$ is obtained by direct substitution $\Psi(\mathbf{x},t)\rightarrow\Psi(\mathbf{x},t)+\hat{\phi}(\mathbf{x},t)$ in the GP equation and expansion up to linear order. In the specific case of an SMBF state discussed in the main text, one uses a decomposition of the type $\Psi(\mathbf{x},t)\rightarrow\left[u(\mathbf{x},t)+\hat{\phi}(\mathbf{x},t)\right]e^{-i\tilde{\mu}t/\hbar}$,
which yields Eq. (\ref{eq:BdGfieldequation}). 

\subsection{Hartree-Fock equations and time-dependent Hartree-Fock approximation}

The Hartree-Fock (HF) equations for fermions are recovered from the MCTDH equations by considering $N$ fermions in $M=N$ orbitals, which means that there is only one Slater determinant present in the expansion of Eq. (\ref{eq:MBAnsatz}). Thus, the one-body and two-body density matrices of Eq. (\ref{eq:MCTDH}) are $\rho_{lk}(t)=\delta_{lk}$ and $\rho_{lk,jm}(t)=\delta_{lk}\delta_{jm}-\delta_{lm}\delta_{jk}$. We define a single-particle mean-field operator, the HF Hamiltonian
\begin{equation}
    h^{\rm{HF}}\equiv h+V^{\rm{HF}},~V^{\rm{HF}}\phi_k=\sum^M_{i=1}W_{ii}\phi_k-W_{ik}\phi_i
\end{equation}
After taking $f=Ph^{\rm{HF}}P$, we recover the usual form of the self-consistent time-dependent Hartree-Fock (TDHF) equations 
\begin{equation}\label{eq:TDHFSC}
    i\hbar\partial_t\left[\phi_{k}(\mathbf{x},t)\right]_\alpha=\sum_\beta \int\mathrm{d}\mathbf{x}'~h_{\alpha\beta}^{\rm{HF}}(\mathbf{x},\mathbf{x}',t)\left[\phi_{k}(\mathbf{x}',t)\right]_\beta=\sum_\beta h_{\alpha\beta}(\mathbf{x})\left[\phi_{a}(\mathbf{x},t)\right]_\beta+\int\mathrm{d}\mathbf{x}'~V_{\alpha\beta}^{\rm{HF}}(\mathbf{x},\mathbf{x}',t)\left[\phi_{k}(\mathbf{x}',t)\right]_\beta
\end{equation}
The explicit expression of the nonlocal time-dependent HF potential $V^{\rm{HF}}_{\alpha\beta}(\mathbf{x},\mathbf{x}',t)$
is given by
\begin{equation}\label{eq:TDHFSCPotential}
    V^{\rm{HF}}_{\alpha\beta}(\mathbf{x},\mathbf{x}',t)= \sum^M_{k=1}\sum_{\lambda\mu}\delta(\mathbf{x}-\mathbf{x}')\int\mathrm{d}\mathbf{x}''~V_{\alpha\beta,\lambda\mu}(\mathbf{x}-\mathbf{x}'')\left[\phi^*_k(\mathbf{x}'',t)\right]_\lambda \left[\phi_k(\mathbf{x}'',t)\right]_\mu- V_{\alpha\mu,\lambda\beta}(\mathbf{x}-\mathbf{x}')\left[\phi_k(\mathbf{x},t)\right]_\mu \left[\phi^*_k(\mathbf{x}',t)\right]_\lambda
\end{equation}
If we assume Floquet solutions $\phi_k(\mathbf{x},t)=e^{-i
\varepsilon_k t/\hbar}u_k(\mathbf{x},t)$, with $u_k(\mathbf{x},t+T)=u_k(\mathbf{x},t)$, the HF Hamiltonian then becomes periodic and the Floquet states are self-consistent solutions of the TDHF equations. This is an explicit example of a fermionic SMBF state.

Moreover, the spectrum of quantum fluctuations, computed from the TDHF approximation (TDHFA), also takes the form of a periodic Floquet equation. First, we move to the full single-particle Floquet basis of the periodic HF Hamiltonian $\phi_k(\mathbf{x},t)=e^{-i
\varepsilon_k t/\hbar}u_k(\mathbf{x},t)$, where in the following indices $k=a,b=1\ldots M$ label occupied states and $k=c,d$ unoccupied ones. Due to the formal analogy between the BdG and the TDHFA equations (see for instance Appendix B of Ref. \citeSupp{deNova2017}), quantum fluctuations can be computed by considering small particle-hole excitations of the form:
\begin{equation}\label{eq:TDHFASMBFLinear}
    \phi_{a}(\mathbf{x},t)\rightarrow e^{-i
\varepsilon_a t/\hbar}\left[u_{a}(\mathbf{x},t)+\sum_c \hat{C}_{ca}(t)u_{c}(\mathbf{x},t)\right],~\hat{C}_{ca}(t)\equiv\hat{c}^\dagger_{a}(t)\hat{c}_{c}(t)
\end{equation}
By expanding the TDHF equations (\ref{eq:TDHFSC}) up to linear order and projecting onto the unoccupied states $u_{c}(\mathbf{x},t)$, we obtain the TDHFA equations
\begin{equation}\label{eq:TDHFAFloquetElectronHole}
      i\hbar\partial_t\hat{C}_{ca}=(\varepsilon_c-\varepsilon_a) \hat{C}_{ca}+\sum_{bd}\left[\bar{V}_{ca,bd}(t)-\bar{V}_{cd,ba}(t)\right]\hat{C}_{db}+\left[\bar{V}_{ca,db}(t)-\bar{V}_{cb,da}(t)\right]\hat{C}^{\dagger}_{db}\equiv \sum_{bd} N_{ac,bd}(t)\hat{C}_{db}+A_{ac,db}(t)\hat{C}^{\dagger}_{db}
\end{equation}
or, in matrix notation,
\begin{equation}\label{eq:TDHFAMatrix}
i\hbar\frac{d\hat{\mathbf{C}}}{dt}=\mathbf{X}(t)\hat{\mathbf{C}},~\hat{\mathbf{C}}=\left[\begin{array}{c}
\hat{C}_{ca}(t)\\
\hat{C}^{\dagger}_{ca}(t)
\end{array}\right],~\mathbf{X}=\left[\begin{array}{c|c}
\mathbf{N} & \mathbf{A} \\
\hline
-\mathbf{A}^* & -\mathbf{N}^*
\end{array}\right]
\end{equation}
where $\bar{V}_{lk,jm}(t)$ are the matrix elements of the interacting potential in terms of the $u_k(\mathbf{x},t)$. Since they are periodic, the matrix $\mathbf{X}(t)$ of the TDHFA equations is periodic and thus, the spectrum is also described in terms of Floquet bands. The BdG/TDHFA correspondence implies that a similar expansion to that of Eq. (\ref{eq:BdGmodesexpansion}) can be made \citeSupp{deNova2017}. A more rigorous derivation of the TDHFA equations involves the Heisenberg equation of motion for the operators $\hat{C}_{ca}(t)$, where pairs of operators are contracted in the spirit of the HF approximation to obtain a linear equation, namely, Eq. (\ref{eq:TDHFAFloquetElectronHole}).

\subsection{Bose-Hubbard model}
We consider a general Bose-Hubbard Hamiltonian of the type
\begin{equation}
    \hat{H}=-\sum_{ij} t_{ij}\hat{c}^\dagger_i\hat{c}_j+\sum_i(\epsilon_i-\mu)\hat{c}^\dagger_i\hat{c}_i+\frac{U}{2}\hat{c}^\dagger_i\hat{c}^\dagger_i\hat{c}_i\hat{c}_i
\end{equation}
with $i,j$ labeling lattice sites, $t_{ij}$ the hopping matrix and $\epsilon_i$ some on-site energy. We first consider the discrete GP equation, obtained by an ansatz of the form
\begin{equation}\label{eq:BHGPAnsatz}
    \ket{\Psi(t)}= \frac{\left[\sum_i\phi_i(t)\hat{c}^{\dagger}_i\right]^{N}}{\sqrt{N!}}\ket{\mathbf{0}}
\end{equation}
After inserted into the Dirac-Frenkel variational principle, it gives 
\begin{equation}\label{eq:discreteGP}
i\hbar\partial_t\psi_i=\sum_{j}H_{ij}\psi_j,~H_{ij}=-t_{ij}+\delta_{ij}\left[(\epsilon_i-\mu)+U|\psi_i|^2\right]
\end{equation}
with $\psi_i\equiv\sqrt{N}\phi_i$. SMBF states emerge once more when one considers Floquet states $\psi_i(t)=e^{-i \tilde{\mu} t/\hbar}u_i(t)$, $u_i(t+T)=u_i(t)$, since the resulting nonlinear operator $H_{ij}$ is self-consistently periodic. Similarly, the associated discrete BdG equations for the quantum fluctuations (obtained after expanding $\psi_i\rightarrow[u_i(t)+\hat{b}_i]e^{-i \tilde{\mu} t/\hbar}$) are periodic:
\begin{equation}
    i\hbar\partial_t\hat{b}_i=\sum_{j}N_{ij}\hat{b}_j+Uu^2_i(t)\hat{b}^{\dagger}_i,~N_{ij}=-t_{ij}+\delta_{ij}\left[(\epsilon_i-\mu)+2U|u_i(t)|^2-\tilde{\mu}\right]
\end{equation}
Thus, the BdG spectrum is also described in terms of quasi-energy bands.



More interestingly, SMBF states can also arise within a Gutzwiller ansatz, which goes beyond the MCTDH framework and is able to describe the superfluid-Mott insulator phase transition [39],
\begin{equation}
    \ket{\Psi(t)}=\prod_i \ket{\phi_i(t)},~\ket{\phi_i(t)}=\sum^{\infty}_{n=0} c_{i,n}(t)\ket{n}_i,~\sum^{\infty}_{n=0} |c_{i,n}(t)|^2=1
\end{equation}
$\ket{n}_i$ being a Fock state of the lattice site $i$. When inserted into the Dirac-Frenkel variational principle, this ansatz yields the following equations of motion
\begin{equation}\label{eq:Gutzwiller}
    i\hbar\frac{dc_{i,n}}{dt}=\sum^{\infty}_{m=0}\mathcal{H}^{(i)}_{nm}(t)c_{i,m},~\mathcal{H}^{(i)}_{nm}=\delta_{nm}\left[n(\epsilon_i-\mu)+U\frac{n(n-1)}{2}\right]-\sum_{j}t_{ij}\left[\delta_{n,m+1}\sqrt{n}\Psi_j+\delta_{n+1,m}\sqrt{m}\Psi^*_j\right]
\end{equation}
with 
\begin{equation}
    \Psi_i(t)=\braket{\hat{c}_i}=\sum^{\infty}_{n=0} \sqrt{n+1}c^*_{i,n}(t)c_{i,n+1}(t)
\end{equation}
Once more, if $\mathcal{H}^{(i)}(t)$ is periodic, one can look for Floquet solutions of Eq. (\ref{eq:Gutzwiller}), $c_{i,n}(t)=e^{-i\varepsilon_it/\hbar}a_{i,n}(t)$, with $a_{i,n}(t)$ periodic. Self-consistently, $\Psi_i(t)$ becomes periodic and, consequently, $\mathcal{H}^{(i)}(t)$ does. The on-site wave functions behave as Floquet states $\ket{\phi_i(t)}=e^{-i\varepsilon_it/\hbar}\ket{u_i(t)}$, and hence the total wave function $\ket{\Psi(t)}$ becomes an SMBF state.

Regarding quantum fluctuations, they are also computed by expanding to linear order as in the previous cases, $c_{i,n}(t)\rightarrow e^{-i\varepsilon_it/\hbar}\left[a_{i,n}(t)+\hat{b}_{i,n}(t)\right]$, which yields periodic BdG equations of the form \citeSupp{Caleffi2020}
\begin{equation}
    i\hbar\partial_t\hat{b}_{i,n}=\sum^{\infty}_{m=0}\sum_{j}N^{ij}_{nm}(t)\hat{b}_{j,m}+A^{ij}_{nm}(t)\hat{b}^{\dagger}_{j,m}
\end{equation}
or, in matrix notation,
\begin{equation}
    i\hbar\frac{d\hat{\mathbf{b}}}{dt}=\mathbf{M}(t)\hat{\mathbf{b}},~\hat{\mathbf{b}}=\left[\begin{array}{c}
\hat{b}_{i,n}\\
\hat{b}^{\dagger}_{i,n}
\end{array}\right],~\mathbf{M}(t)=\left[\begin{array}{c|c}
\mathbf{N} & \mathbf{A} \\
\hline
-\mathbf{A}^* & -\mathbf{N}^*
\end{array}\right]
\end{equation}
where

\begin{eqnarray}
    \nonumber N^{ij}_{nm}(t)&\equiv&{H}^{(i)}_{nm}(t)\delta_{ij}-t_{ij}\left[\sqrt{n}\sqrt{m}a_{n-1,i}(t)a^*_{m-1,j}(t)+\sqrt{n+1}\sqrt{m+1}a_{n+1,i}(t)a^*_{m+1,j}(t)\right]\\
    A^{ij}_{nm}(t)&\equiv&-t_{ij}\left[\sqrt{n}\sqrt{m+1}a_{n-1,i}(t)a_{m+1,j}(t)+\sqrt{n+1}\sqrt{m}a_{n+1,i}(t)a_{m-1,j}(t)\right]
\end{eqnarray}
are periodic operators, and thus, $\mathbf{M}(t)$ is periodic.

\begin{figure}[t]
\begin{tabular}{@{}ccc@{}}
    \includegraphics[width=0.33\columnwidth]{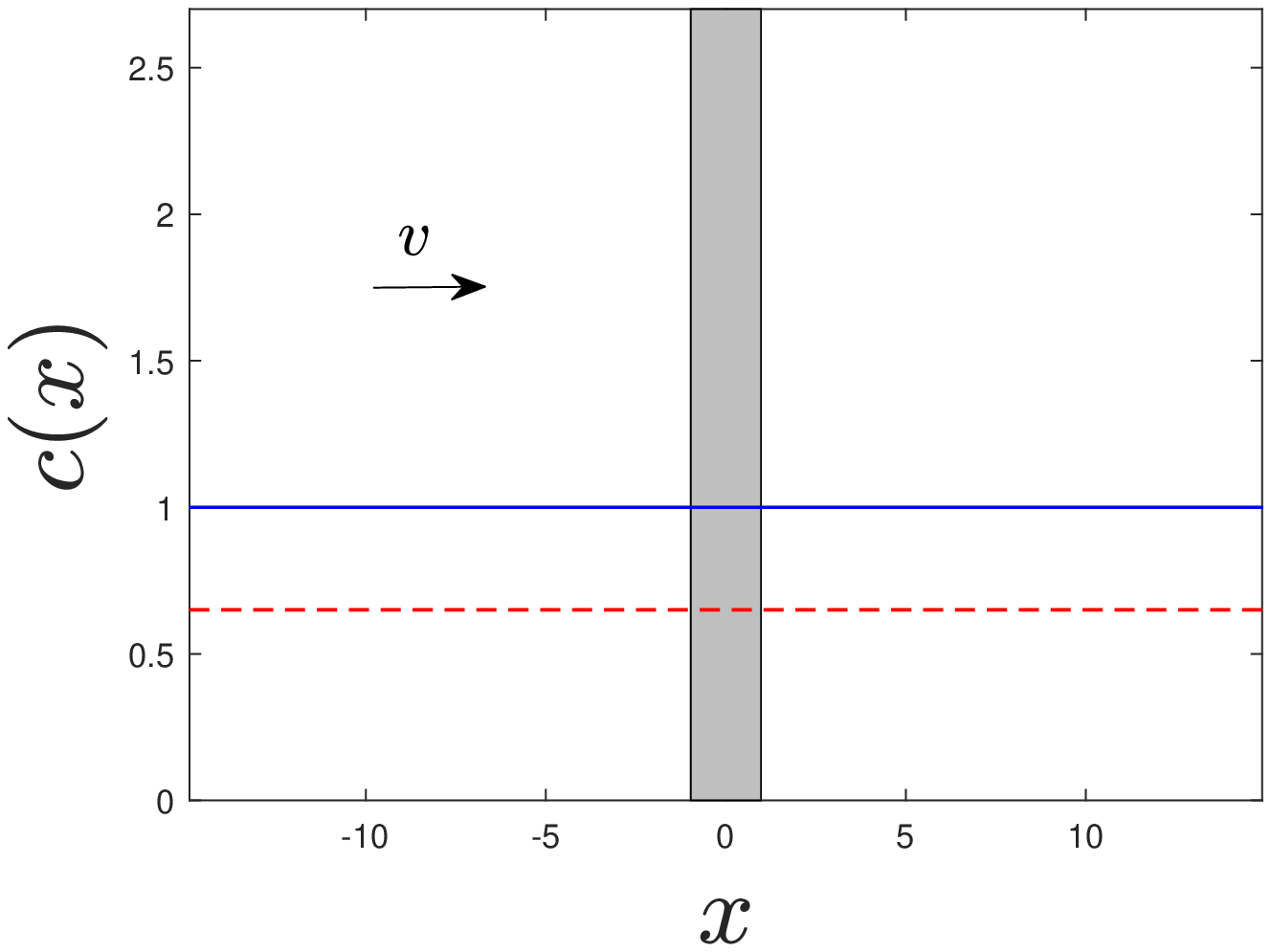} & 
    \includegraphics[width=0.33\columnwidth]{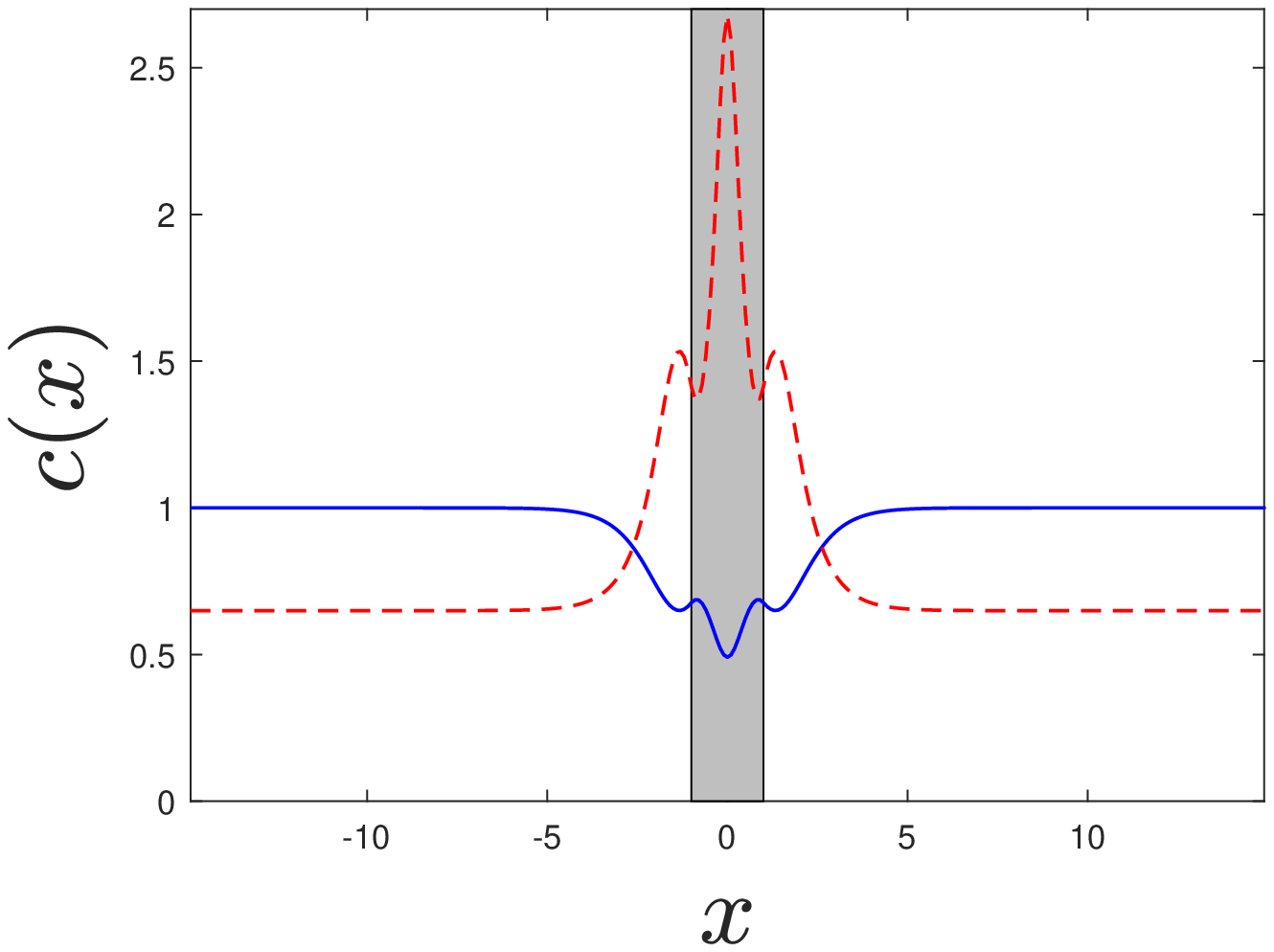} &
    \includegraphics[width=0.33\columnwidth]{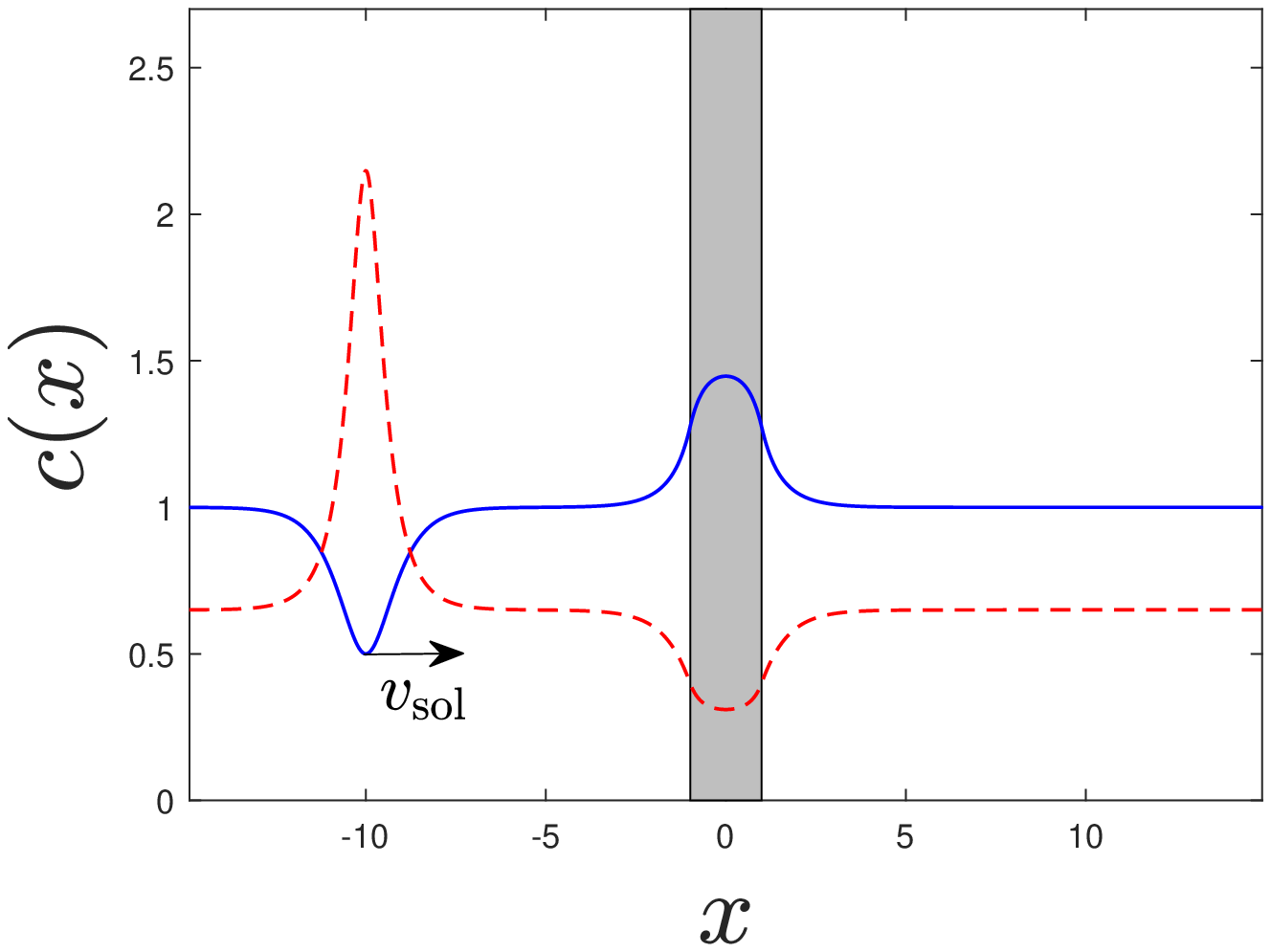} \\ \\ \\
    \includegraphics[width=0.33\columnwidth]{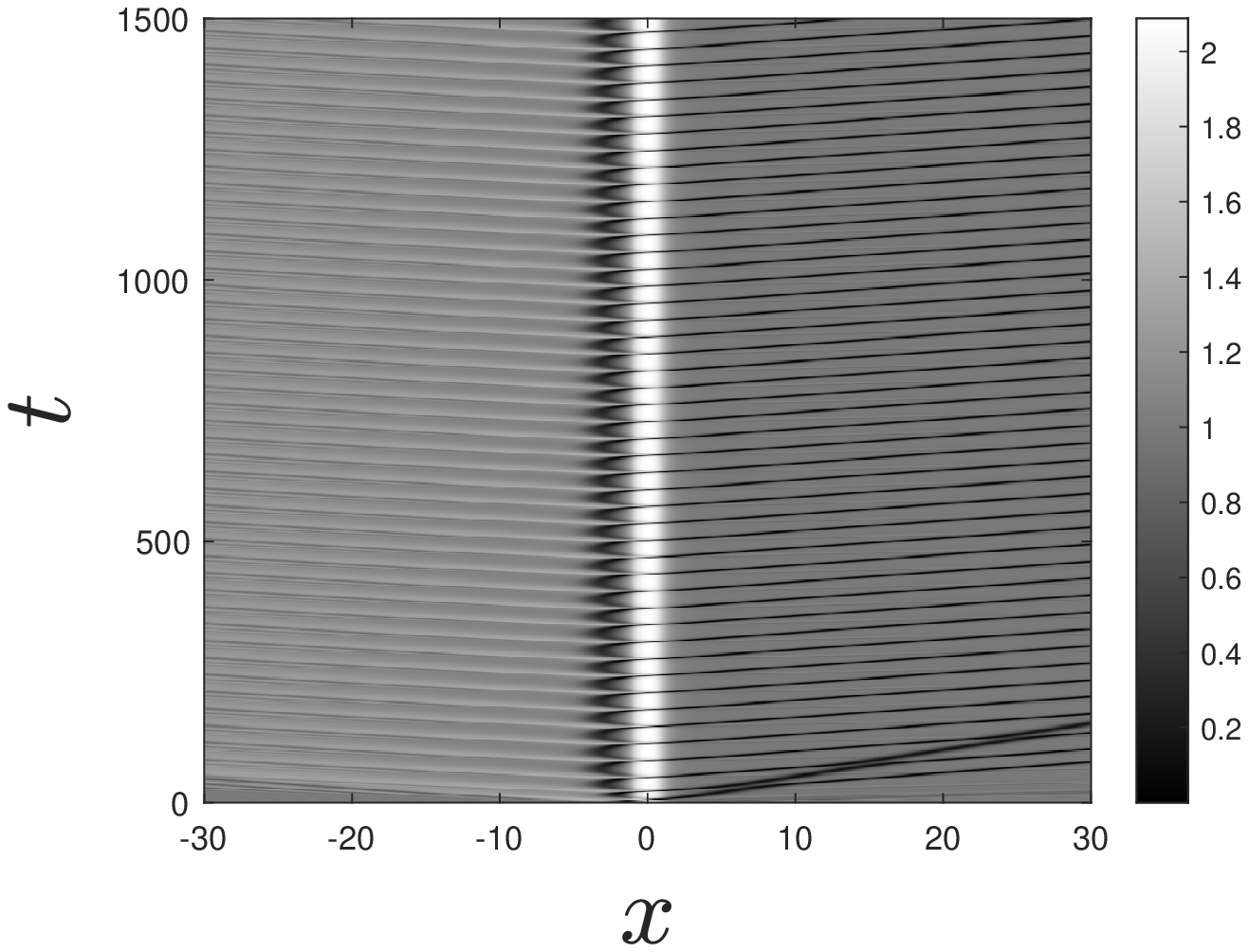} & 
    \includegraphics[width=0.33\columnwidth]{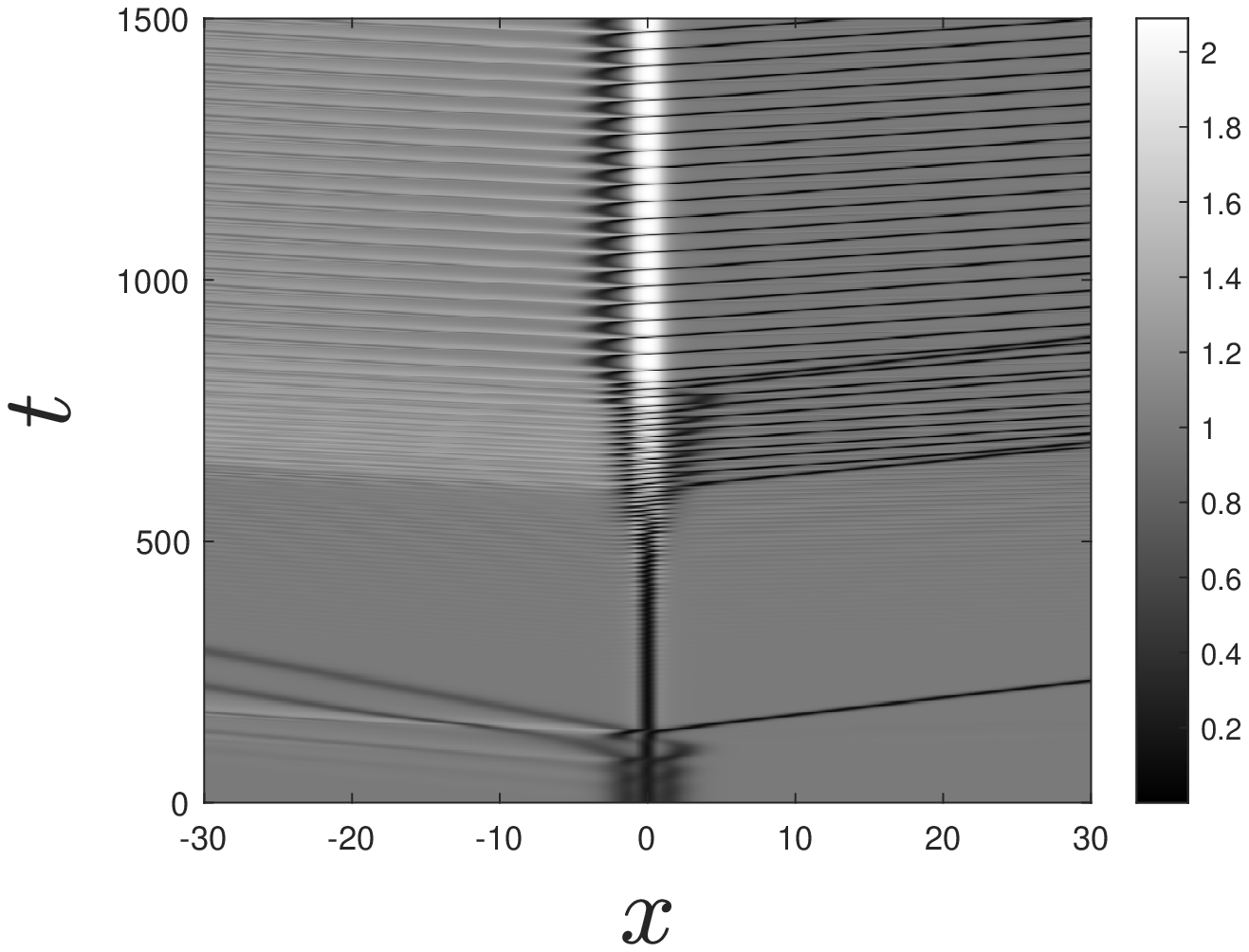} &
    \includegraphics[width=0.33\columnwidth]{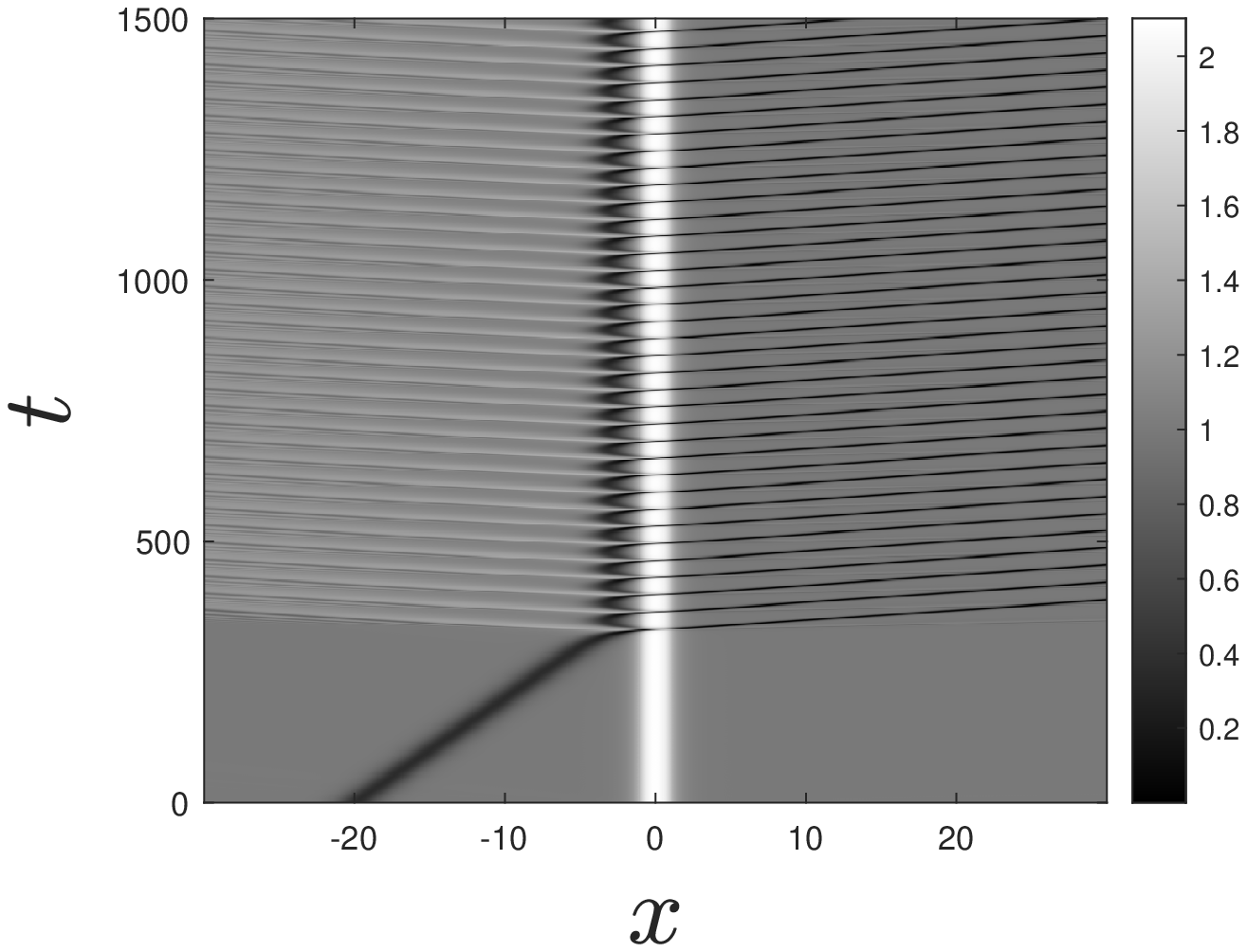}
\end{tabular}
\caption{Upper row: Spatial profile of sound (solid blue) and flow (dashed red) velocities for different types of initial condition, where the shaded area represents the region in which the attractive constant potential $V(x)=-V_0$ is placed. Lower row: 2D plot of $|\Psi(x,t)|^2$ for the time evolution of each initial condition above, where in all cases $v=0.65,~V_0=1,~X=2$. Left column: IHFC. Central column: BHL. Right column: SGS.} 
\label{fig:Model}
\end{figure}

\begin{figure}[t]\includegraphics[width=0.5\columnwidth]{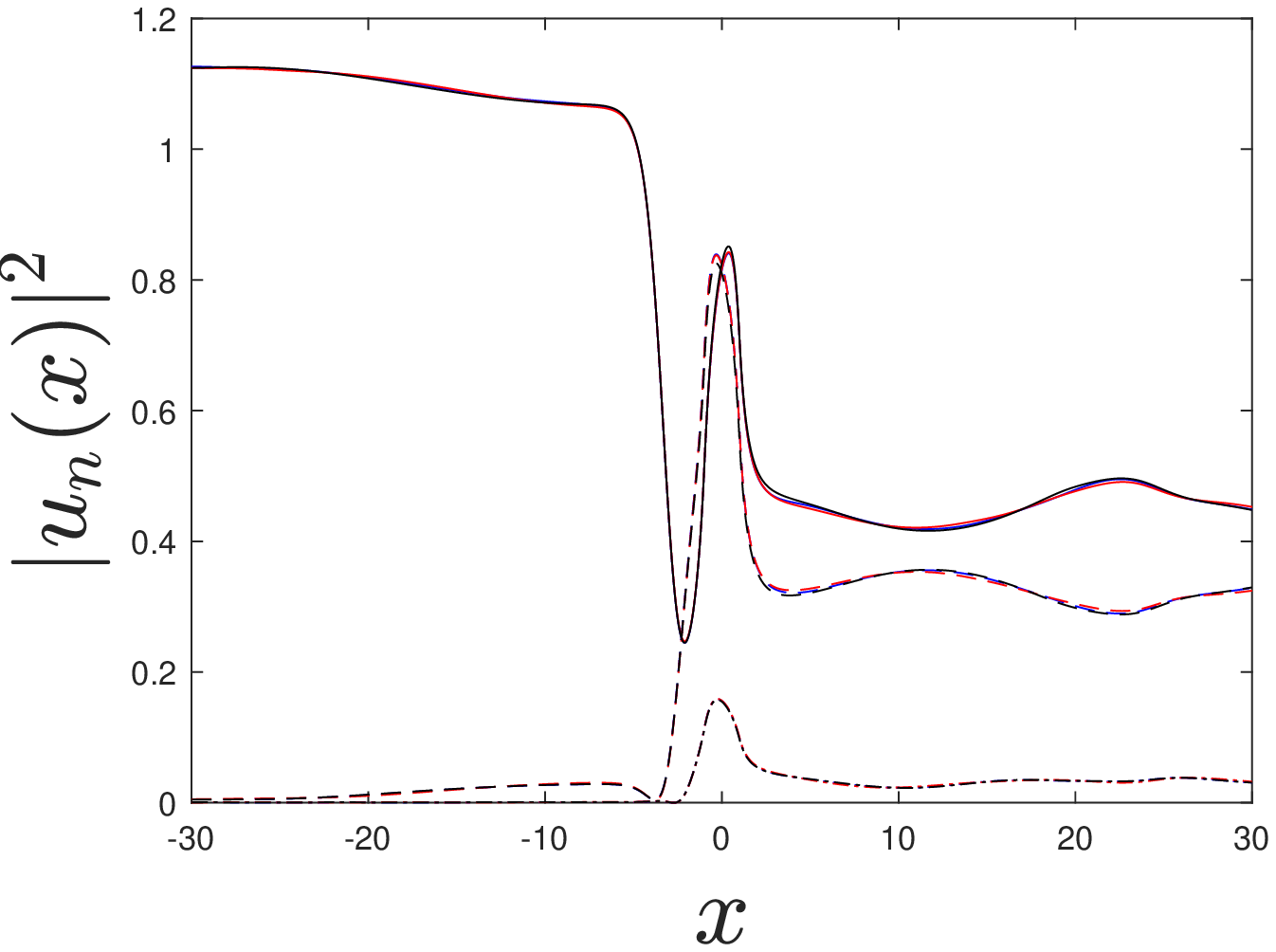}
\caption{Comparison of the Floquet components $u_0(x)$ (solid line), $u_{-1}(x)$ (dashed line) and $u_{-2}(x)$ (dashed-dotted line) of the SMBF state for the three simulations of Fig. \ref{fig:Model}. Color code: IHFC (blue), BHL (red), SGS (black).} 
\label{fig:FloquetComparison}
\end{figure}

\section{Robustness of the CES time crystal}

We provide here the technical details of the analysis of the robustness of the CES time crystal.

\subsection{Independence of the initial state and transient}

We examine the independence of the SMBF state with respect to the initial condition and the transient details. In particular, we consider three different types of initial condition at $t=0$ that asymptotically match (upstream and downstream) a subsonic plane wave $\Psi(x,0)\sim e^{ivx}$, and whose evolution is described by the same Hamiltonian (including the same attractive square-well potential). Therefore, their evolution is characterized by the same set of parameters $(v,X,V_0)$. A schematic representation of these initial conditions is provided in Fig. \ref{fig:Model}.

The first type of initial condition is the main model considered in this work, starting from an IHFC in which we introduce an attractive square well potential at $t=0$ (left column in Fig. \ref{fig:Model}). In order to analyze the robustness against the transient details, the square well is not suddenly quenched but instead introduced within a time scale $\tau$.

The second type of initial condition is an unstable stationary black-hole laser (BHL) solution within an attractive square well potential [51] (central column in Fig. \ref{fig:Model}). At $t=0$, some small noise is placed on top of it, triggering the dynamical instabilities that eventually will grow up to the nonlinear saturation regime, where the system will reach either the CES state or the GS [40,41].

Finally, we consider an initial condition in which we directly start from the GS, localized around the attractive square well. However, in the upstream region, we introduce a solitonic defect that travels with velocity $v_{\rm{sol}}$ towards the GS, eventually destabilizing it (right column in Fig. \ref{fig:Model}). We refer to this model as the soliton-ground state (SGS) model. We recall that the IHFC and BHL states were already considered in Ref. [41], while the SGS model is novel of this work.


In all cases, for a fixed set of parameters $(v,V_0,X)$ that lies within the CES region of the phase diagram, the same CES state is eventually reached. A comparison of their time evolution is shown in the bottom row of Fig. \ref{fig:Model}, where we observe that, after some transient that strongly depends on the initial condition, they all enter the CES state. Within numerical accuracy, the three CES states oscillate with the same period. Moreover, we have extracted the Floquet components $u_n(x)$ of each CES state and compared them in Fig. \ref{fig:FloquetComparison}, finding an excellent agreement. 

When reached, the CES state is also quite insensitive to the details of the transient. Specifically, it is independent 1) for an IHFC, from the time scale $\tau$ at which the potential is introduced (except for sufficiently adiabatic rates, in whose case the system always evolves towards GS); 2) for a BHL solution, from the stochastic initial noise; 3) for the SGS model, from the velocity of the launched soliton $v_{\rm{sol}}$ (except for sufficiently fast solitons, whose short passage through the localized GS is not enough to push it to the CES state).



\begin{figure}[t]
\begin{tabular}{@{}ccc@{}}
    \includegraphics[width=0.33\columnwidth]{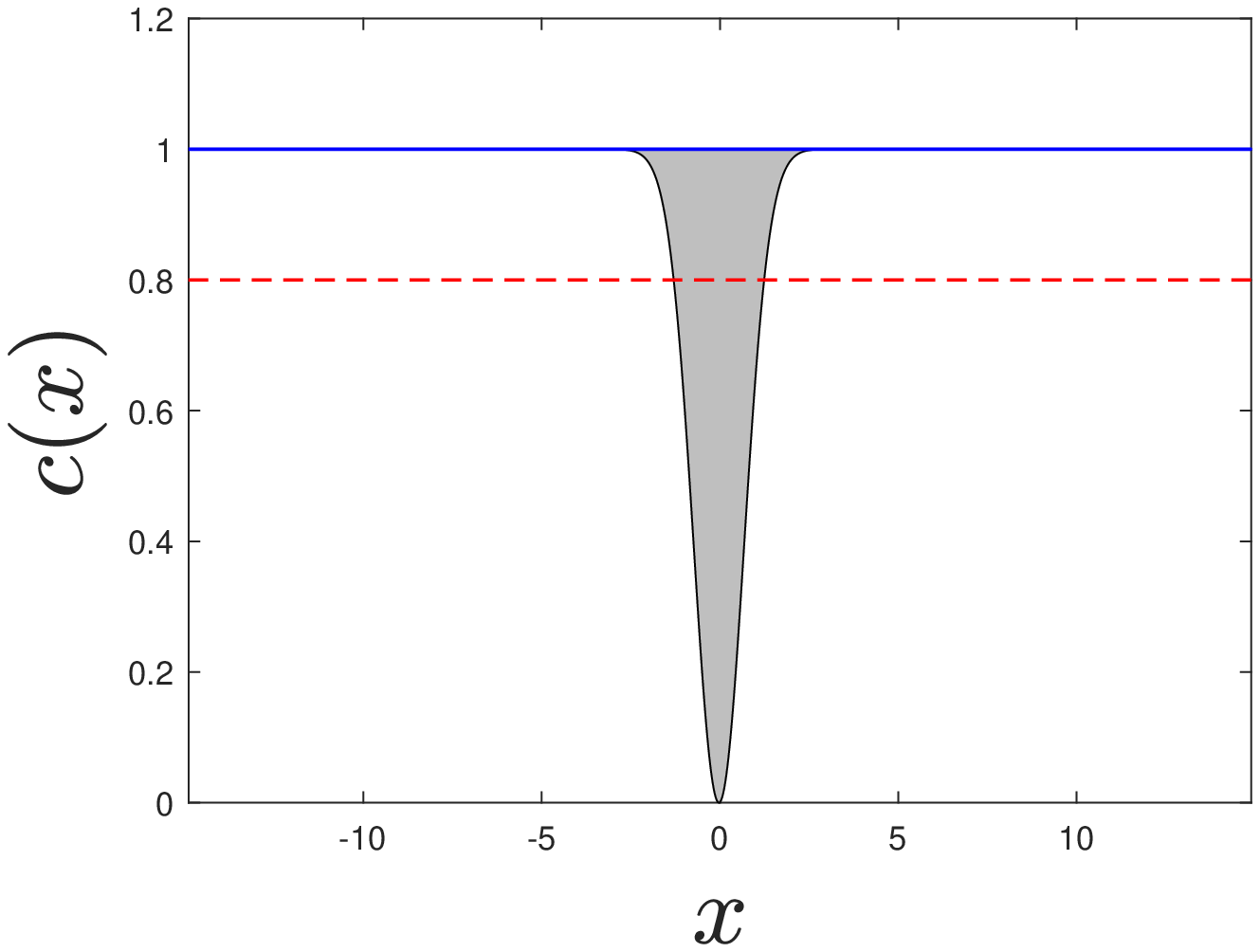} & 
    \includegraphics[width=0.33\columnwidth]{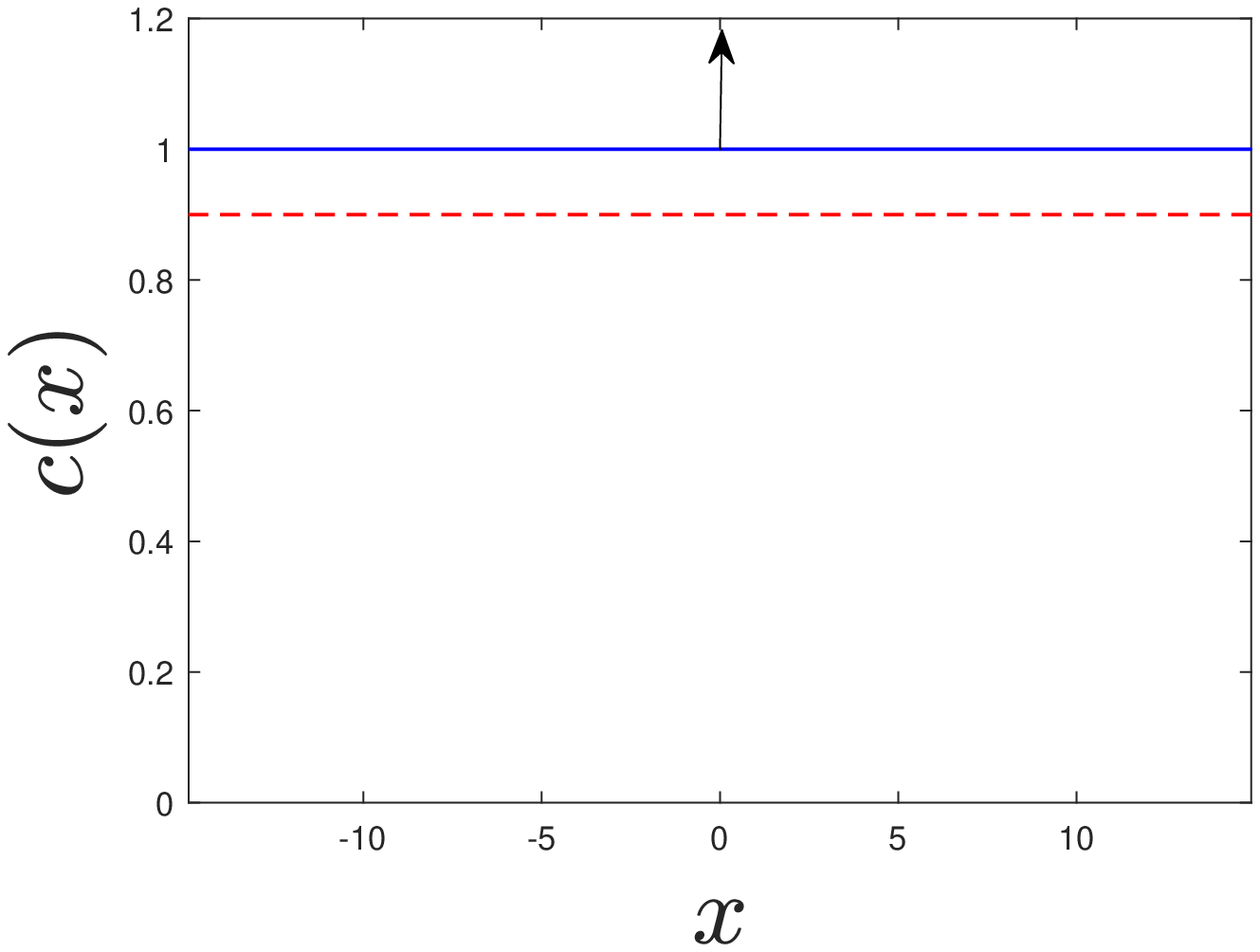} &
    \includegraphics[width=0.33\columnwidth]{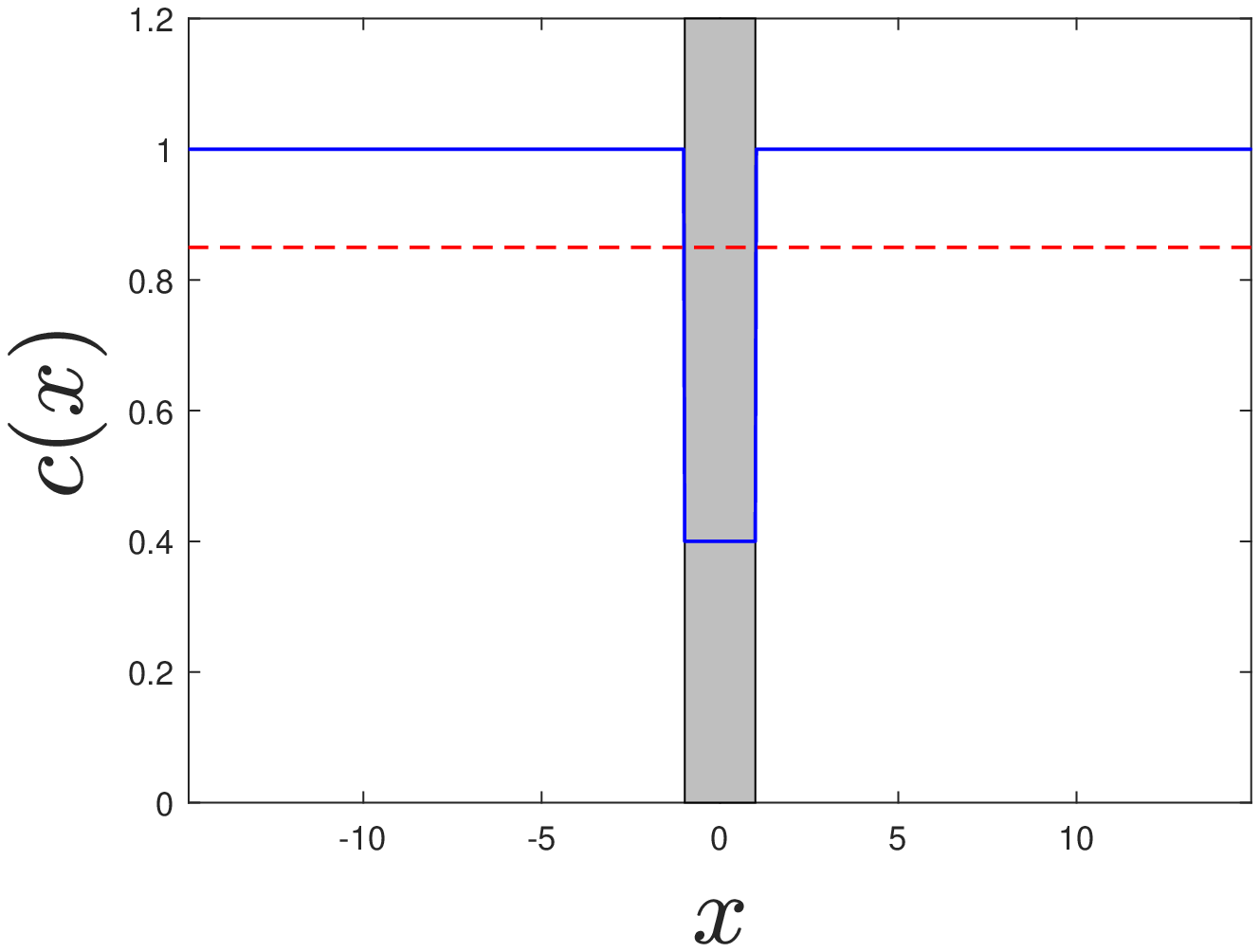} \\ \\ \\
    \includegraphics[width=0.33\columnwidth]{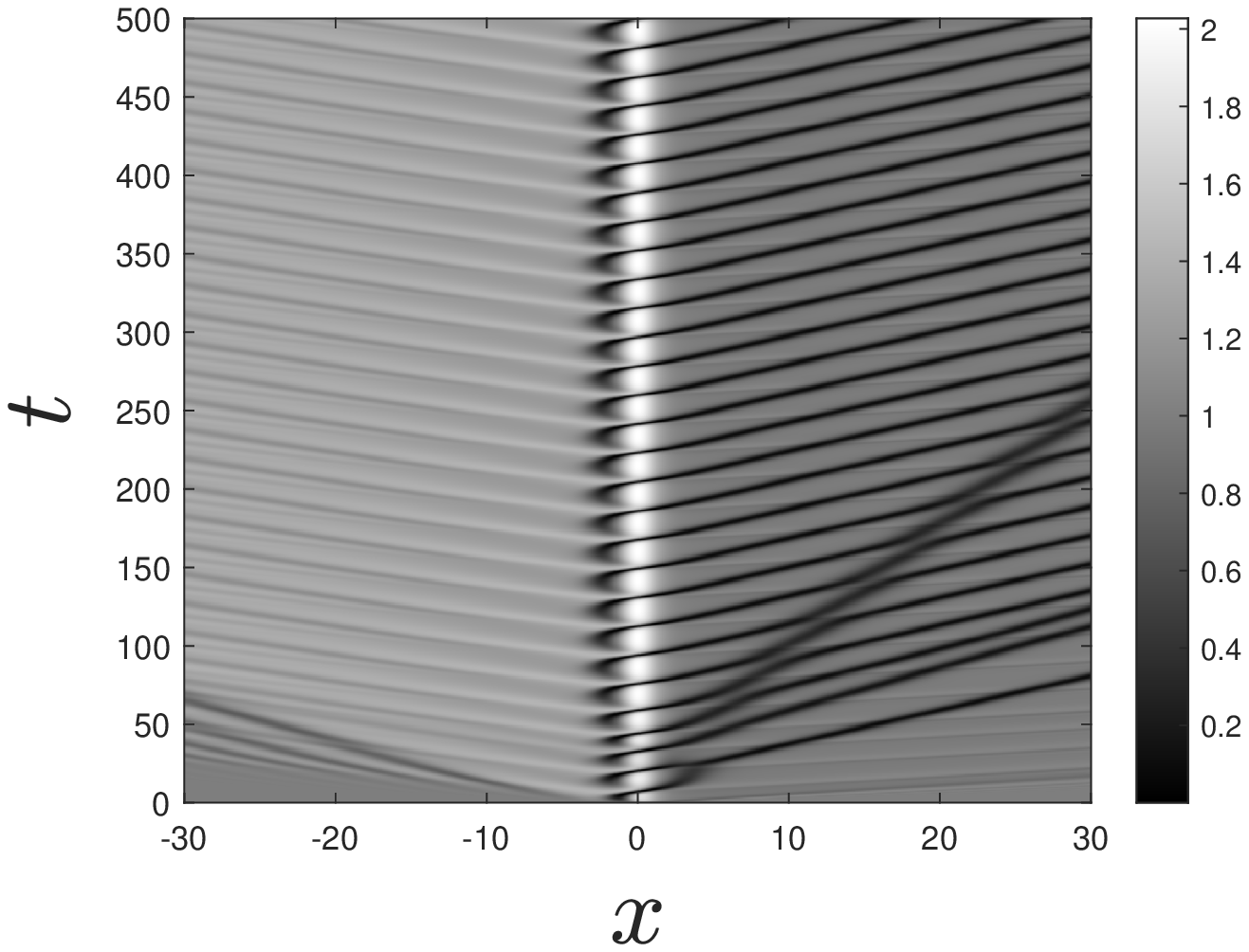} & 
    \includegraphics[width=0.33\columnwidth]{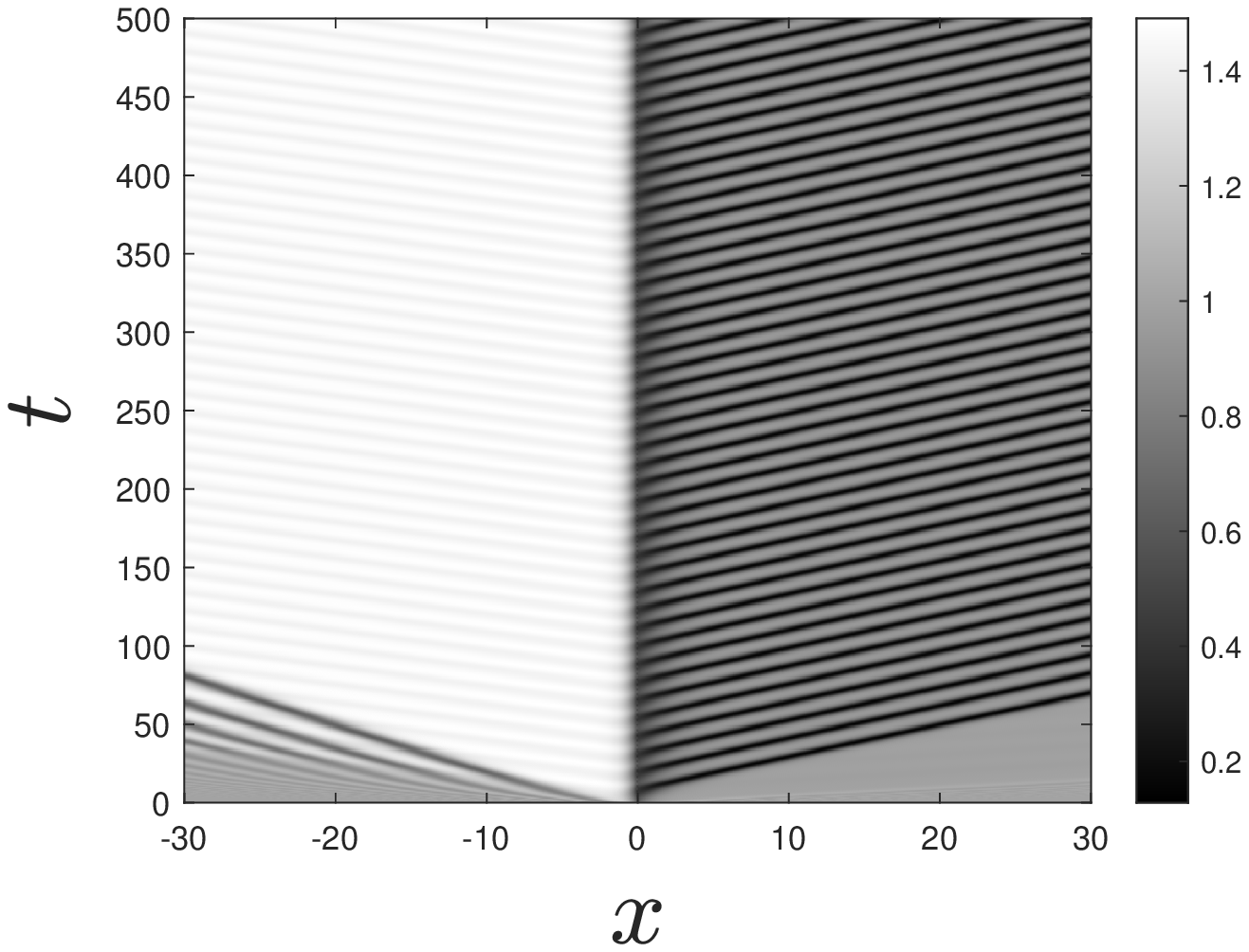} &
    \includegraphics[width=0.33\columnwidth]{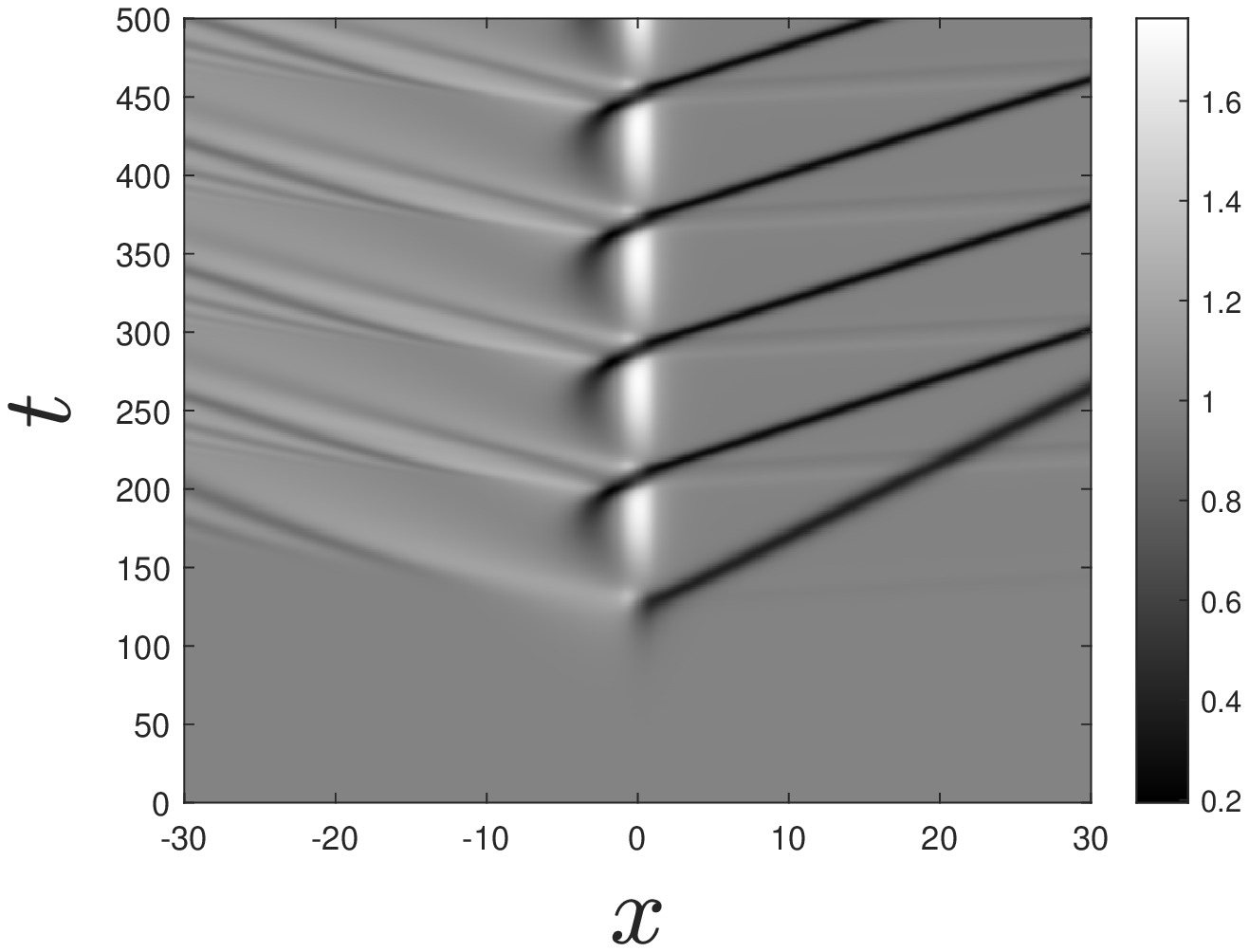}
\end{tabular}
\caption{Upper row: Spatial profile of sound (solid blue) and flow (dashed red) velocities for different configurations at $t=0$. Left: Attractive Gaussian well of Eq. (\ref{eq:GaussianPotential}), where the shaded area represents the potential profile. The velocity is $v=0.8$ and the potential parameters are $V_0=1, \sigma=1$. Center: Repulsive delta barrier of Eq. (\ref{eq:DeltaPotential}). The parameters are $Z=0.5$ and $v=0.9$. Right: Flat-profile BHL of Eq. (\ref{eq:FlatProfile}). Background parameters are $v=0.85,~c_2=0.4,~X=2$. Lower row: 2D plot of $|\Psi(x,t)|^2$ for the time evolution of the initial conditions of the upper row.} 
\label{fig:Universality}
\end{figure}

\begin{figure}[t]
\begin{tabular}{@{}cc@{}}
    \includegraphics[width=0.5\columnwidth]{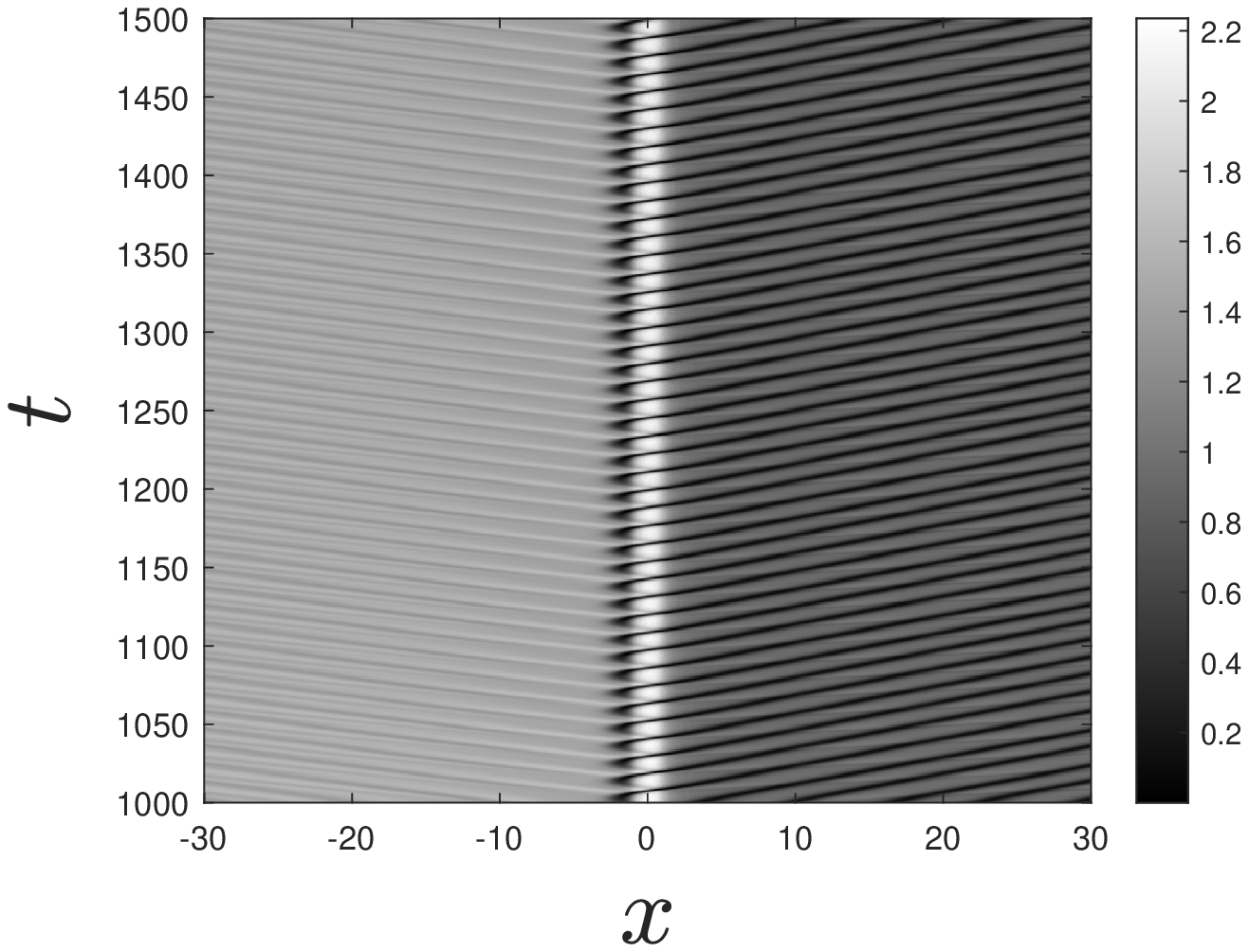} & 
    \includegraphics[width=0.5\columnwidth]{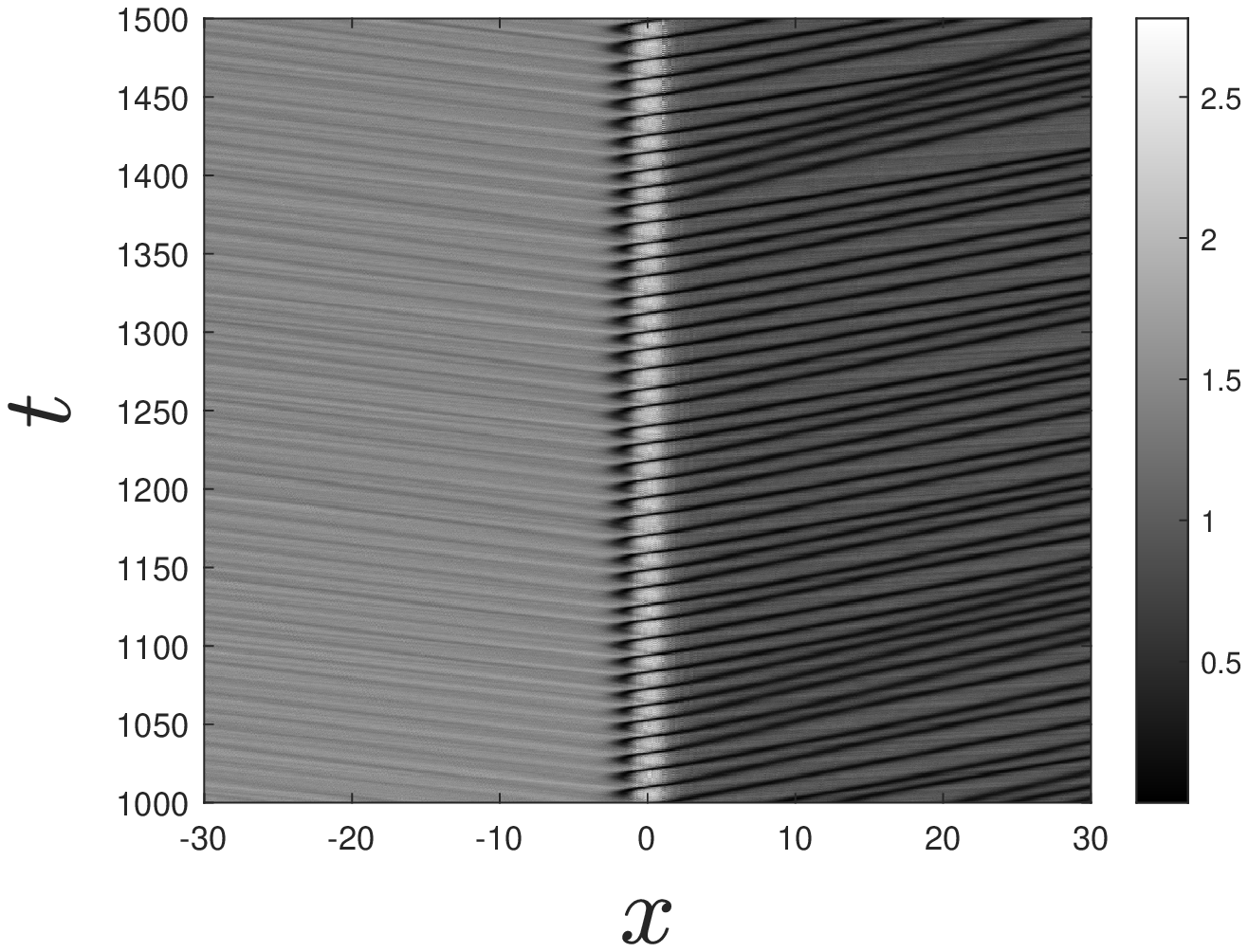} \\ \\ \\ 
    \includegraphics[width=0.5\columnwidth]{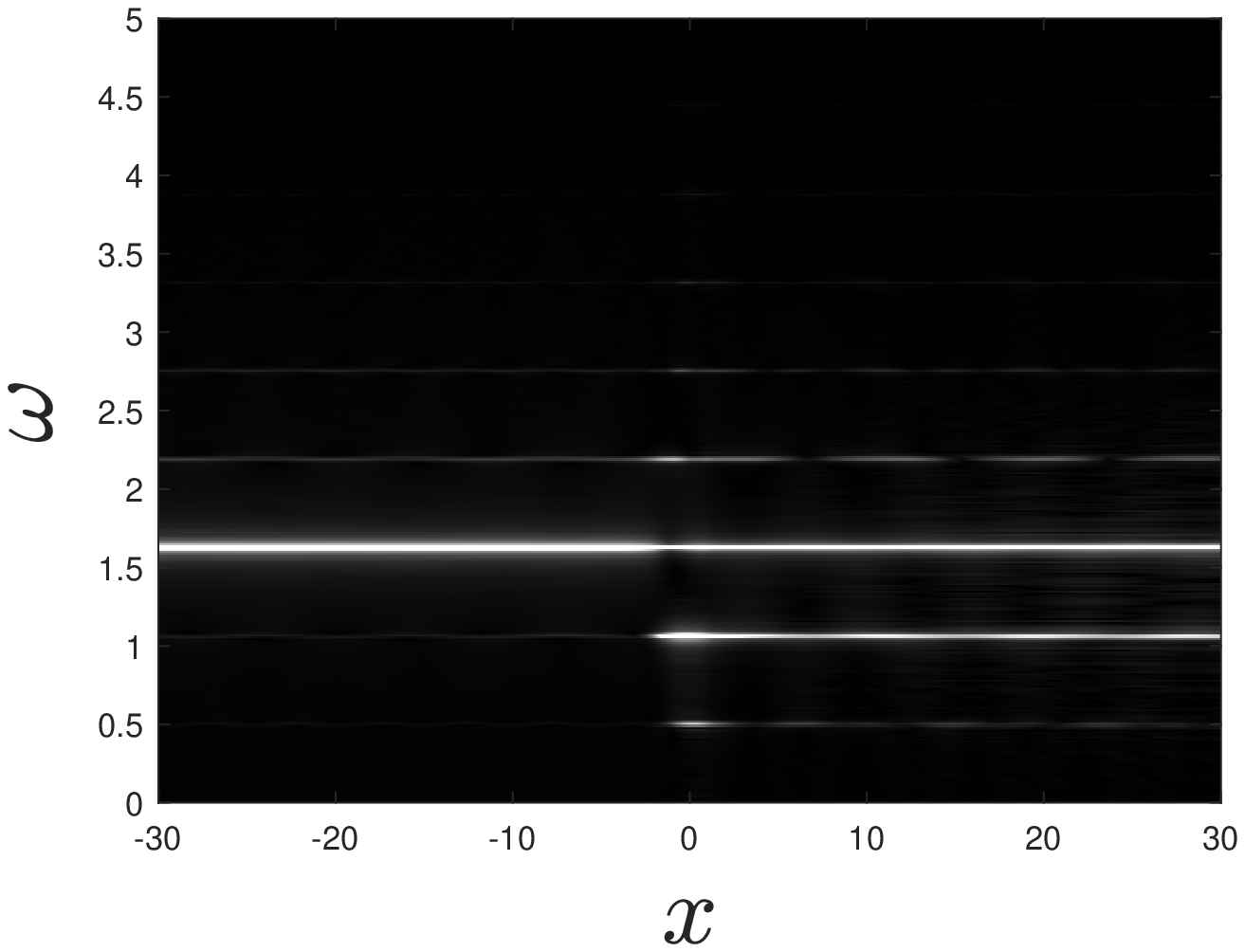} & 
    \includegraphics[width=0.5\columnwidth]{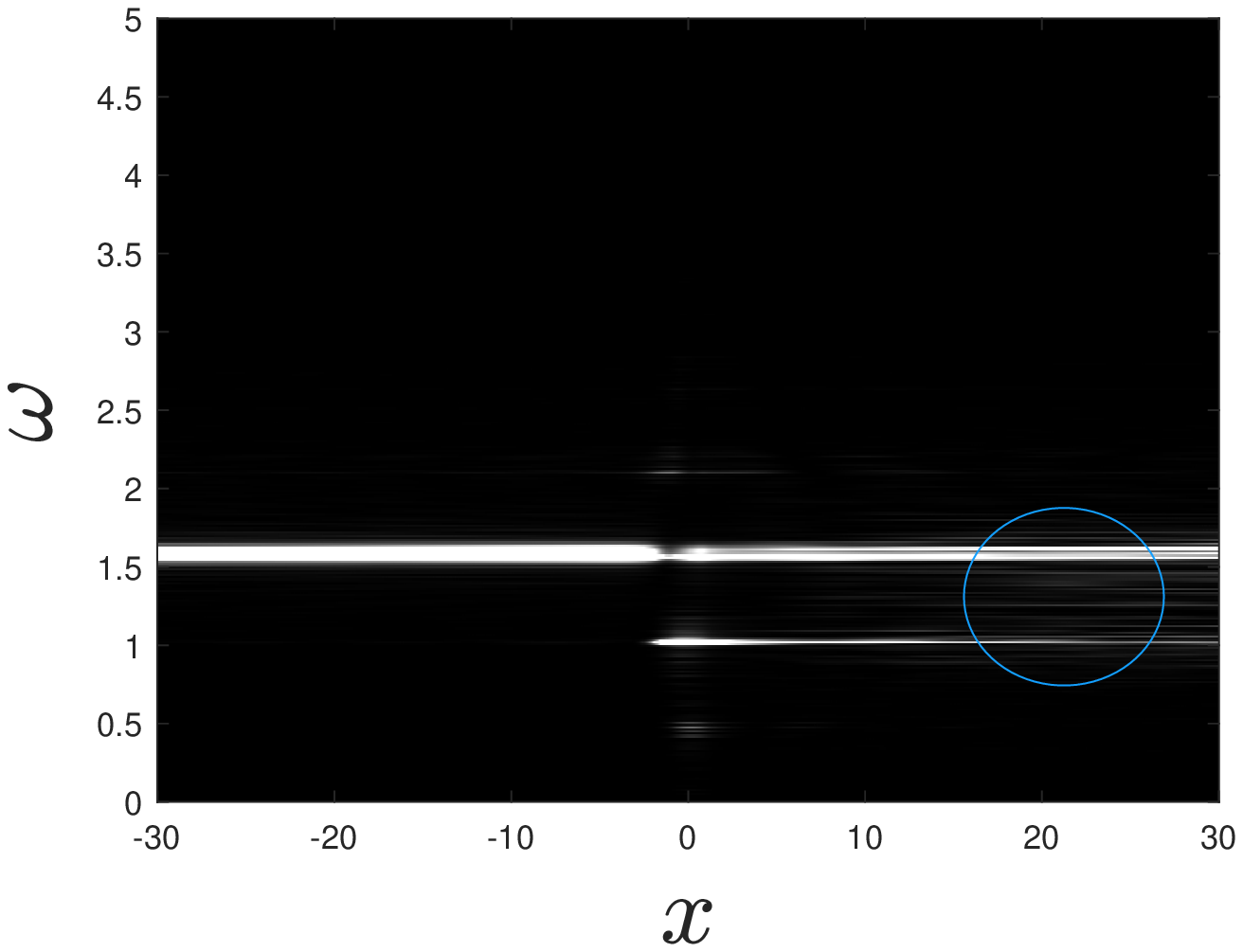} 
\end{tabular}
\caption{Time evolution of an IHFC in which the quenched attractive well is stochastically modulated in time through Eq. (\ref{eq:Disordered}). The background parameters of the flow are  $v=0.95,~V_0=1,~X=2$. Upper row: 2D plot of $|\Psi(x,t)|^2$. Lower row: Fourier spectrum of upper row,  $|\Psi(x,\omega)|^2$. Left column: $\epsilon=0.1$. Right column: $\epsilon=0.7$.} 
\label{fig:Disorder}
\end{figure}

\begin{figure}[t]
\begin{tabular}{@{}cc@{}}
    \includegraphics[width=0.5\columnwidth]{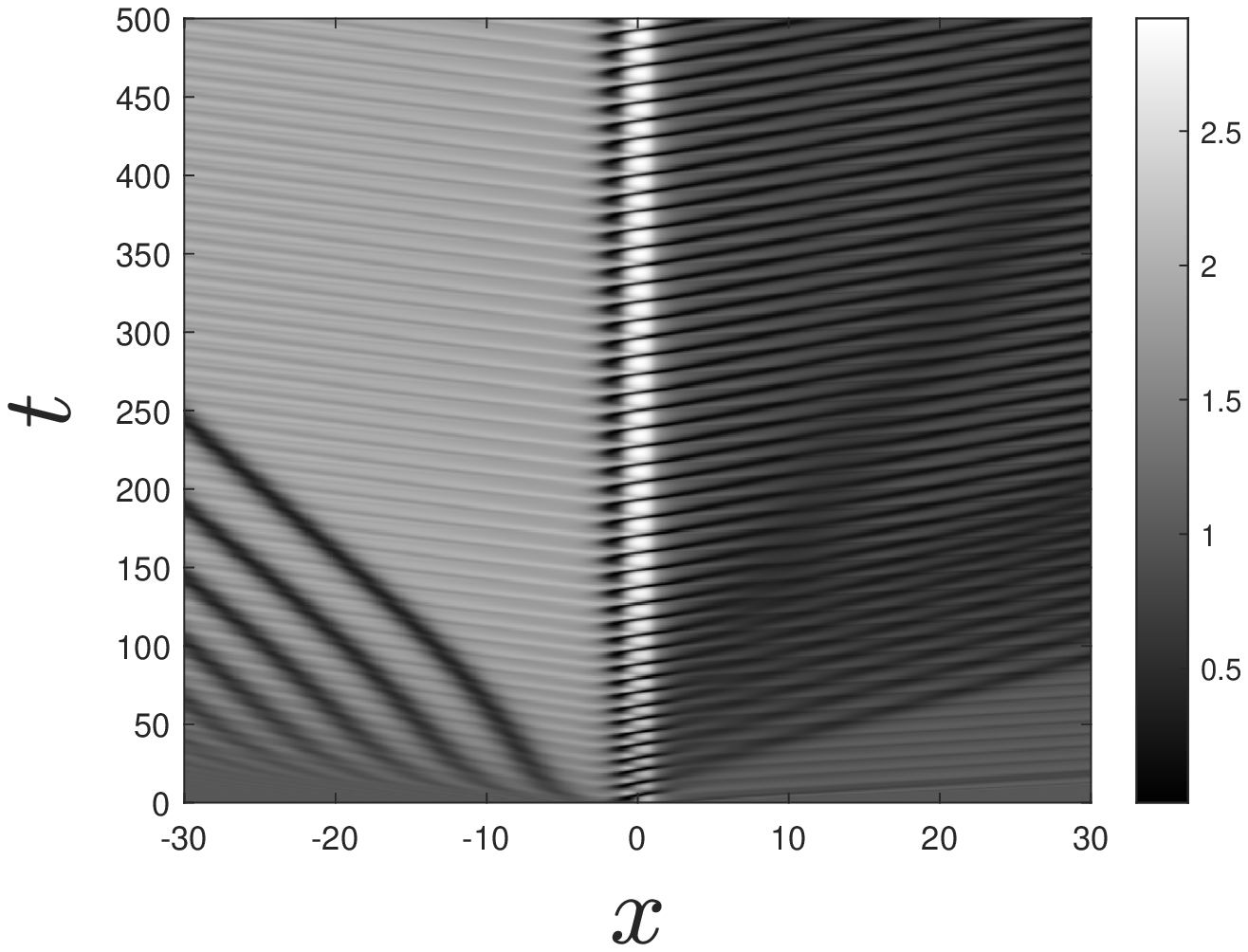} & 
    \includegraphics[width=0.5\columnwidth]{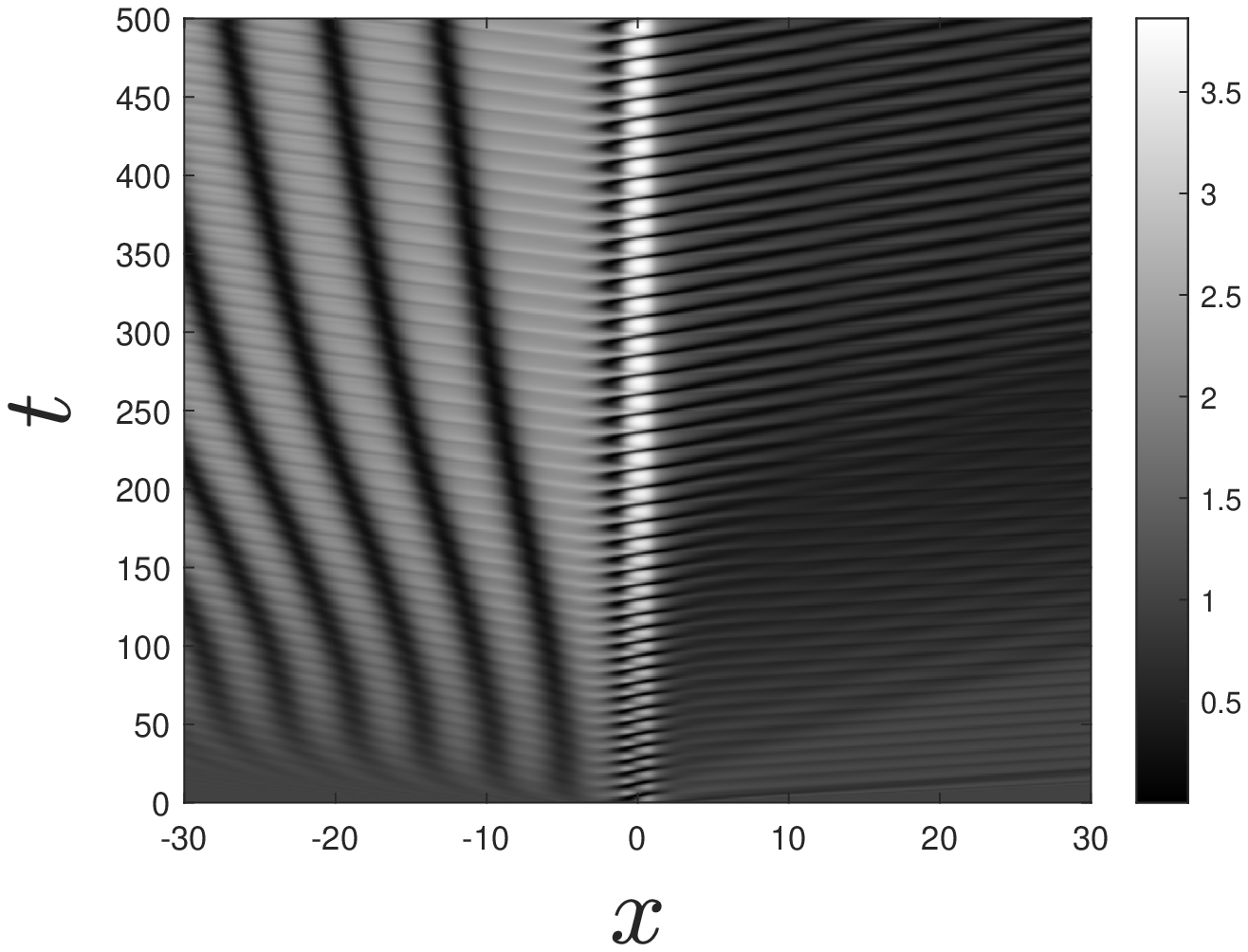} \\ \\ \\
    \includegraphics[width=0.5\columnwidth]{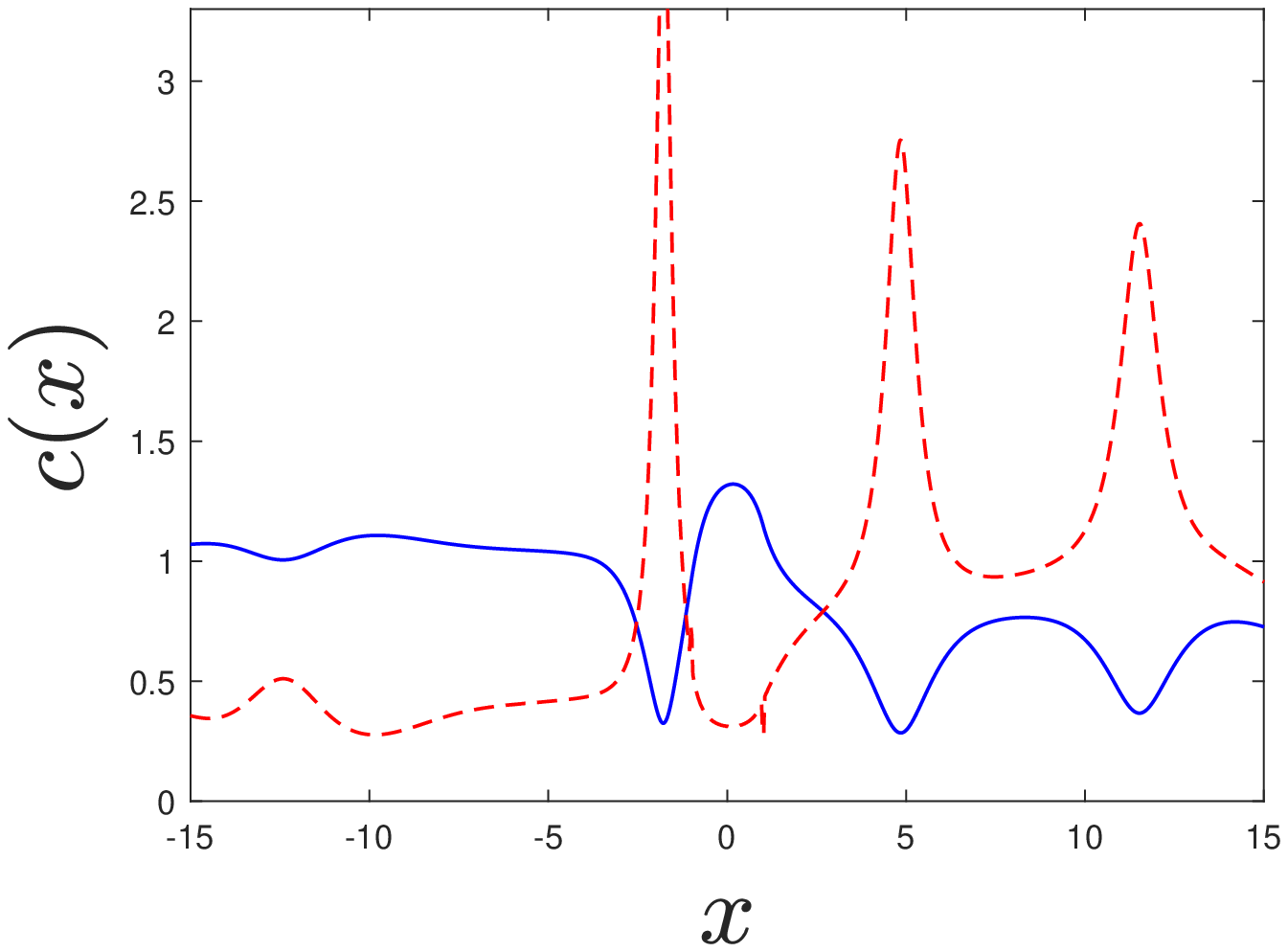} & 
    \includegraphics[width=0.5\columnwidth]{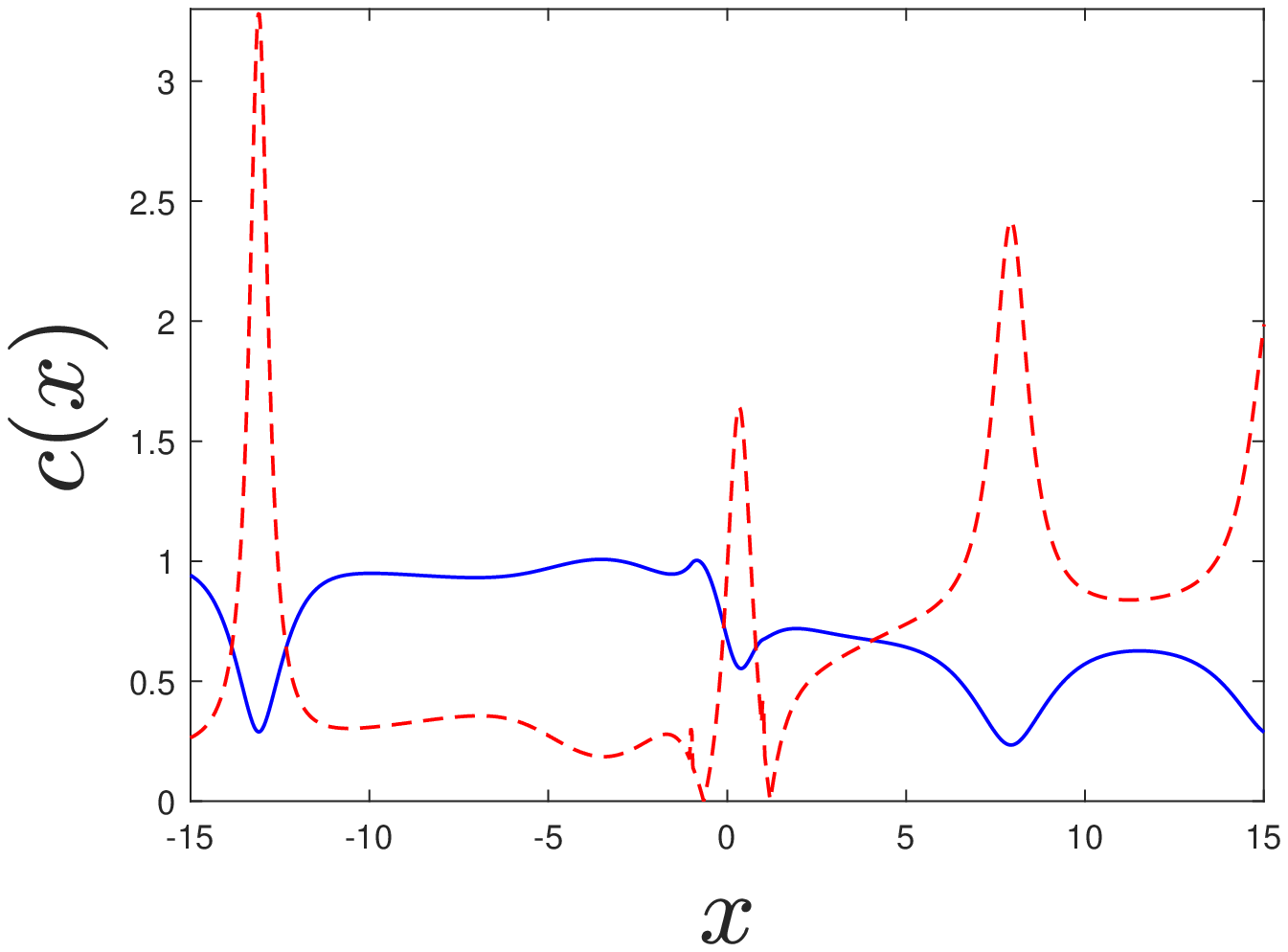} 
\end{tabular}
\caption{Analysis of the role of interactions as a function of the parameter $\lambda$ for an IHFC with background parameters $v=0.95,~V_0=1,~X=2$. Upper row: 2D plot of $|\Psi(x,t)|^2$. Lower row: 1D snapshot of the sound (solid blue) and flow (dashed red) velocities for the latest time of upper row. 
Left: $\lambda=0.6$. Right: $\lambda=0.4$.}
\label{fig:Interactions}
\end{figure}

\begin{figure}[t]
\begin{tabular}{@{}cc@{}}
    \includegraphics[width=0.5\columnwidth]{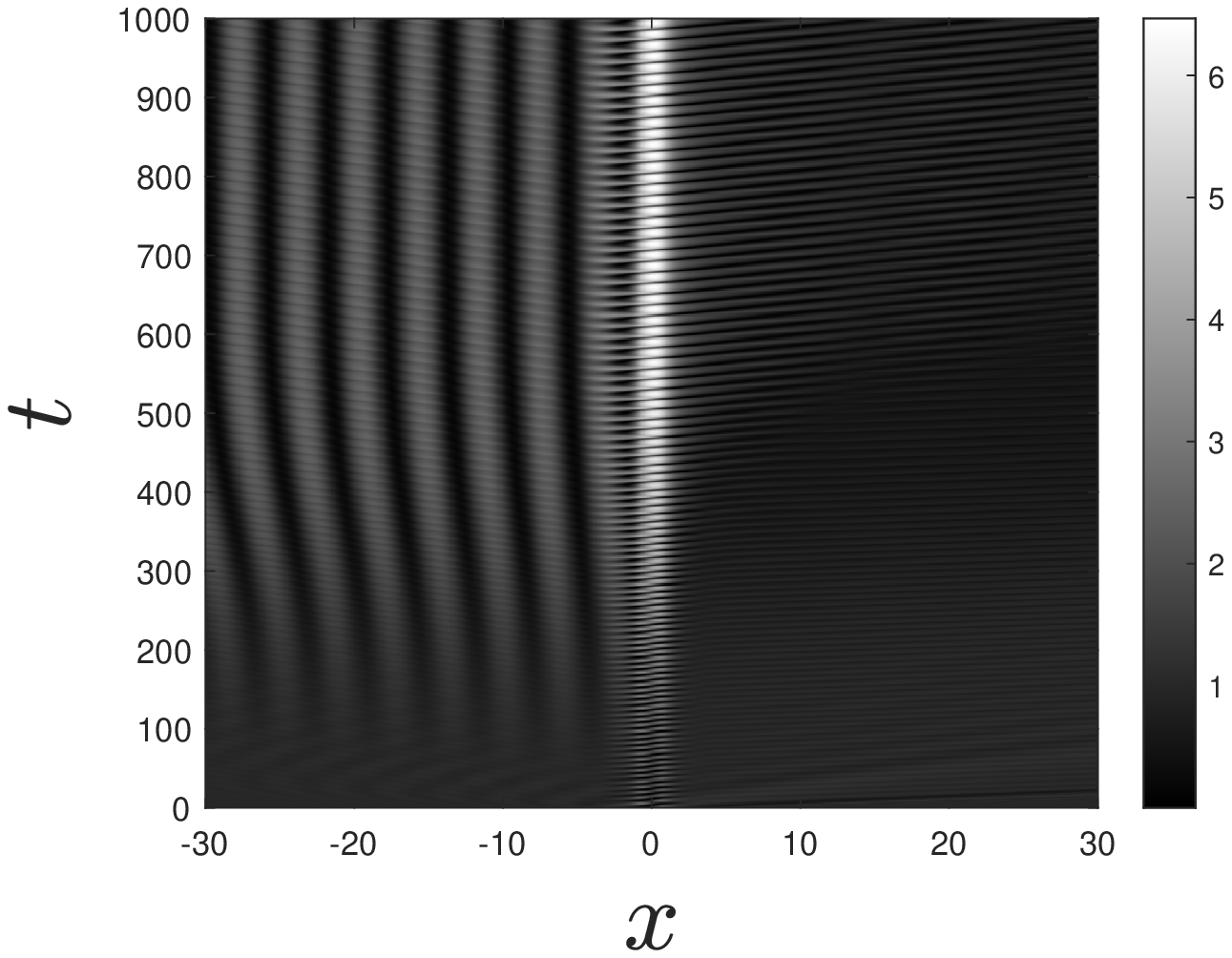} & 
    \includegraphics[width=0.5\columnwidth]{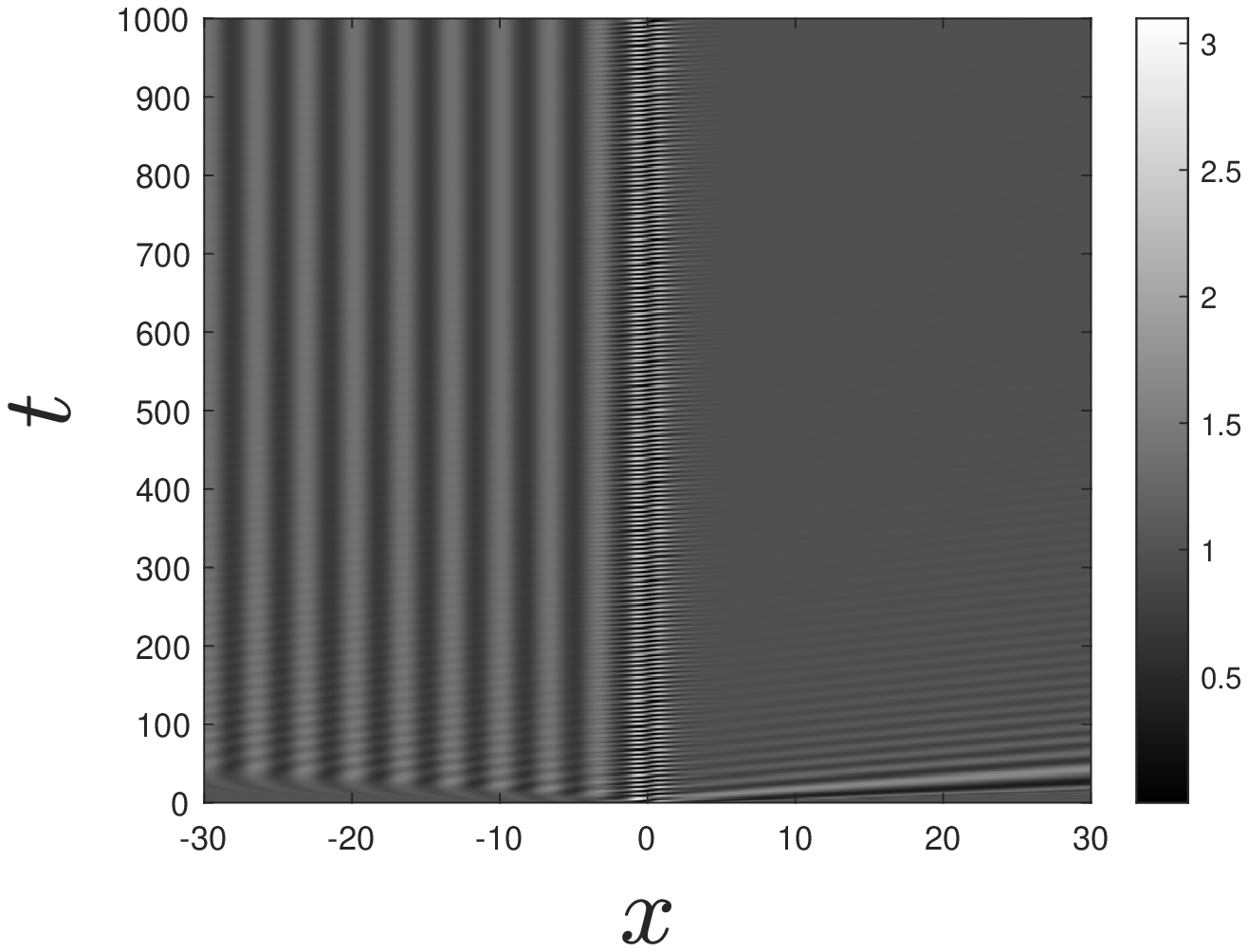} \\ \\ \\
    \includegraphics[width=0.5\columnwidth]{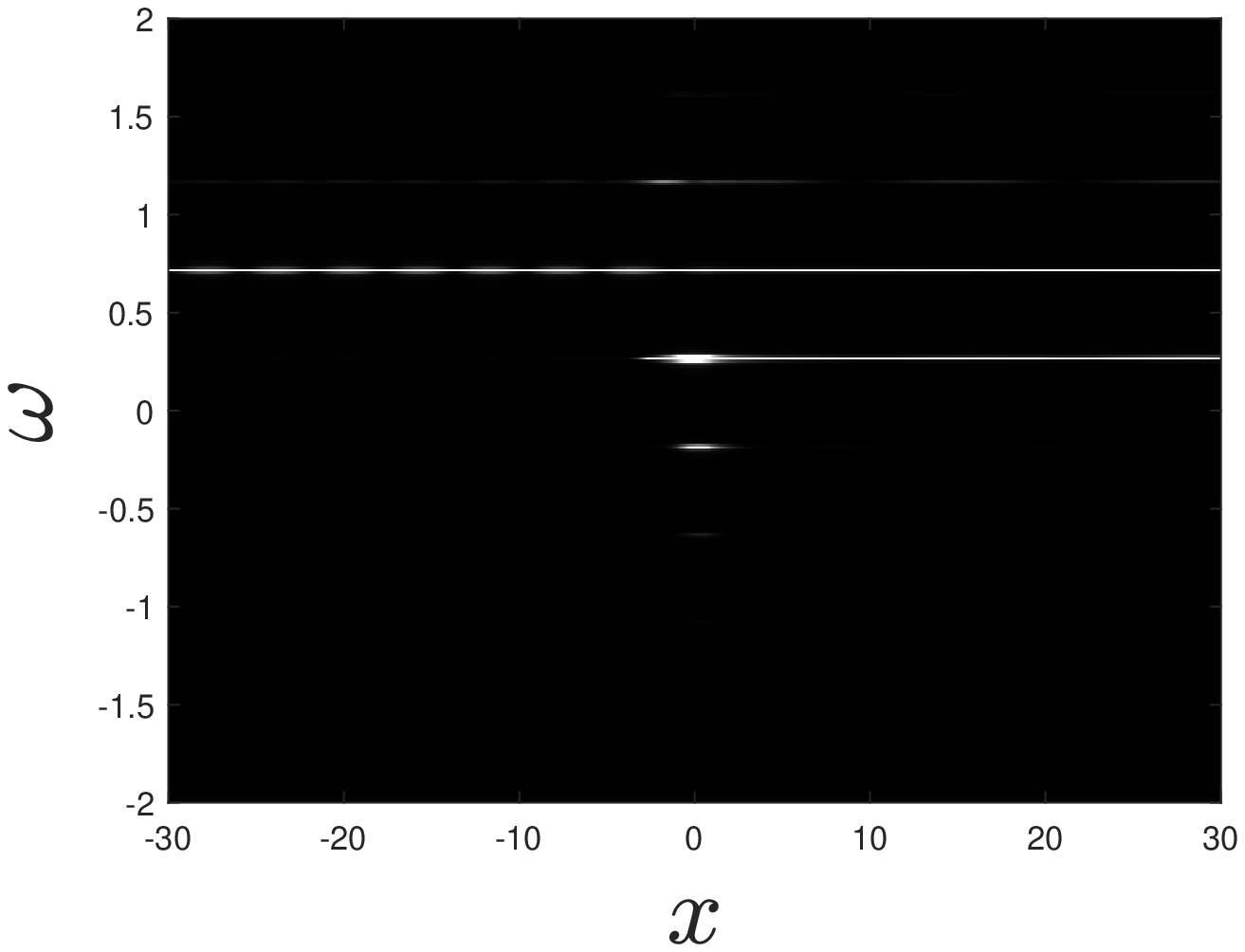} & 
    \includegraphics[width=0.5\columnwidth]{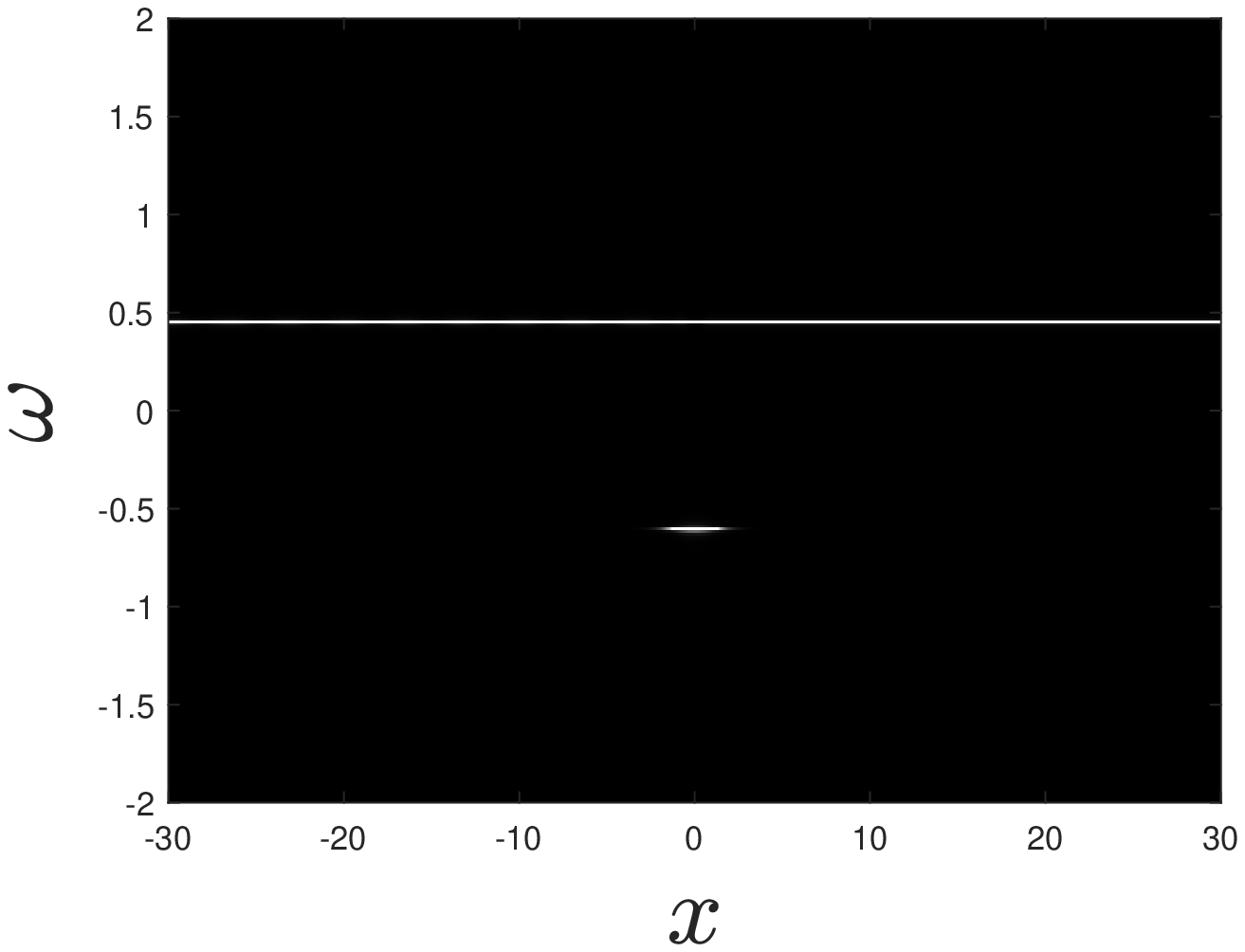} 
\end{tabular}
\caption{Analysis of the role of interactions as a function of the parameter $\lambda$ for an IHFC with background parameters $v=0.95,~V_0=1,~X=2$. Upper row: 2D plot of $|\Psi(x,t)|^2$. Lower row: 2D plot of $|\Psi(x,\omega)|^2$ for the simulations of upper row, once in the stationary regime. Left: $\lambda=0.2$. Right: $\lambda=0$.}
\label{fig:Scrodinger}
\end{figure}

\begin{figure}[t]
\begin{tabular}{@{}cc@{}}
    \includegraphics[width=0.5\columnwidth]{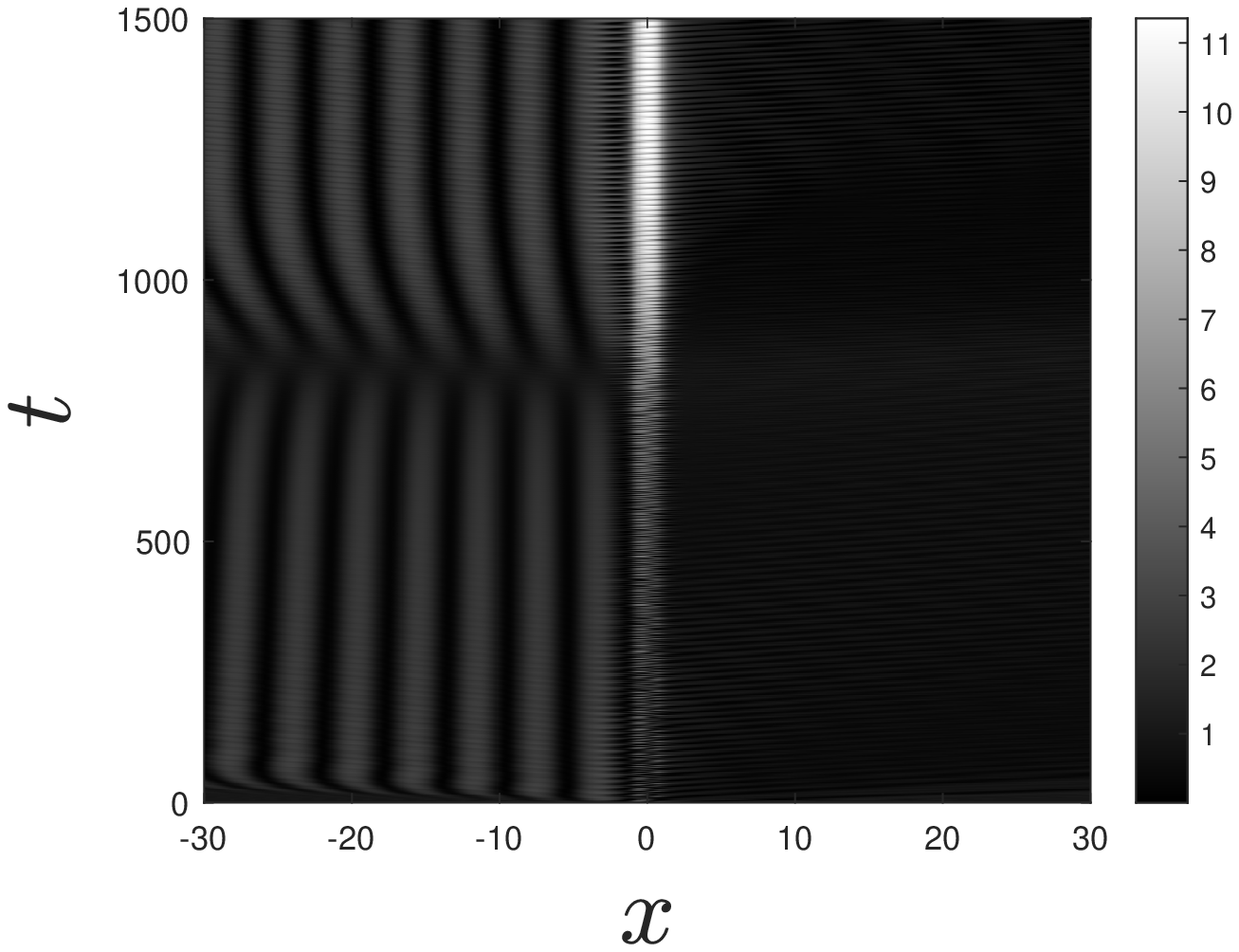} & 
    \includegraphics[width=0.5\columnwidth]{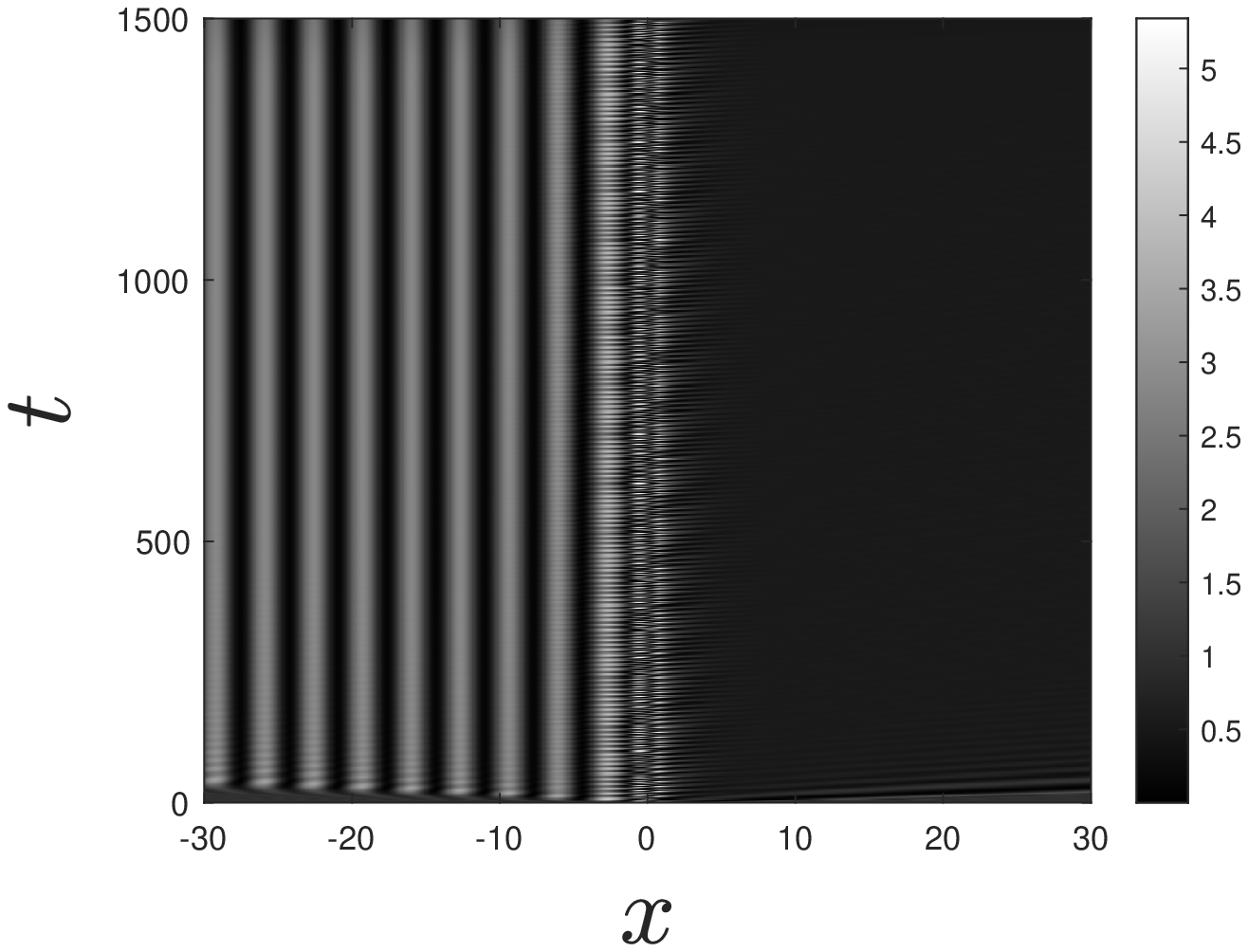} \\ \\ \\
    \includegraphics[width=0.5\columnwidth]{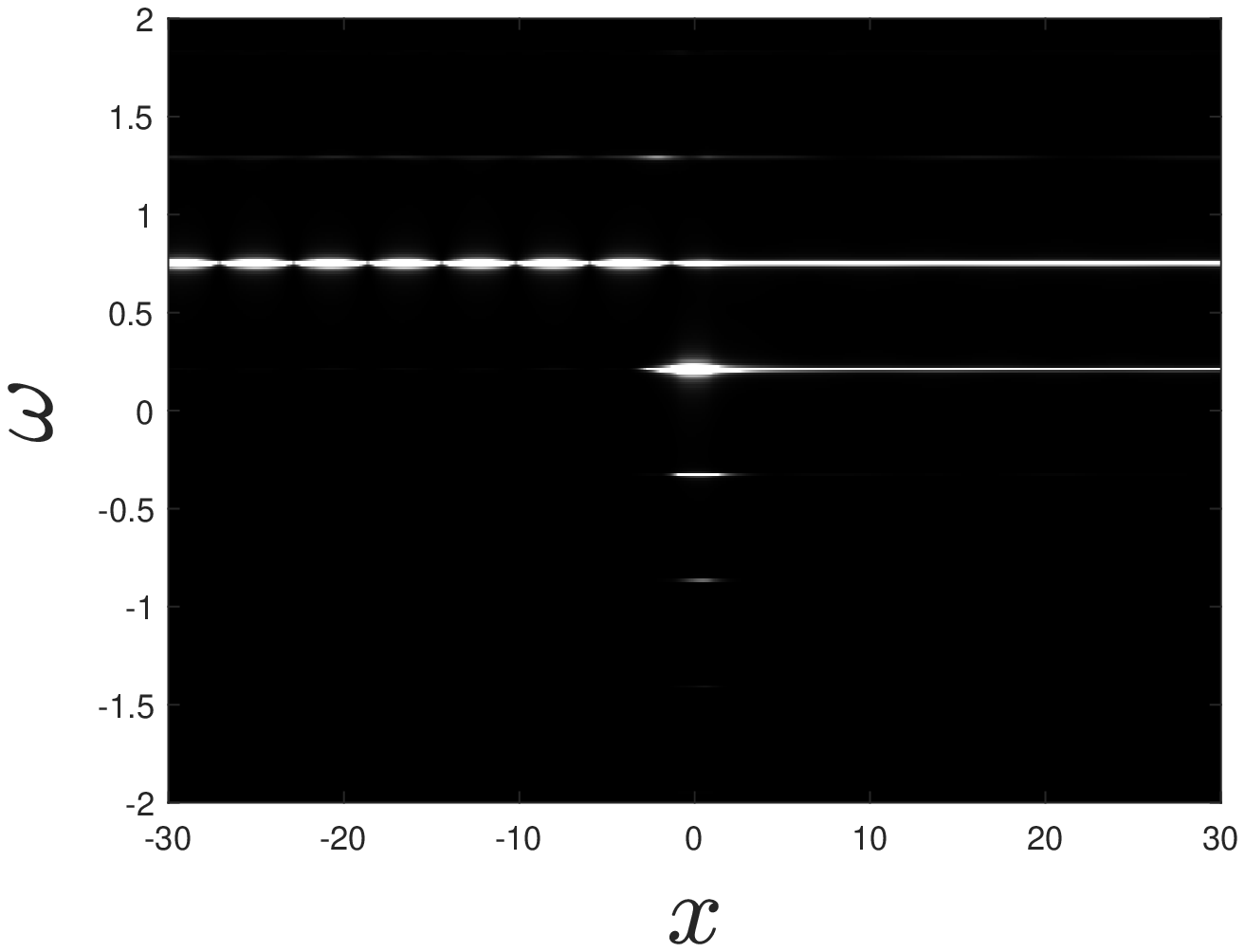} & 
    \includegraphics[width=0.5\columnwidth]{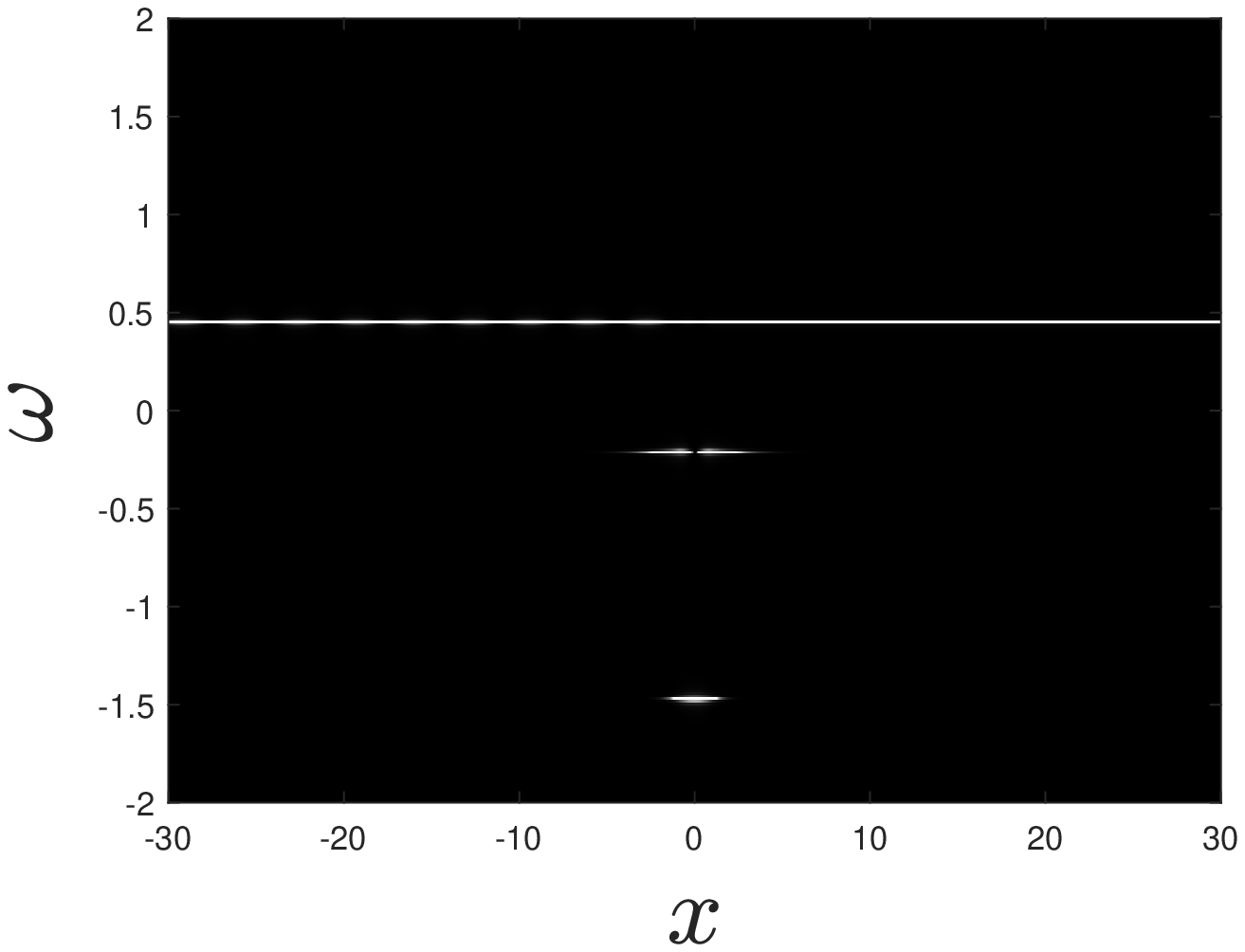} 
\end{tabular}
\caption{Same as Fig. \ref{fig:Scrodinger} but for $V_0=2$.}
\label{fig:Scrodinger2}
\end{figure}

\subsection{Universality}

We examine the universality of the proposed SMBF state by considering other obstacles instead of the idealized attractive square well, finding in all cases that an SMBF state is reached provided that the background flow velocity is sufficiently high. Here we describe in detail three of those models, represented in upper row of Fig. \ref{fig:Universality}.

First, we consider that the potential quenched at $t=0$ is a more realistic Gaussian attractive well
\begin{equation}\label{eq:GaussianPotential}
    V(x)=-V_0e^{-\frac{x^2}{\sigma^2}}
\end{equation}

We also consider a repulsive delta barrier, similarly to Ref. [44],
\begin{equation}\label{eq:DeltaPotential}
    V(x)=Z\delta(x)
\end{equation}

Finally, we consider the flat-profile BHL model of Ref. [40], where the quench involves both the external potential and the coupling constant in such a way that
\begin{equation}\label{eq:FlatProfile}
    g(x)n_0+V(x)=E_b
\end{equation}
with $E_b$ some constant. In this way, an IHFC of density $n_0$ remains as a stationary solution after the quench. In particular, $g(x)$ is chosen as a piecewise function so that the speed of sound satisfies $c(x)=1$ for $|x|>X/2$, and $c(x)=c_2<v$ for $|x|<X/2$. As a result, the central finite region of size $X$ becomes supersonic and, consequently, the IHFC after the quench is an unstable stationary BHL solution. Once more, some small noise is added to trigger the dynamical instabilities.

For all the described models, we have found that an SMBF state can be eventually reached. Examples are displayed in lower row of Fig. \ref{fig:Universality}.

\subsection{Temporal disorder}

We analyze the stability of the time crystal against time-dependent disorder. Specifically, we consider stochastic perturbations resulting from driving the square well potential with a time-dependent modulation
\begin{equation}\label{eq:Disordered}
    V(x,t)=V(x)\left[1+\epsilon h(t)\right]
\end{equation}
with $V(x)$ the square well potential, $h(t)$ normally distributed white noise, and $\epsilon$ the strength of the modulation. We have found that the CES state is quite robust against stochastic perturbations, as can be seen in Fig. \ref{fig:Disorder}. In the left column, we analyze the case of an appreciable modulation with $\epsilon=0.1$, once the system is well in the CES state. We observe that the CES state keeps its periodicity (upper left panel) as well as its Floquet character, as seen from its Fourier spectrum (lower left panel). 

Only in the presence of strong disorder is the periodicity spoiled, as seen in right column of Fig. \ref{fig:Disorder}, where we examine an extreme case $\epsilon=0.7$. The distortion of the periodicity (upper right panel) is translated into a blurring of the Floquet components (lower right panel), especially in the downstream region (blue circle).

We have also considered a stochastic driving in the coupling constant $g(t)=g\left[1+\epsilon h(t)\right]$, finding qualitatively similar results.

\subsection{Role of interactions}

Interactions are an essential ingredient for the formation of an SMBF state. In our model, they are hidden behind the system of units, $gn_0=1$. In order to keep track the role of interactions, we start from a regular subsonic IHFC with $v<1$, and introduce a dimensionless parameter $0\leq\lambda\leq1$ that reduces the strength of the interactions as $\lambda|\Psi(x,t)|^2$ (note that an IHFC is a stationary solution of the homogeneous GP equation for $t<0$ regardless the strength of the interaction). The sound speed then becomes $c(x,t)=\sqrt{\lambda|\Psi(x,t)|^2}$ and, in particular, the initial value of the homogeneous speed of sound is $c_0=\sqrt{\lambda}$. Therefore, whenever $\sqrt{\lambda}<v$, the IHFC is supersonic, which is known to be an energetically unstable flow. 

We study the formation of an SMBF state starting from a supersonic IHFC in Fig. \ref{fig:Interactions}. For moderate values of the initial supersonic Mach number $M\equiv v/\sqrt{\lambda}$, the same type of SMBF state (upper left panel) is still reached. In the process, the system increases its upstream average density until it becomes subsonic (lower left panel). Since particle number must be conserved, this increase is reached at the expense of the downstream region, which reduces its average density, keeping its initially supersonic character. However, for increasing values of $M$ (right column), more upstream density increase is required to become subsonic. This is translated into an enlarged transient, as seen by the long-lasting presence of the upstream emitted train of solitons (black fringes in the upstream region of upper right panel). 

The limit of small $\lambda$ is analyzed in detail Fig. \ref{fig:Scrodinger}, in which the system cannot become subsonic and the upstream black fringe pattern remains stationary, at least within the time scope of the simulations (upper left panel). Nevertheless, even in that regime, the system still behaves as an SMBF state, as seen by its Fourier spectrum (lower left panel). Finally, the case of $\lambda=0$ is well understood since it corresponds to the usual Schr\"odinger equation (upper right panel). Here, what we have for long times is simply the scattering of the incident wave packet $e^{ivx}$ from the left, giving rise to the interference between the incident and the reflected wave (upstream black fringe pattern) and to the transmission of the wave packet (homogeneous downstream density), plus the oscillation of the localized bound state of the well. This is clearly observed in the Fourier spectrum (lower right panel), which only displays two lines at the frequency of the kinetic energy of the scattering state associated to the incident wave, $\omega=v^2/2\approx 0.4512$, and at the negative energy $E_0$ of the only bound state of the well, $\omega=E_0\approx -0.6039$. We recall that the binding energies $E_n<0$ of an attractive square well are computed from the zeros of the transcendental equations
\begin{equation}
    \sqrt{-\frac{E_n+V_0}{E_n}}=\frac{1}{\tan\sqrt{\frac{V_0+E_n}{2}}X},~\sqrt{-\frac{E_n+V_0}{E_n}}=-\tan\sqrt{\frac{V_0+E_n}{2}}X
\end{equation}
where the number of bound states is
\begin{equation}
    N=\left\lfloor{\frac{\sqrt{2V_0}X}{\pi}}\right\rfloor+1
\end{equation}
with $\left\lfloor x \right\rfloor$ the floor function. The trivial linear combination of two eigenstates with different frequencies indeed results in a periodic oscillation of the density, but cannot be regarded as a Floquet state. 

An ever more spectacular result is presented in Fig. \ref{fig:Scrodinger2}, where we analyze the case of a square well with larger amplitude $V_0$, resulting in the presence of two bound states with energies $E_0\approx-0.2035$ and $E_1\approx -1.4697$ in the Schr\"odinger case $\lambda=0$. As a result, the non-interacting case (upper right panel) has not even well-defined frequency for the density oscillations since the differences between the frequencies of the spectrum are incommensurate (lower right panel). However, when switching on interactions, we find that, after some larger transient, the system reaches again an SMBF state (upper left panel), where interactions have stabilized the frequency of oscillation and the Floquet character of the Fourier spectrum is recovered (lower left panel). This demonstrates the key role played by interactions in the formation of the SMBF state.

\section{Long-range order in time}

We study the presence of long-range order in time for the one-body correlation function of a condensate. The calculation proceeds along the same lines as the usual study of long-range order in space [52,53], explicitly accounting now for the time dependence. We first consider the case of a 1D condensate. In order to study this problem, a phase-density decomposition of the field operator $\hat{\Psi}(x,t)$ is typically used due to the infrared divergence of the phase fluctuations, 
\begin{equation}\label{eq:phasedensity}
 \hat{\Psi}(x,t)=\Psi(x,t)+\hat{\phi}(x,t)=e^{i[\theta(x,t)+\delta\hat{\theta}(x,t)]}\sqrt{n(x,t)+\delta\hat{n}(x,t)}
\end{equation}
with $\hat{\phi}(x,t)$ the quantum fluctuations around the mean-field GP wave function $\Psi(x,t)=\sqrt{n(x,t)}e^{i\theta(x,t)}$. By absorbing the phase of the condensate into the field operator as $\hat{\phi}(x,t)\equiv e^{i\theta(x,t)}\hat{\chi}(x,t)$, we obtain after expanding to linear order in Eq. (\ref{eq:phasedensity}) that
\begin{eqnarray}
\delta\hat{\theta}(x,t)&=&\frac{\hat{\chi}(x,t)-\hat{\chi}^{\dagger}(x,t)}{2i\sqrt{n(x,t)}}\\
\nonumber \delta\hat{n}(x,t)&=&\sqrt{n(x,t)}\left[\hat{\chi}(x,t)+\hat{\chi}^{\dagger}(x,t)\right]
\end{eqnarray}
from where we can express the density and phase fluctuations in terms of the field fluctuations, whose dynamics is in turn described by the BdG equations. It is immediate to see that density and phase fluctuations obey canonical commutation relations 
\begin{equation}
    [\delta\hat{n}(x,t),\delta\hat{\theta}(x',t)]=i\delta(x-x')
\end{equation}
since $[\hat{\chi}(x,t),\hat{\chi}^{\dagger}(x',t)]=\delta(x-x')$.

In order to study the long-range order of the one-body correlation function $G(x,x',t,t')=\braket{\hat{\Psi}^{\dagger}(x,t)\hat{\Psi}(x',t')}$, we focus on the simple case of a homogeneous condensate, $\Psi(x,t)=\sqrt{n_0}e^{i\theta_0}e^{-i\mu t}$ where, through the rest of this section, we set $\hbar=m=c_0=k_B=1$ unless otherwise specified, with $c_0=\sqrt{gn_0/m}$ the condensate speed of sound. The fluctuations of the field operator, $\hat{\phi}(x,t)=\hat{\chi}(x,t)e^{i\theta_0}e^{-i\mu t}$, are described by the usual Bogoliubov expansion:
\begin{equation}\label{eq:fieldfluctuations}
\hat{\chi}(x,t)=\frac{1}{\sqrt{L}}\sum_k \hat{\alpha}_{k} u_ke^{i(kx-\Omega_kt)}+\hat{\alpha}^{\dagger}_k v^*_ke^{-i(kx-\Omega_kt)}
\end{equation}
with $\hat{\alpha}_{k}$ the bosonic operator describing a quasiparticle, $L$ the size of the condensate, $u_k,v_k$ the usual Bogoliubov components
\begin{equation}
u_k=\frac{\frac{k^2}{2}+\Omega_k}{\sqrt{2k^2\Omega_k}},~
v_k=\frac{\frac{k^2}{2}-\Omega_k}{\sqrt{2k^2\Omega_k}}
\end{equation}
and $\Omega_k$ the Bogoliubov dispersion relation
\begin{equation}
\Omega_k=\sqrt{k^2+\frac{k^4}{4}}    
\end{equation}
where the healing length satisfies $\xi=\hbar/mc_0=1$ in these units. The density and phase fluctuations admit similar expansions 
\begin{eqnarray}\label{eq:phasedensityfluctuations}
\nonumber \delta\hat{n}(x,t)&=&\sqrt{\frac{n_0}{L}}\sum_k\hat{\alpha}_{k} \rho_ke^{i(kx-\Omega_kt)}+\hat{\alpha}^{\dagger}_k \rho^*_ke^{-i(kx-\Omega_kt)},~\rho_k=u_k+v_k=\sqrt{\frac{k^2}{2\Omega_k}}\\
\delta\hat{\theta}(x,t)&=&\frac{1}{\sqrt{n_0L}}\sum_k\hat{\alpha}_{k} \theta_ke^{i(kx-\Omega_kt)}+\hat{\alpha}^{\dagger}_k \theta^*_ke^{-i(kx-\Omega_kt)},~\theta_k=\frac{u_k-v_k}{2i}=-i\sqrt{\frac{\Omega_k}{2k^2}}
\end{eqnarray}
Since $\Omega_k\sim |k|$ for low $k$, then $\rho_k\sim |k|^{\frac{1}{2}}$ and $\theta_k\sim |k|^{-\frac{1}{2}}$, which explicitly shows that phase fluctuations are dominant in the infrared limit. However, in the ultraviolet limit of large $k$, both fluctuations are of the same order, since $\Omega_k\simeq k^2/2$ and then $|\rho_k|\simeq 1$,~$|\theta_k|\simeq 1/2$.

Thus, when computing the one-body correlation function, we concentrate on the critical infrared contribution of phase fluctuations and neglect density fluctuations
\begin{equation}\label{eq:OneBodyDensityPhaseFluctuations}
    G(x,x',t,t')\simeq \Psi^*(x,t)\Psi(x',t')\braket{e^{-i\delta\hat{\theta}(x,t)}e^{i\delta\hat{\theta}(x',t')}}
\end{equation}
We recall that the commutator between phase fluctuations at different times is a finite c-number, $\left[\delta\hat{\theta}(x,t),\delta\hat{\theta}(x',t')\right]=C(x-x',t-t')$, with 
\begin{equation}\label{eq:PhaseCommutator}
    C(x,t)=-\frac{2i}{n_0L}\sum_k |\theta_k|^2e^{ikx}\sin\Omega_kt=-\frac{i}{\pi n_0}\int^{\infty}_{-\infty}\mathrm{d}k~ |\theta_k|^2e^{ikx}\sin\Omega_kt
\end{equation}
By employing the usual relation for two operators $A,B$ that satisfy $[A,[A,B]]=[B,[A,B]]=0$,
\begin{equation}
    e^{A}e^{B}=e^{A+B}e^{\frac{[A,B]}{2}}
\end{equation}
we get
\begin{equation}\label{eq:OneBodyExpectationValuePhase}
    G(x,x',t,t')=\Psi^*(x,t)\Psi(x',t')e^{\frac{C(x-x',t-t')}{2}}\braket{e^{i\left[\delta\hat{\theta}(x',t')-\delta\hat{\theta}(x,t)\right]}}
\end{equation}
The expectation value is computed from the general case of the expectation value in a Gaussian state of the imaginary exponential of a Hermitian operator $\hat{A}$ that is linear in annihilation and creation operators, and with vanishing expectation value $\braket{\hat{A}}=0$. Expanding the exponential yields
\begin{equation}
    \braket{e^{i\hat{A}}}=\sum^{\infty}_{n=0}\frac{\braket{i^n\hat{A}^n}}{n!}=
    \sum^{\infty}_{n=0}(-1)^n\frac{\braket{\hat{A}^{2n}}}{2n!}
\end{equation}
while Wick's theorem gives
\begin{equation}
    \braket{\hat{A}^{2n}}=\frac{2n!}{2^nn!}\braket{\hat{A}^2}
\end{equation}
where the prefactor takes into account the number of different pairs that can be taken from $2n$ elements. divided by $n!$ to avoid double-counting. Therefore,
\begin{equation}\label{eq:exponentialwick}
    \braket{e^{i\hat{A}}}=
    \sum^{\infty}_{n=0}(-1)^n\frac{\braket{\hat{A}^{2}}}{2^n n!}=e^{-\frac{\braket{\hat{A}^{2}}}{2}}
\end{equation}

Since phase fluctuations satisfy
\begin{eqnarray}\label{eq:PreGaussian}
\braket{\delta\hat{\theta}(x',t')-\delta\hat{\theta}(x,t)}&=&0 \\
\nonumber \braket{\left[\delta\hat{\theta}(x',t')-\delta\hat{\theta}(x,t)\right]^2}&\equiv&\Theta(x,x',t,t')=\braket{\delta\hat{\theta}^2(x',t')+\delta\hat{\theta}^2(x,t)-\delta\hat{\theta}(x',t')\delta\hat{\theta}(x,t)-\delta\hat{\theta}(x,t)\delta\hat{\theta}(x',t')}
\end{eqnarray}
and our state is Gaussian (either a $T=0$ ground state or a $T>0$ thermal state), we can make use of Eq. (\ref{eq:exponentialwick}), finding
\begin{equation}\label{eq:OneBodyCorrelationFunctionPhaseCorrelations}
G(x,x',t,t')=\Psi^*(x,t)\Psi(x',t')e^{\frac{C(x-x',t-t')}{2}}e^{-\frac{\Theta(x,x',t,t')}{2}}
\end{equation}
Thus, we only need to compute the phase correlations. Proceeding in a similar way to Eq. (\ref{eq:PhaseCommutator}) yields
\begin{eqnarray}\label{eq:phasecorrelation}
 \braket{\delta\hat{\theta}(x,t)\delta\hat{\theta}(x',t')}=\frac{1}{2\pi n_0}\int^{\infty}_{-\infty}\mathrm{d}k~ |\theta_k|^2\left[e^{ik(x-x')}e^{-i\Omega_k(t-t')}+2n_k\cos\left[k(x-x')-\Omega_k(t-t')\right]\right]
\end{eqnarray}
with $n_k$ the Bose occupation factor
\begin{equation}\label{eq:OccupationNumber}
    n_k=\frac{1}{e^{\frac{\Omega_k}{T}}-1}
\end{equation}
which results from considering a thermal Gaussian state with temperature $T$. By combining Eq. (\ref{eq:phasecorrelation}) with Eq. (\ref{eq:PreGaussian}), it is immediate to see that $\Theta(x,x',t,t')=\Theta(x-x',t-t')$, with $\Theta(x,t)=\Theta_0(x,t)+\Theta_T(x,t)$, where $\Theta_0(x,t)$ is the zero-temperature contribution
\begin{eqnarray}\label{eq:CorrelationZeroTemperature}
\Theta_0(x,t)=\frac{1}{\pi n_0}\int^{\infty}_{-\infty}\mathrm{d}k~ |\theta_k|^2\left[1-\cos\left(kx-\Omega_kt\right)\right]
\end{eqnarray}
and $\Theta_T(x,t)$ is the thermal contribution
\begin{eqnarray}\label{eq:CorrelationFiniteTemperature}
\Theta_T(x,t)=\frac{2}{\pi n_0}\int^{\infty}_{-\infty}\mathrm{d}k~ |\theta_k|^2n_k\left[1-\cos\left(kx-\Omega_kt\right)\right]
\end{eqnarray}
In the following, as we are examining the existence of long-range order, we focus on the asymptotic behavior for large $x$ and $t$. Due to dispersive effects, using a saddle-point approximation for large $t$ gives $C(x,t)\sim |t|^{-1/2}\simeq 0$, so phase fluctuations commute. Moreover, we note that for deriving Eq. (\ref{eq:OneBodyCorrelationFunctionPhaseCorrelations}) we have neglected density fluctuations. This implicitly assumes a long-wavelength approximation, since at high momentum the magnitude of both fluctuations is similar. Thus, we must cut all the integrals at values $|k|\simeq 1$. In the same spirit, we approximate all quantities by their low $k$ expansions in order to study possible divergences:
\begin{equation}
    \Omega_k\simeq |k|,~|\theta_k|^2\simeq\frac{1}{2|k|},~n_k\simeq\frac{1}{e^{\frac{|k|}{T}}-1}\simeq \frac{T}{|k|}
\end{equation}
We consider first the $T=0$ case,
\begin{eqnarray}\label{eq:CorrelationZeroTemperatureCut}
\Theta_0(x,t)&\simeq&\frac{1}{\pi n_0}\int^{1}_{-1}\mathrm{d}k~ |\theta_k|^2\left[1-\cos\left(kx-\Omega_kt\right)\right]
\simeq\frac{1}{2\pi n_0}\int^{1}_{-1}\mathrm{d}k~ \frac{1-\cos\left(kx-\Omega_kt\right)}{|k|}\\
\nonumber &=&\frac{1}{2\pi n_0}\left[\int^{1}_{0}\mathrm{d}k~ \frac{1-\cos k(x-t)}{k}+\int^{1}_{0}\mathrm{d}k~ \frac{1-\cos k(x+t)}{k}\right]
=\frac{1}{2\pi n_0}\left[f(x-t)+f(x+t)\right]
\end{eqnarray}
where we have defined the function
\begin{equation}
    f(x)\equiv\int^{1}_{0}\mathrm{d}k~ \frac{1-\cos kx}{k}=
    \int^{x}_{0}\mathrm{d}z~ \frac{1-\cos z}{z}
\end{equation}
This function has a logarithmic divergence for large $x$, $f(x)\simeq \ln |x|$. As a result, long-range phase-fluctuations diverge logarithmically as
\begin{eqnarray}\label{eq:CorrelationZeroTemperatureDivergence}
\Theta_0(x,t)&=&\frac{1}{2\pi n_0}\left[f(x-t)+f(x+t)\right]
\simeq\frac{1}{2\pi n_0}\left[\ln|x-t|+\ln|x+t|\right]
=\frac{\ln \left|x^2-t^2\right|}{2\pi n_0}
\end{eqnarray}

Gathering all together, and after momentarily restoring units, we find that the one-body correlation function at $T=0$ decays algebraically as
\begin{eqnarray}\label{eq:CorrelationFunctionZeroTemperatureLongRange}
    G(x,x',t,t')&\simeq&\Psi^*(x,t)\Psi(x',t')\left|\frac{(x-x')^2-c^2_0(t-t')^2}{\xi^2}\right|^{-\frac{1}{4\pi n_0\xi}}
\end{eqnarray}

The $T>0$ contribution for the phase correlations gives
\begin{equation}
\Theta_T(x,t)\simeq\frac{2}{\pi n_0}\int^{1}_{-1}\mathrm{d}k~ |\theta_k|^2n_k\left[1-\cos\left(kx-\Omega_kt\right)\right]\simeq
\frac{T}{\pi n_0}\int^{1}_{-1}\mathrm{d}k~ \frac{1-\cos\left(kx-|k|t\right)}{k^2}=\frac{T}{\pi n_0}\left[g(x-t)+g(x+t)\right]
\end{equation}
where 
\begin{equation}
    g(x)\equiv\int^{1}_{0}\mathrm{d}k~ \frac{1-\cos kx}{k^2}=x\int^{x}_{0}\mathrm{d}z~ \frac{1-\cos z}{z^2}
\end{equation}
In the limit of large $x$, it behaves as
\begin{equation}
    g(x)\simeq |x|\int^{\infty}_{0}\mathrm{d}z~ \frac{1-\cos z}{z^2}=\frac{\pi}{2}|x|
\end{equation}
Hence, phase fluctuations diverge linearly for $T>0$:
\begin{eqnarray}\label{eq:CorrelationFiniteTemperatureLinear}
\Theta_T(x,t)&=&\frac{T}{\pi n_0}\left[g(x-t)+g(x+t)\right]
\simeq\frac{T}{2 n_0}\left[|x-t|+|x+t|\right]
\end{eqnarray}
After restoring units once more, this results in a exponential decay of the long-range order of the one-body correlation function:
\begin{eqnarray}\label{eq:CorrelationFunctionZeroTemperatureLongRange}
    G(x,x',t,t')&\simeq&\Psi^*(x,t)\Psi(x',t')e^{-\frac{mk_BT}{4 n_0\hbar^2 }\left[|(x-x')-c_0(t-t')|+|(x-x')+c_0(t-t')|\right]}
\end{eqnarray}
The phase-coherence length of the condensate is thus
\begin{equation}
    L_{\theta}=\frac{4n_0\hbar^2}{mk_BT}=\frac{4\hbar^2}{m\xi^2k_BT}(n_0\xi)\xi=
    4\frac{\mu_0}{k_BT}(n_0\xi)\xi
\end{equation}
with $\mu_0=gn_0$ the chemical potential. 

In summary, long-range order in time is suppressed in the same manner as long-range order in space in 1D: 
\begin{eqnarray}
    G(x,x',t,t')&\sim& \left|\frac{(x-x')^2-c^2_0(t-t')^2}{\xi^2}\right|^{-\frac{1}{4\pi n_0\xi}},~T=0\\
\nonumber G(x,x',t,t')&\sim& e^{-\frac{mk_BT}{4 n_0\hbar^2 }\left[|(x-x')-c_0(t-t')|+|(x-x')+c_0(t-t')|\right]},~T>0    
\end{eqnarray}
i.e., algebraically at $T=0$ and exponentially for $T>0$.

We can repeat the same calculations for 2D. We have that phase correlations then read
\begin{eqnarray}\label{eq:phasecorrelation2D}
 \braket{\delta\hat{\theta}(\mathbf{x},t)\delta\hat{\theta}(\mathbf{x}',t')}=\frac{1}{(2\pi)^2 n_0}\int\mathrm{d}^2\mathbf{k}~ |\theta_\mathbf{k}|^2\left[e^{i\mathbf{k}(\mathbf{x}-\mathbf{x}')}e^{-i\Omega_\mathbf{k}(t-t')}+2n_\mathbf{k}\cos\left[\mathbf{k}(\mathbf{x}-\mathbf{x}')-\Omega_\mathbf{k}(t-t')\right]\right]
\end{eqnarray}
After cutting the integrals for values $k=|\mathbf{k}|<1$, for $T=0$ we have
\begin{eqnarray}\label{eq:CorrelationZeroTemperature2D}
\Theta_0(\mathbf{x},t)=\frac{1}{4\pi^2 n_0}\int^1_0\mathrm{d}k\int^{2\pi}_0\mathrm{d}\phi~ \left[1-\cos\left(kr\cos\phi-\Omega_kt\right)\right]
\end{eqnarray}
while for $T>0$,
\begin{eqnarray}\label{eq:CorrelationFiniteTemperature2D}
\Theta_T(\mathbf{x},t)=\frac{T}{2\pi^2 n_0}\int^1_0\mathrm{d}k\int^{2\pi}_0\mathrm{d}\phi~ \frac{1-\cos\left(kr\cos\phi-\Omega_kt\right)}{k}
\end{eqnarray}
$\phi$ being the polar angle between $\mathbf{k}\mathbf{x}=kr\cos\phi$ and $r=|\mathbf{x}|$.

\textit{Mutatis mutandis}, we find that at $T=0$ there is indeed long-range order in time for the one-body correlation function, while for $T>0$ it decays algebraically with the same exponent as in space,
\begin{equation}
    \eta=\frac{k_BT}{\hbar c_02\pi n_0\xi}=\frac{1}{2\pi n_0\xi^2}\frac{k_BT}{\mu_0}
\end{equation}

In summary, long-range order in time for the one-body correlation function behaves in exactly the same way as in space for the corresponding spatial dimension. Naively, one could think that, since time alone is 1D, it should always behave like 1D in space. However, since the divergence is controlled by the infrared behavior of the density of states, set by the number of spatial dimensions, and that in this limit $\Omega_k\sim k$, long-range order in time presents the same dependence as in space.


\section{Truncated Wigner calculation of quantum fluctuations}

We explore the robustness of the CES time crystal against quantum fluctuations via the Truncated Wigner method [54,55]. The Truncated Wigner approximation computes symmetric-ordered expectation values from ensemble averages of integrations of the GP equation
\begin{equation}
    i\hbar\partial_t\Psi_W(x,t)=\left[-\frac{\hbar^2}{2m}\partial_x^2+V(x)+g|\Psi_W(x,t)|^2\right]\Psi_W(x,t)
\end{equation}
where the stochastic initial condition is
\begin{equation}
\Psi_W(x,0)=\sqrt{n_0(x)+\delta n_W(x)}e^{i\theta_0(x)}e^{i\delta \theta_W(x)}
\end{equation}
with $\Psi_0(x)=\sqrt{n_0(x)}e^{i\theta_0(x)}$ the GP wave function describing the initial condensate at equilibrium and
\begin{eqnarray}
\delta\theta_W(x)&=&\frac{\chi_W(x)-\chi^*_W(x)}{2i\sqrt{n_0(x)}}\\
\nonumber \delta n_W(x)&=&\sqrt{n_0(x)}\left[\chi_W(x)+\chi^*_W(x)\right]
\end{eqnarray}
where $\chi_W(x)e^{i\theta_0(x)}=\phi_W(x,0)$ describes the quantum fluctuations of the field operator, sampled from the Wigner distribution describing the condensate. Due to the critical role of phase fluctuations, we are employing a quasi-condensate description to describe the present physics \citeSupp{Martin2010PRL,Martin2010NJP}.

In the particular case of this work, $\Psi_0(x)=e^{ivx}$, where we are recovering the system of units of the main text. Then, we have that
\begin{equation}\label{eq:TWExpansionNotConserving}
\Psi_W(x,0)=\sqrt{1+\frac{\delta n_W(x)}{n_0}}e^{ivx}e^{i\delta \theta_W(x)}
\end{equation}
The field fluctuations are written similarly to Eq. (\ref{eq:fieldfluctuations}),
\begin{equation}
\chi_W(x)=\frac{1}{\sqrt{L}}\sum_k \alpha_{k} u_ke^{ikx}+\alpha^*_k v^*_ke^{-ikx}
\end{equation}
as well as the density and phase fluctuations,
\begin{eqnarray}
\nonumber \delta n_W(x)&=&\sqrt{\frac{n_0}{L}}\sum_k \alpha_{k} \rho_ke^{ikx}+\alpha^*_k \rho^*_ke^{-ikx}\\
\delta \theta_W(x)&=&\frac{1}{\sqrt{n_0L}}\sum_k \alpha_{k} \theta_ke^{ikx}+\alpha^*_k \theta^*_ke^{-ikx}
\end{eqnarray}
The amplitudes $\alpha_k$ of each mode are stochastic variables that are sampled from the Wigner distribution of the initial equilibrium state. Our initial quantum state is the $T=0$ ground state in the comoving frame of the condensate, where it is at rest. Hence, our stochastic initial condition is sampled from a Gaussian distribution characterized by
\begin{equation}
\braket{\alpha_k}=0,~\braket{\alpha_{k'}\alpha_k}=\braket{\alpha^*_{k'}\alpha^*_k}=0,~\braket{\alpha^*_{k'}\alpha_k}=\frac{\delta_{kk'}}{2}
\end{equation}

Since our goal is the study of long-range order in time and space of the one-body correlation function $G(x,x',t,t')$, we restrict to large values of $|x-x'|,|t-t'|$, so 
\begin{equation}
\braket{\Psi^*_W(x,t)\Psi_W(x',t')}=\frac{\braket{\hat{\Psi}^{\dagger}(x,t)\hat{\Psi}(x',t')}+\braket{\hat{\Psi}(x',t')\hat{\Psi}^{\dagger}(x,t)}}{2}\simeq\braket{\hat{\Psi}^{\dagger}(x,t)\hat{\Psi}(x',t')}=G(x,x',t,t')
\end{equation}
where the commutator for sufficiently separated points in space-time
\begin{equation}\label{eq:fieldcommutator}
 \left[\hat{\Psi}(x',t'),\hat{\Psi}^{\dagger}(x,t)\right]=\left[\hat{\phi}(x',t'),\hat{\phi}^{\dagger}(x,t)\right]=
\frac{1}{2\pi}\int^{\infty}_{-\infty}\mathrm{d}k~ e^{ik(x-x')}e^{-i\Omega_k(t-t')}-\frac{i}{\pi }\int^{\infty}_{-\infty}\mathrm{d}k~ e^{ik(x-x')}\sin\Omega_k(t-t')|v_k|^2
\end{equation}
is neglected on similar grounds to those invoked in the case of phase fluctuations [see discussion after Eq. (\ref{eq:CorrelationFiniteTemperature})].

\begin{figure}[t]
\begin{tabular}{@{}cc@{}}
    \includegraphics[width=0.5\columnwidth]{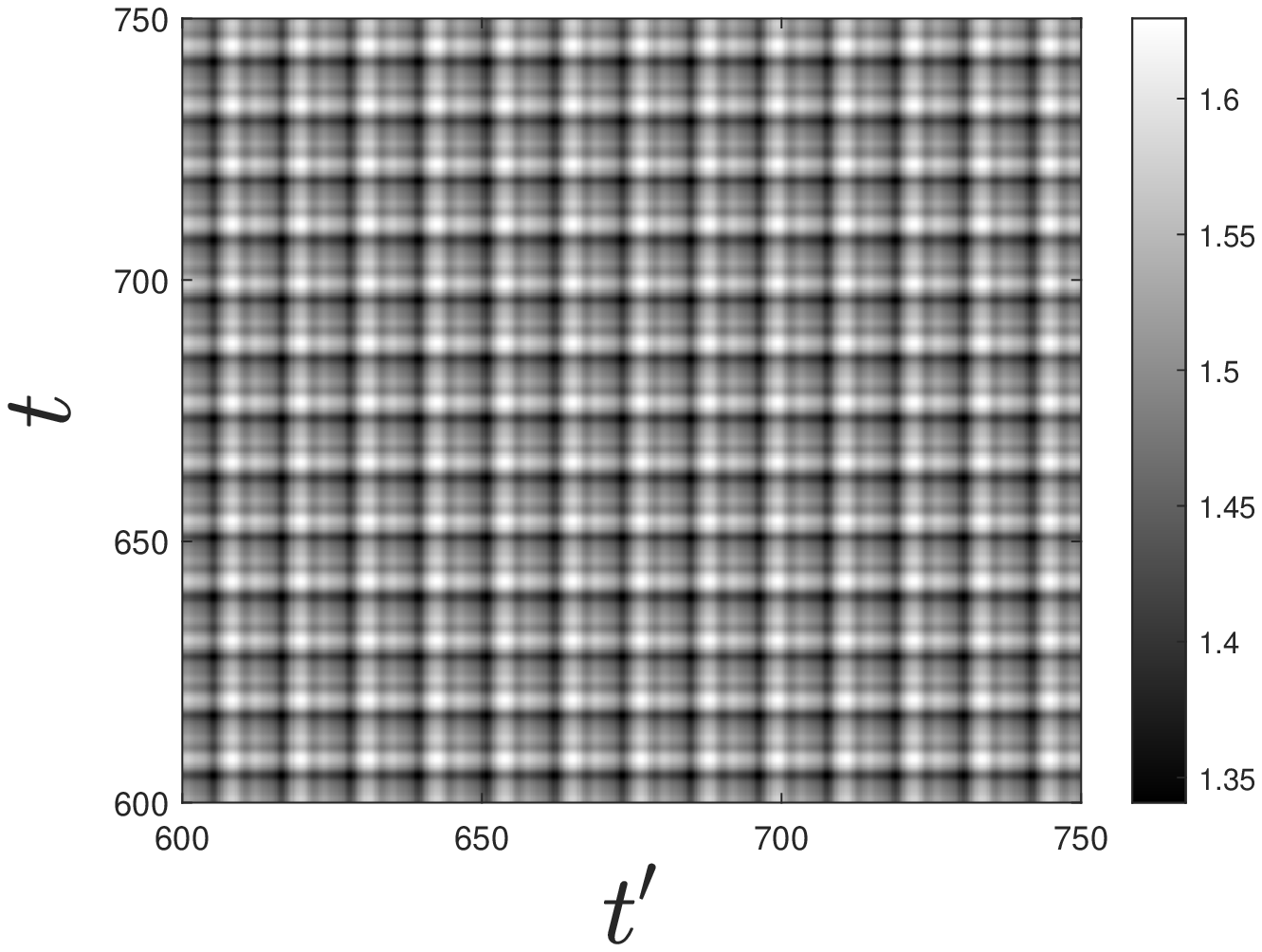} &  
    \includegraphics[width=0.5\columnwidth]{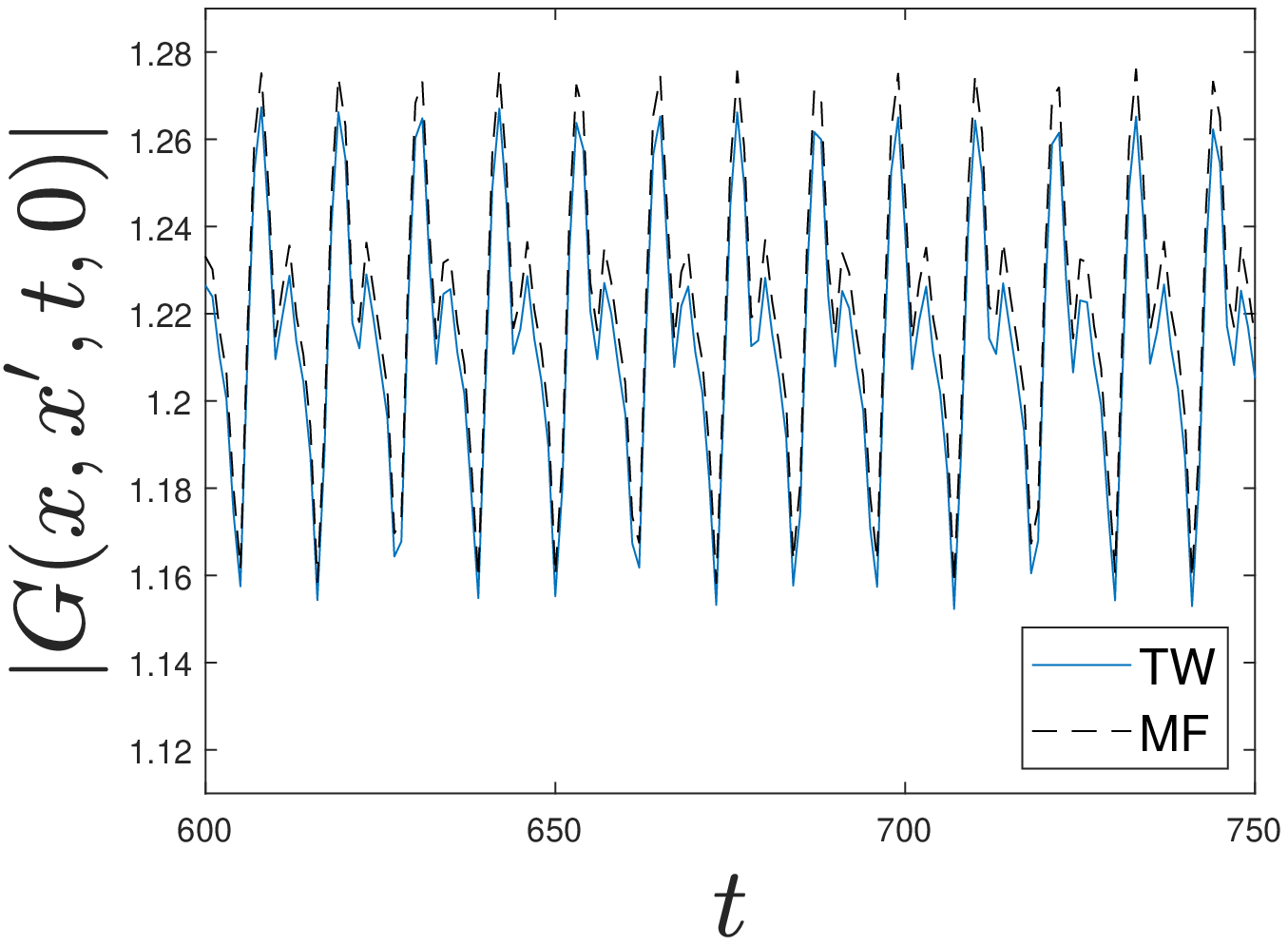}
\end{tabular}
\caption{Absolute value of the correlation function $|G(x,x',t,t')|$. Left: Mean-field value of $|G(x,x',t,t')|$ for $x=-30$ and $x'=30$ as a function of $(t,t')$. Right: Comparison of $|G(x,x',t,0)|$ for $-x=x'=100$ as a function of $t$ between a Truncated Wigner simulation that includes quantum fluctuations (solid blue) and the mean-field prediction (dashed-black).} 
\label{fig:TWA}
\end{figure}

For the simulations, we have chosen as mean-field parameters those of lower Fig. 2 of the main text, $v=0.95,~V_0=1,~X=2$. With respect to the specific parameters of the Truncated Wigner simulation, we have used $L\simeq 1885$ and $N=10^8$, with $n_0=N/L$ the condensate density. The number of modes is $N_m=3000$, which corresponds to a cut-off in $k$ space of $|k|<5$. Expectation values are evaluated after ensembles of $1000$ simulations.

Left Fig. \ref{fig:TWA}  shows the time dependence of the mean-field value of $|G(x,x',t,t')|$ for fixed spatial points far away, one upstream and one downstream. The one-body correlation function clearly exhibits long-range time-periodicity in both $t,t'$, revealing the time crystal character of the CES state. Quantum fluctuations, included via the Truncated Wigner method, do not destroy the time-periodic long-range order for typical times of experiments, thus allowing for a potential experimental observation. This can be seen in right Fig. \ref{fig:TWA}, where we compare the mean-field result for $|G(x,x',t,0)|$ with a Truncated Wigner simulation for long times. 

Technically, one should use a number conserving approximation in the computations, taking the number of particles in the condensate as $N_0=N-N_{\rm{NC}}$, with $N_{\rm{NC}}$ the number of non-condensed particles
\begin{equation}
    N_{\rm{NC}}=\sum_k \left(|u_k|^2+|v_k|^2\right)\braket{\hat{\alpha}_k^{\dagger}\hat{\alpha}_k}+|v_k|^2=\sum_k \left(|u_k|^2+|v_k|^2\right)\left(|\alpha_k|^2-\frac{1}{2}\right)+|v_k|^2
\end{equation}
This is translated into a modified version of Eq. (\ref{eq:TWExpansionNotConserving}),
\begin{equation}\label{eq:TWExpansionConserving}
\Psi_W(x,0)=\sqrt{1-\frac{N_{\rm{NC}}}{N}+\frac{\delta n_W(x)}{n_0}}e^{ivx}e^{i\delta \theta_W(x)}
\end{equation}
For the parameters considered, we have not found any significant effect resulting from the use of a non-conserving approximation.

\bibliographystyleSupp{apsrev4-1}
\bibliographySupp{Hawking}

\end{document}